\documentclass[aps,physrev,groupedaddress]{revtex4-2}
\usepackage[utf8]{inputenc}
\usepackage[margin=1.0in]{geometry}
\usepackage{xcolor}
\usepackage{algorithm}
\usepackage{algpseudocode}
\usepackage{mathtools}
\usepackage[draft,authormarkup=none]{changes}
\usepackage{amsmath}
\usepackage{subcaption}
\usepackage[citecolor=blue]{hyperref}
\usepackage{multirow}
\usepackage{hhline}
\usepackage{booktabs}
\usepackage{chngcntr} %for appendix figures

\usepackage{subeqnarray}
\usepackage{amsmath, amssymb}

\usepackage{caption}
\usepackage{graphicx}
\usepackage{epsfig, epstopdf}
\usepackage{overpic}
\usepackage[usestackEOL]{stackengine}
\usepackage{float}
\usepackage[section]{placeins}
\usepackage{svg}
\hypersetup{
    colorlinks=true,
    linkcolor=blue,
    filecolor=magenta,      
    urlcolor=blue,
    pdftitle={Overleaf Example},
    pdfpagemode=FullScreen,
    }

\newcommand{\rmH}{\mathrm{H}}

\newcommand{\rmL}{\mathrm{L}}
\newcommand{\Reynolds}{\mathrm{Re}}
\newcommand{\Weber}{\mathrm{We}}
\newcommand{\caseHH}{Re$_\rmH$We$_\infty$}
\newcommand{\caseHM}{Re$_\rmH$We$_{20}$}
\newcommand{\caseHL}{Re$_\rmH$We$_{10}$}
\newcommand{\caseLH}{Re$_\rmL$We$_\infty$}
\newcommand{\caseLM}{Re$_\rmL$We$_{20}$}
\newcommand{\caseLL}{Re$_\rmL$We$_{10}$}

\usepackage[draft,authormarkup=none]{changes}
\usepackage{amsmath}

\usepackage{subcaption}

\usepackage[citecolor=blue]{hyperref}
\usepackage{multirow}

\usepackage{tikz}
\usetikzlibrary{positioning}
\usepackage{transparent}
\usetikzlibrary{backgrounds, shadows.blur}

\usepackage[most]{tcolorbox}

\newif\ifdraftnotes
\draftnotestrue
\begin{document}
\preprint{APS}

%\title{Surface-based sensing of near-surface turbulence statistics in free-surface turbulence using a shallow recurrent decoder}
\title{
Predicting 
%Reconstructing 
subsurface turbulence statistics from free-surface observations across flow regimes
 using a shallow neural network
}
\author{Kristoffer S. Moen$^1$}
\author{Simen {\AA}. Ellingsen$^1$}
\author{J. Nathan Kutz$^2$}
\author{J{\o}rgen R. Aarnes$^1$}
\affiliation{$^1$Norwegian University of Science and Technology, Trondheim, Norway}
\affiliation{$^2$Autodesk Research, 6 Agar Street, London, UK}

\begin{abstract}

There is considerable interest in the remote sensing of natural flows from surface observations, due to the cost and difficulty of making measurements in situ. 
Here we demonstrate that key statistical properties of the turbulent flow beneath a free surface may be predicted from observations of the free surface only, by applying the SHallow REcurrent Decoder (SHRED).
%We reconstruct time-resolved subsurface turbulence statistics from free-surface observations 
%using the SHallow REcurrent Decoder (SHRED), evaluating temporally separated test intervals and generalization capabilities across flow conditions. 
%Reliable cost-efficient inference of turbulence statistics beyond the training data remains insufficiently established, limiting their use in remote characterization of near-surface motions relevant to gas and heat exchange. 
Using six datasets from direct numerical simulations spanning two Reynolds numbers and three Weber numbers, we infer plane-averaged statistics of horizontal divergence, vertical vorticity and turbulent kinetic energy from the surface curvature and flow structures detected at the free surface. Near-surface divergence and vorticity statistics are accurately inferred, while mean TKE is somewhat harder to reconstruct. The accuracy generally decreases with depth.
%The three turbulence quantities are inferred accurately near the surface, while temporal correlation generally decreases with depth.
When models are deployed to new parameter regimes, near-surface performance retains high accuracy, in particular for temporal correlations. An observed increase in normalized mean-square error mainly reflects an offset between the reconstructed and true interval means. The magnitude of the offset depends on the specific flow the model is trained on and deployed to, respectively, rather than the direction of change in Reynolds or Weber number. %Reconstruction retains substantial skill across the investigated flows, even for flows in new parameter regimes. 
Temporal correlation varies mainly from one target interval to another and tends to be higher in intervals with greater relative fluctuations in mean-square surface curvature. %, whereas amplitude errors show a larger source-model contribution. %NMSE also depends on source--target interactions. 
Data assimilation at inference time slightly reduces this mean offset. %is applied to improve the across-case reconstructions, which yields moderate overall improvements.% data assimilation improves variance fidelity more consistently than temporal correlation or NMSE, with near-surface horizontal divergence gains associated with stronger and more temporally variable surface-curvature signals. Its effects remain mixed across quantities and depths. These results demonstrate the potential for surface-based reconstruction of evolving subsurface turbulence statistics and identify conditions associated with successful temporal reconstruction and beneficial adaptation across flows.

\end{abstract}

\maketitle

%%%%%%%%%%%%%%%%%%%%%%%%%%%%%%%%%%%%%%%%%%%
%%%%%%%%%%%%%% S E C T I O N %%%%%%%%%%%%%%
%%%%%%%%%%%%%%%%%%%%%%%%%%%%%%%%%%%%%%%%%%%

\section{Introduction}
%%%%%%%%%%%%%%%%%%%%%%%%%%%%%%%%%%%%%%%%%%%%%%%%%%%%%
%%%%%%%%%%%%%% I N T R O D U C T I O N %%%%%%%%%%%%%%
%%%%%%%%%%%%%%%%%%%%%%%%%%%%%%%%%%%%%%%%%%%%%%%%%%%%%

% establishing air-water exchange

Turbulence near the air--water interface plays an important role in the exchange of gases and heat between water bodies and the atmosphere. By transporting fluid towards and away from the interface, turbulent motions renew the near-surface fluid and influence the gradients through which molecular transfer occurs \cite{Kermani_2011,Turney_Banerjee_2013}. Quantifying the turbulence near and at the surface is therefore important for understanding interfacial exchange, but direct measurements of relevant flow dynamics at, and especially beneath, the water surface are challenging. However, as recent research has demonstrated, surface deformations produced by the underlying turbulence offer a possible means of observing the otherwise hidden dynamics on the water side of an air--water interface \cite{Babiker_2023,babiker2026,ruth2026}.

% Surface divergence in particular

One particularly useful measure of the water-side turbulence is the surface divergence, and its subsurface variant, the horizontal divergence of the fluid velocities (definitions in \S\ref{sec:methodology}), which measure how fluid is locally transported upwards or downwards in the water column and renews the surface with water from  the bulk. The root-mean-square of the surface divergence is used in surface-divergence models to estimate the gas-transfer velocity at the interface, providing a quantitative connection between turbulence statistics and interfacial transport \cite{banerjee_2004,Turney_Smith_Banerjee_2005,asher_2012}. Moreover, the mean-square of the horizontal divergence is among the fields which has been coupled directly to observables at the surface: 
Babiker et al.~\cite{Babiker_2023} linked the number of dimples (surface-attached vertical vortices) to the mean-square surface divergence. Subsequent experiments showed that relationships between the spatial aggregate of dimples and scars (valley-like patterns at the surface, originating from horizontal vortices below \cite{aarnes2025}) and the subsurface horizontal divergence can persist with depth, even as local correspondence weakens \cite{babiker2026}. Ruth and Coletti \cite{ruth2026} further related the surface curvature to near-surface velocity gradients and found a scaling between root-mean-square curvature and surface-divergence variance. In turn, surface divergence and vertical vorticity relate to the turbulent dissipation rate \cite{Qi2025}. With a well-chosen window for moving averaging \cite{aarnes2025,kjellevold2026}, maps of estimated subsurface features could be drawn based on surface measurements alone, holding promise for remote sensing applications. Together, these findings motivate investigating whether surface observations can recover the evolving intensity of subsurface turbulence through spatially averaged statistics.

In the present study, we consider the sensing problem of determining what turbulence statistics observations of surface deformations reveal, how deep beneath the surface this information extends, and whether the learned dynamics remains useful across different flow conditions. As observables for the surface, we use the mean-square surface curvature and the fraction of the surface area occupied by dimples and scars, the latter likely being readily inferable from simple optical surface measurements with computer vision \cite{Babiker_2023,aarnes2025,babiker2026}. From these, we infer three complementary measures of subsurface turbulence: mean-square horizontal divergence, mean-square vertical vorticity and mean (i.e., plane averaged) turbulent kinetic energy. Mean-square horizontal divergence measures the intensity of horizontal spreading and convergence, which near the surface are associated with upwelling and downwelling motions; mean-square vertical vorticity characterizes rotation in the horizontal plane; mean turbulent kinetic energy measures the energy of velocity fluctuations. Each quantity is averaged over horizontal planes but retains its dependence on time and depth, so that the reconstruction aims to follow fluctuations in turbulence intensity rather than only recover a long-time mean profile.

The physical relationships between surface signatures and subsurface statistics motivate our sensing approach but do not by themselves establish how reliably the temporal evolution of the turbulence statistics can be recovered. Doing so requires a method of reconstruction and inference that maps the information available in surface observations to the unmeasured subsurface quantities. Related problems have been studied in wall-bounded turbulence, where neural networks have been used to reconstruct velocity fluctuations away from the wall using wall pressure and shear stresses \cite{Guastoni2021, Cuellar2024}. For free-surface flows, Xuan and Shen \cite{xuan2023} reconstructed three-dimensional subsurface velocity fields from surface elevation and velocity measurements using a convolutional neural network (CNN) that outperformed linear stochastic estimation. However, training a CNN to reconstruct full velocity fields can be computationally expensive. When only turbulence statistics are needed, estimating them directly could reduce both training cost and inference time.

One should bear in mind that surface information alone need not uniquely determine the full subsurface flow.. 
In potential flow, exact knowledge of the velocity field on the boundaries determines the flow everywhere---the uniqueness theorem (e.g.\ \cite{Lamb1932})---whereas the same is not the case when the flow has vorticity. 
Turbulent structures in the bulk of the fluid may leave only a weak imprint on the surface, making them difficult to identify from surface observations alone. The question we consider here is how reliably the temporal evolution of plane-averaged turbulence statistics at different depths can nevertheless be reconstructed from those observations. The non-uniqueness makes the problem we consider a particularly interesting case for exploring the capabilities of machine learning.

%SHRED introduction

Beyond the spatial information available in individual observations, measurement histories can reveal the evolution of the underlying dynamics. The SHallow REcurrent Decoder (SHRED), introduced by Williams et al.\ \cite{williams2024}, uses temporal information in measurement histories to reconstruct unmeasured quantities from a limited number of sensors. A recurrent network processes the sensor histories into a latent representation, which a shallow decoder maps to the output statistics, either directly or through a reduced representation \cite{tomasetto2025reducedPub}. The measured and targeted quantities need not be the same physical variable, allowing accessible observations to reveal other coupled components of the system. Its compact architecture and demonstrated low training cost make SHRED attractive for reconstruction from sparse measurements.

Its capabilities have been explored in several multiphysics applications, including plasma dynamics \cite{Faraji2025}, nuclear-reactor state estimation \cite{riva2025}, experimental thermal-hydraulic sensing \cite{introini2025modelsexperimentsshallowrecurrent}, free-surface turbulence \cite{moen2026} and magnetohydrodynamic flows in fusion-reactor blankets \cite{loverso2026}. These studies demonstrate sensing of coupled physical quantities from limited observations, with the latter also examining reconstruction under previously unseen magnetic-field configurations. Extensions of SHRED have also addressed dynamical modeling and uncertainty in fluid reconstruction \cite{gao2026, uqshred2026}.

We have previously demonstrated that SHRED can reconstruct subsurface velocity fields from the histories of as few as three surface-elevation sensors in simulations and experiments of free-surface turbulence \cite{moen2026}. SHRED far outperformed sparse linear reconstruction, and also improved on dense linear reconstruction in experiments and at depth in simulations. However, successful reconstruction relied on temporally interspersed training and test samples, while reconstruction on temporally separated test intervals remained challenging. This limitation matters for deployment, where we rely on surface observations only, and subsurface measurements, if available at all, seldom overlap in time with the interval to be reconstructed. For applications where paired surface and subsurface data are insufficient for training a new model, a further challenge is whether models trained under different flow conditions can be reused.  
Such generalization across flow configurations and physical parameters remains a central issue for machine learning in fluid dynamics \cite{Taira2025}. 

In the present study we move from an interpolative towards extrapolative paradigm, using SHRED to predict the spatially averaged turbulence statistics for free-surface turbulent flows different from flows used in training. Limiting the targeted outputs to averaged statistics enables us to take on generalizations across temporal domains and flow configurations, since the targets retain information about the evolving intensity of the flow without requiring recovery of the detailed instantaneous velocity field. Our results thus represent a major step towards practical employment of pre-trained models for remote sensing use, a notoriously difficult challenge for turbulent flow.

We make use of direct numerical simulation data of free-surface turbulent flow, from simulations at the University of Minnesota \cite{guoInteractionDeformableFree2010, xuanConservativeSchemeSimulation2019}, previously published by Aarnes et al.\ \cite{aarnes2025b}.Two dimensionless numbers are varied across the six simulations: the Reynolds number  ($\Reynolds$) and the Weber number ($\Weber$)

\begin{equation}
    \Reynolds=UL/\nu; ~~ We= U^2L/\Upsilon,
\end{equation}
($U$: characteristic velocity scale, $L$: characteristic length scale, $\nu$: kinematic viscosity, $\Upsilon$: kinematic surface tension),
representing the balance of inertia to viscous damping, and the ability of surface tension to keep the surface flat ($\Weber=\infty$ means no surface tension, the surface motion being restricted by gravity only, characterized by the Froude number, which is the same in all our cases).  
All datasets are in the ``weak turbulence'' regime in the taxonomy of Brocchini \& Peregrine \cite{brocchini2001}, which means the free surface is only mildly deformed by sub-surface turbulence. Our results are not limited to the lowest-turbulence flows, however---similarly strong surface-to-subsurface correlations were found in experiment with $\Reynolds$ up to two orders of magnitude higher, inspiring confidence that the results we obtain herein are indicative of the level of success one can reasonably hope for also in natural turbulent free-surface flow, bearing in mind that such upscaling is a formidable next task and full-scale training data remains extremely scarce. Notwithstanding, as proof of concept, our results here are encouraging.

We examine in detail whether the temporal evolution of the three target quantities can be inferred from aggregate surface observations of the flows on temporally separated test intervals, and whether the learned relationship transfers across Reynolds- and Weber-number regimes without retraining. Using six direct numerical simulations spanning two Reynolds numbers and three Weber numbers, we evaluate inferences to temporally separated data from the same simulations (in-case reconstruction) and also deploy pretrained models to different datasets (across-case reconstruction). For across-case reconstruction, we assume that reference ranges of the surface and subsurface quantities are available from separate time intervals in the new flow. We examine how performance changes with the choice of depth and reconstructed quantity, and distinguish variability associated with the source-model realization, the target interval, and their interaction. We also investigate whether inference-time data assimilation of the observed surface quantities through data-assimilated SHRED (DA-SHRED) \cite{Bao2025} improves reconstruction in previously unseen flows while keeping the trained network parameters fixed. 

The paper is organized as follows: In \S \ref{sec:methodology} we present the reconstruction problem, the datasets, the SHRED method, and the metrics we use to assess its performance. In \S \ref{sec:resultsAnddiscussion}, we present the results for in-case and across-case reconstructions, the latter with and without data assimilation. We present our conclusions in \S \ref{sec:conclusions}, followed by appendices with additional details. 

\FloatBarrier

%%%%%%%%%%%%%%%%%%%%%%%%%%%%%%%%%%%%%%%%%%%
%%%%%%%%%%%%%% S E C T I O N %%%%%%%%%%%%%%
%%%%%%%%%%%%%%%%%%%%%%%%%%%%%%%%%%%%%%%%%%%

%%%%%%%%%%%%%%%%%%%%%%%%%%%%%%%%%%%%%%%%%%%
%%%%%%%%%%%%%% S E C T I O N %%%%%%%%%%%%%%
%%%%%%%%%%%%%%%%%%%%%%%%%%%%%%%%%%%%%%%%%%%
\section{Data and methods}
\label{sec:methodology}
%%%%%%%%%%%%%%%%%%%%%%%%%%%%%%%%%%%%%%%%%%%%%%%%%%%
%%%%%%%%%%%%%% M E T H O D O L O G Y %%%%%%%%%%%%%%
%%%%%%%%%%%%%%%%%%%%%%%%%%%%%%%%%%%%%%%%%%%%%%%%%%%

%%%%%%%%%%%%%%%%%%%%%%%%%%%%%%%%%%%%%%%%%%%%%%%%%%%%%%%%%%%%%%%%%%%%%%%%%%%%%%%%
%%%%%%%%%%%%%% R E C O N S T R U C T I O N     P R O B L E M %%%%%%%%%%%%%%%%%%%
%%%%%%%%%%%%%%%%%%%%%%%%%%%%%%%%%%%%%%%%%%%%%%%%%%%%%%%%%%%%%%%%%%%%%%%%%%%%%%%%
\subsection{Reconstruction problem}
\label{sec:reconstruction_problem}
\begin{figure}[t!]
    \centering
    \begin{overpic}[
        width=0.98\linewidth,
        trim={0.0cm 0cm 0.0cm 0cm},
        clip
        ]{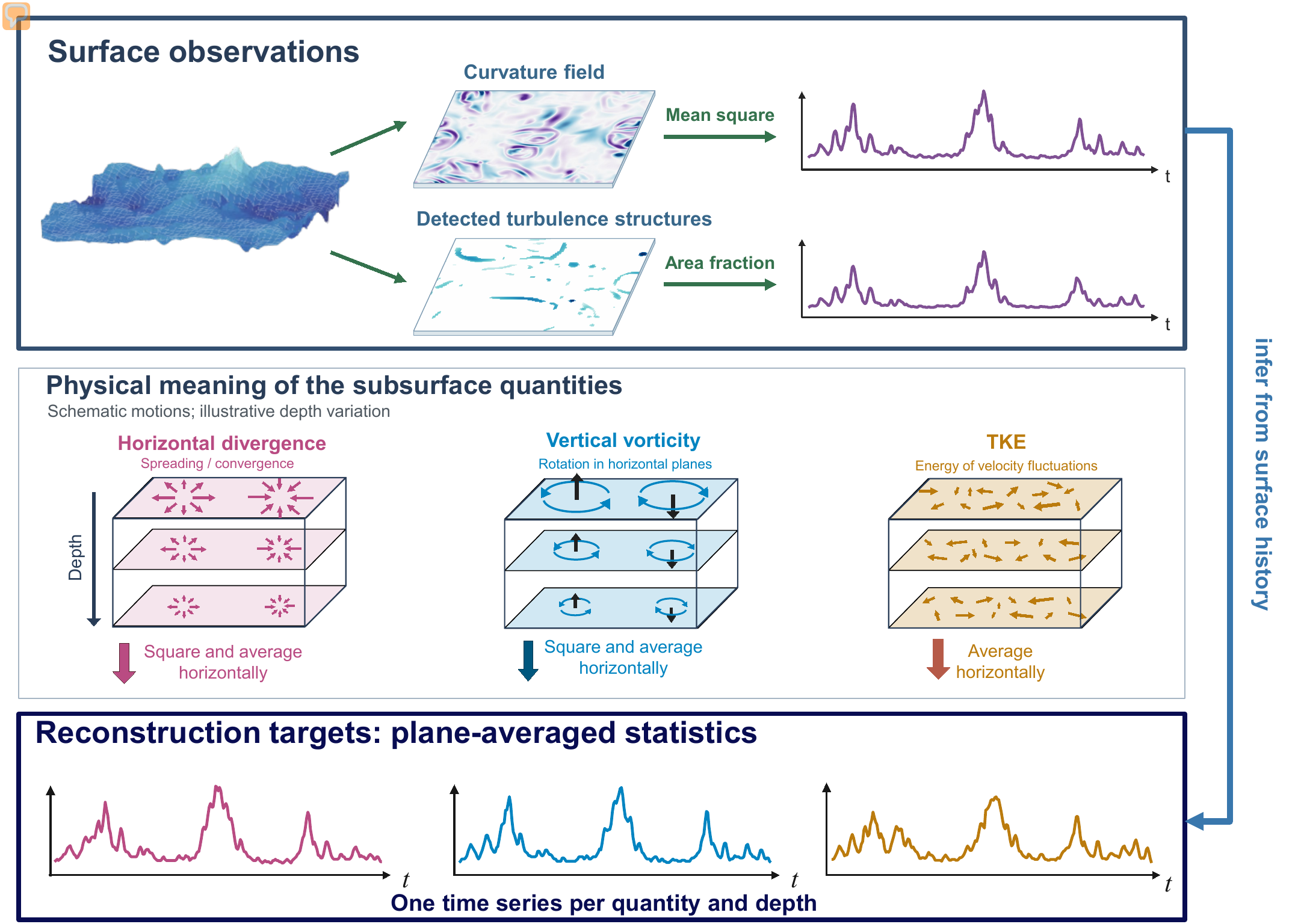}

        % ---- labels ----

        \put(64,65.3){ $i_{\kappa}(t)  = \left\langle \kappa^2(x,y,t) \right\rangle $}

        \put(64,52.3){ $i_s(t)  = \langle M_s(x,y,t)\rangle $}
        
        \put(5,11.3){ $q_{\beta}(z,t)  = \left\langle \beta^2(x,y,z,t) \right\rangle $}
        \put(35.5,11.3){ $q_{\omega}(z,t)  = \left\langle \omega_z^2(x,y,z,t) \right\rangle$}
        \put(64,11.3){ $q_k(z,t)  = \left\langle k(x,y,z,t) \right\rangle$}

    \end{overpic}
    \caption{A general flow chart showing the reconstruction of subsurface turbulence statistics from surface inputs.}
    \label{fig:shred_MS_reconstruction_scheme}
\end{figure}

Figure \ref{fig:shred_MS_reconstruction_scheme} illustrates the turbulence sensing problem we discuss in this paper. The top row depicts the input: We start from the full, time-resolved surface elevation field $\eta(x,y,t)$, from which we compute two time-resolved fields, the curvature field and the binary field of detected surface structures (dimples and scars). The curvature field emphasizes local deformation more strongly than the elevation itself and has been coupled to the water side turbulent dynamics, particularly surface divergence \cite{ruth2026}. The dimples and scars are closely connected to the subsurface vortical structures \cite{aarnes2025} and have shown close correlation with both the horizontal divergence and the vertical vorticity field near the surface \cite{babiker2026, kjellevold2026}. Since the surface slopes are small ($|\nabla_h\eta|^2 \ll 1$, where $\nabla_h$ denotes the horizontal gradient), the curvature is well approximated by $\kappa(x,y,t) \simeq \nabla_h^2 \eta(x,y,t)$. The surface structures were detected using a wavelet based algorithm that identifies persistent surface indentations and sorts them as dimples and scars according to eccentricity, see \cite{Babiker_2023} and \cite{aarnes2025} for details. 
Next, the input mean-square (MS) curvature and the area fraction of surface structures are computed according to
\begin{align}
    \label{eq:input_statistics}
    i_{\kappa}(t)  = \left\langle \kappa^2(x,y,t) \right\rangle \, , \qquad
    i_s(t)  = \langle M_s(x,y,t)\rangle \, ,
\end{align}
where $M_s(x,y,t)$ is a binary mask of the detected structures, the operator $\langle ...\rangle$ denotes taking the mean over the horizontal plane. We now have two time series that make up the input channels to our application of the SHRED network. 

The reconstructed statistics consist of three depth-resolved turbulence statistics that represent divergent, rotational, and energetic motions in the subsurface flow. The full-field versions of these are the horizontal divergence, $\beta(x,y,z,t) = \partial u / \partial x +\partial v / \partial y$, the vertical vorticity component, $\omega_z(x,y,z,t) = \partial v / \partial x - \partial u / \partial y$, and turbulent kinetic energy, $k(x,y,z,t) = \left( u'^2+v'^2 + w'^2\right)/2$. Note that for computation of the turbulent kinetic energy we first subtract the instantaneous plane mean, i.e. we do a Reynolds decomposition arriving at the fluctuating velocity components $u' = u - \left\langle u \right\rangle$ etc. 
This is done to counteract any mean flow modes that may show up in the datasets due to the periodic domain in horizontal directions of the simulations. 
At each depth $z$ and time $t$, the corresponding instantaneous field is reduced to a scalar quantity by averaging over the horizontal plane. This yields three time series as target statistics for each reconstruction plane:
\begin{align}
    \label{eq:output_statistics}
    q_{\beta}(z,t)  = \left\langle \beta^2(x,y,z,t) \right\rangle \, , \qquad
      q_{\omega}(z,t)  = \left\langle \omega_z^2(x,y,z,t) \right\rangle \, , \qquad
    q_k(z,t)  = \left\langle k(x,y,z,t) \right\rangle \, ,
\end{align}
where the $z$ coordinates follow the convention defined in \cite{aarnes2025}. For compactness, these targets are henceforth denoted MS horizontal divergence, MS vertical vorticity and mean TKE.
%\FloatBarrier

%%%%%%%%%%%%%%%%%%%%%%%%%%%%%%%%%%%%%%%%%%%%%%%%%%%%
%%%%%%%%%%%%%% F L O W   D A T A  %%%%%%%%%%%%%%%%%%
%%%%%%%%%%%%%%%%%%%%%%%%%%%%%%%%%%%%%%%%%%%%%%%%%%%%

%%%%%%%%%%%%%%%%%%%%%%%%%%%%%%%%%%%%%%%%%%%%%%%
%%%%%%% FLOW DATA AND SURFACE OBSERVABLES %%%%%
%%%%%%%%%%%%%%%%%%%%%%%%%%%%%%%%%%%%%%%%%%%%%%%

\subsection{Data}%Flow data and reconstruction quantities}
\label{sec:flow_data_main_method}

\subsubsection{DNS datasets}
\label{sec:DNS_cases}

The DNS datasets stem from simulations that mimic experiments of free-surface homogeneous turbulence where turbulence is generated far beneath a water surface and diffuses towards the surface through self-interaction of turbulence vortices and viscous diffusion, mimicking the set-up in a number of experiments \citep[see, e.g.,][]{brumley1987, herlina_experiments_2008, babiker2026, ruth2024}.
%Lacassagne2017, Variano2013, Jamin2024, 
%, ruth2026,Berhanu26}. 
The DNS datasets are available online for public use \cite{aarnes2025b} 
%(in partial format, due to storage limitations), 
and have seen use in several recent studies \cite{aarnes2025, moen2026, babiker2026}. We only include a brief description of the DNS here, and refer the reader to the original publication for details \cite{aarnes2025}.

In the simulations, the incompressible Navier--Stokes equations and continuity equations were solved on a three-dimensional domain. The boundary conditions were periodic in horizontal directions, free-slip at the bottom boundary and a free-surface tracked by an undulating grid on the top; the latter boundary enforced by the dynamic and kinematic boundary conditions. Turbulence was generated in the center region of the domain by random linear forcing, yet only data from the region nearest the free surface where no forcing occurs is used in the analysis. 

The DNS data consists of a total of six datasets, with two nominal Reynolds numbers and three nominal Weber numbers, all with the same Froude number. The mesh sizes of the datasets are \(128 \times 128 \times 348\) for the lower and \(256 \times 256 \times 660\) for the higher Reynolds number. While the nominal values are set from initial conditions, the Reynolds numbers and scales (integral length, velocity, etc.) of the turbulence are computed \emph{a posteriori}. An overview of the properties of the datasets can be found in Table \ref{tab:flowProp}. All properties in Table \ref{tab:flowProp} are computed from measurements outside the surface-influenced region. For our analysis, the most important properties are the integral length scale $L_\infty$, the integral time scale $T_\infty = L_\infty/u_{\rm rms}$, and the turbulent Reynolds number $\mathrm{Re}_\infty = 2L_\infty u_{\rm rms}/\nu$, where $\nu$ is the kinematic viscosity, $u_{\rm rms}$ the root-mean-square velocity fluctuations, $l=2L_\infty$ the turbulent macroscale \cite{tennekes1972first}. The turbulent property conventions and measurement depths follow \cite{aarnes2025}, which may be consulted for details; a different convention, more appropriate for direct comparison with experimental data, is described in \cite{babiker2026}.

\begin{table}[t]
    \begin{center}
    \caption{Flow and turbulence properties. From left: Dataset identifier, nominal Reynolds number, nominal Weber number, nominal Froude number, turbulent Reynolds number, Taylor Reynolds number, integral length scale, Taylor length scale, Kolmogorov length scale and integral time scale. Length and time scales are normalized by the simulation's characteristic length and velocity. For a discussion of comparison of nominal dimensionless numbers with dimensional units, see Appendix A of \cite{babiker2026}.}
    \begin{tabular}{c c c c c c c c c c c}
    \hline
    Dataset name & $\mathrm{Re}$  & $\mathrm{We}$ & $\mathrm{Fr}$   & $\mathrm{Re}_\infty$   & $\mathrm{Re}_\lambda$ &$L_\infty$     & $\lambda_T$ & $L_K$ & $T_\infty$\\ 
    \hline
    \caseLL  & 1000  & 10        & 0.1   &  150  & 47 & 0.59 & 0.37 & 0.027 & 4.6\\
    \caseLM  & 1000  & 20        & 0.1   &  164  & 50 & 0.61 & 0.37 & 0.027 & 4.6\\ 
    \caseLH  & 1000  & $\infty$  & 0.1   &  131  & 44 & 0.54 & 0.36 & 0.028 & 4.4\\   
    \caseHL  & 2500  & 10        & 0.1  &  452  & 82 & 0.63 & 0.23 & 0.013 & 4.4\\
    \caseHM  & 2500  & 20        & 0.1   &  368  & 74 & 0.57 & 0.23 & 0.0135 & 4.4\\
    \caseHH  & 2500  & $\infty$  & 0.1   &  433  & 81 & 0.64 & 0.24 & 0.0135 & 4.7\\
    \hline
    \end{tabular}
    \label{tab:flowProp}
    \end{center}
\end{table}

\subsubsection{Dataset statistics}
\label{sec:data_description}

Table~\ref{tab:global_dataset_statistics} presents a characterization of the surface observables and the subsurface target statistics of each dataset, by three descriptors computed over the full time series of the plane-averaged quantities.
Here, $\mu^*$ is the temporal mean of each quantity relative to that of the corresponding quantity for the reference dataset \caseHH. Note that values for different quantities can be compared as relative changes between flows, but not as magnitudes. 
The coefficient of variation, $\mathrm{CV}=\sigma/\mu$, is the size of temporal fluctuations (measured by temporal standard deviation $\sigma$) relative to the temporal mean, $\mu$. 
Finally, $P_{99}/P_{95}$ is the ratio of the 99th to the 95th percentile, measuring how far the most extreme events exceed the merely active periods. For the output quantities, the descriptors are computed at a reference depth of $z=-\pi/3$, fixed for all datasets (see \cite{aarnes2025}).
%each depth and averaged over $-\mathrm{XX} \le z/L_\infty \le 0$.

\newcommand{\rmd}{\mathrm{d}}
\newcommand{\Var}{\mathrm{Var}}
\newcommand{\Std}{\mathrm{Std}}

The statistical descriptors identify statistical differences between the six datasets, particularly for MS curvature at the surface and the area fraction of surface structures. The former shows pronounced variations in magnitude, relative variability, and upper-tail amplification, while the latter is generally larger for the lower--Reynolds number flows and, within each Reynolds-number group, for the datasets with zero surface tension ($\Weber=\infty$). For the subsurface quantities, the largest mean levels are generally found for the high-$\Reynolds$ cases, while their CV and upper-tail descriptors remain broadly similar. Mean TKE exhibits lower relative temporal variability and weaker upper-tail amplification than either horizontal divergence or vertical vorticity. This statistical characterization becomes useful when we consider model performance in across-case reconstructions. 

\begin{table*}[htb!]
\centering
\small
\begingroup
\renewcommand{\arraystretch}{1.16}
\setlength{\tabcolsep}{5pt}
\caption{Global statistical characterization of the six flow datasets, calculated at reference depth. For each quantity $q$, the temporal mean $\mu$ is given as normalized by the corresponding $\mu$ of the \caseHH{} dataset. The coefficient of variation $\mathrm{CV}=\sigma/\mu$ describes temporal variability relative to the mean, while $P_{99}/P_{95}$ characterizes the increase from high-activity to far-upper-tail values.}

\label{tab:global_dataset_statistics}
\begin{tabular}{llcccccc}
\toprule
Quantity & Descriptor & \caseLL{} & \caseLM{} & \caseLH{} & \caseHL{} & \caseHM{} & \caseHH{} \\ 
\midrule
Surface MS curvature  & $\mu^*$ & 0.687 & 0.652 & 0.891 & 0.264 & 0.204 & 1.000\\
 & CV & 2.601 & 1.512 & 1.777 & 1.180 & 1.435 & 1.220\\
 & $P_{99}/P_{95}$ & 4.294 & 2.525 & 1.917 & 1.873 & 3.164 & 1.400\\
%\hline
\addlinespace[5pt]
Structure area fraction & CV & 1.184 & 0.968 & 1.201 & 0.879 & 1.034 & 0.918\\
 & $P_{99}/P_{95}$ & 1.756 & 1.402 & 1.400 & 1.570 & 2.133 & 1.411\\
\midrule
MS horizontal divergence & $\mu^*$ & 0.371 & 0.416 & 0.356 & 1.228 & 0.993 & 1.000\\
 & CV & 0.488 & 0.435 & 0.533 & 0.454 & 0.474 & 0.428\\
 & $P_{99}/P_{95}$ & 1.434 & 1.272 & 1.320 & 1.280 & 1.287 & 1.375\\
%\hline
\addlinespace[5pt]
MS vertical vorticity & $\mu^*$ & 0.345 & 0.380 & 0.330 & 1.231 & 0.986 & 1.000\\
 & CV & 0.512 & 0.457 & 0.548 & 0.456 & 0.478 & 0.434\\
 & $P_{99}/P_{95}$ & 1.550 & 1.579 & 1.377 & 1.194 & 1.301 & 1.408\\
%\hline
\addlinespace[5pt]
mean TKE &  $\mu^*$ & 0.828 & 0.946 & 0.833 & 1.127 & 0.930 & 1.000\\
 & CV & 0.372 & 0.348 & 0.437 & 0.338 & 0.378 & 0.374\\
 & $P_{99}/P_{95}$ & 1.391 & 1.343 & 1.275 & 1.166 & 1.304 & 1.146\\
\bottomrule
\end{tabular}
\endgroup
\end{table*}

%%%%%%%%%%%%%%%%%%%%%%%%%%%%%%%%%%%%%%%%%%%%%%%%%%%%%%%%%%%%%%%%%%%%%%%%%%%%%%%%%%%%%%%%%%%%%%%%%%
%%%%%%%%%%%%%%%%%%%%%%%%%%%%%%%%%%%%%%%%   S H R E D   %%%%%%%%%%%%%%%%%%%%%%%%%%%%%%%%%%%%%%%%%%%
%%%%%%%%%%%%%%%%%%%%%%%%%%%%%%%%%%%%%%%%%%%%%%%%%%%%%%%%%%%%%%%%%%%%%%%%%%%%%%%%%%%%%%%%%%%%%%%%%%

\subsection{Reconstruction model}
\subsubsection{Shallow recurrent decoder}
\label{sec:shred}

\begin{figure}[h!]
    \centering
    %trim={2.7cm 0cm 0cm 1cm}
    \begin{overpic}[
        width=0.99\linewidth,
        trim={0cm 0cm 0cm 0cm},
        clip
    ]{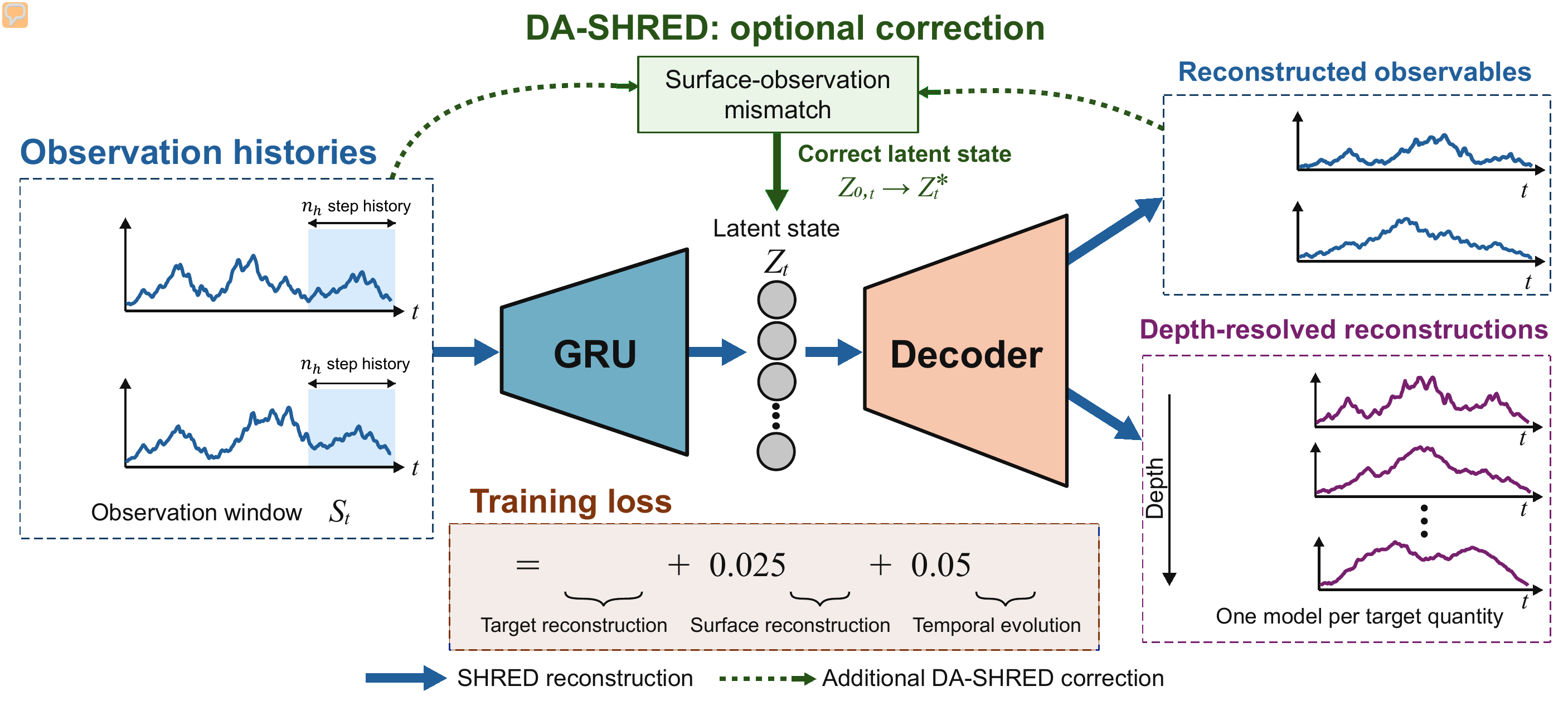}
    
    \put(2.3,27.9){ $i_{\kappa}(t) $}
    \put(2.3,17.4){ $i_s(t) $}

    %\put(19.5,32){\fontsize{7}{8}\selectfont $n_h$}
    %\put(19.5,22){\fontsize{7}{8}\selectfont $n_h$}
    
    \put(76,35.7){ $\hat{i}_{\kappa}(t) $}

    \put(76,29.5){ $\hat{i}_s(t) $}

    \put(76,19.5){\fontsize{7}{8}\selectfont $\hat{q}(z_0,t)$}
    \put(76,15.0){\fontsize{7}{8}\selectfont $\hat{q}(z_1,t)$}
    \put(75.2,9.0){\fontsize{7}{4}\selectfont $\hat{q}(z_{N_z-1},t)$}

     \put(29.7,8.7){\large $\mathcal{L}$}
    \put(36.2,8.7){\large $\mathcal{L}_{\mathrm{out}}$}
    \put(50.7,8.7){\large $\mathcal{L}_{\mathrm{in}}$}
    \put(62.4,8.7){\large $\mathcal{L}_{\mathrm{\Delta t}}$}
    \end{overpic}
    \caption{ Flow chart showing the general SHRED architecture for sensing turbulence statistics. Observable time series inputs are fed to the GRU recurrent network with a time-lagged embedding. The learned temporal dynamics is then mapped to a latent space representation $\mathbf{Z}_t$. A shallow decoder then maps the latent space to the target statistics at each depth, i.e. predicted turbulent quantities, following the blue path. To further optimize predictions at inference, a data-assimilation procedure is employed, using observable inputs as optimization targets fed back to the latent space to update the latent variables based on test-time surface observations. The updated latent state is then mapped to the predicted turbulence quantities. This data-assimilation feedback loop is represented by the green path. }
    \label{fig:shred_flowchart}
\end{figure}

The \emph{SHallow REcurrent Decoder} (SHRED) is a lightweight sensing architecture that maps the temporal history of a limited set of observable quantities onto a larger state of interest \cite{williams2024}. SHRED combines a recurrent neural network, which encodes the learned temporal dynamics of the observations into a latent representation, with a shallow decoder that maps this representation onto the reconstructed state. A more detailed theoretical description of SHRED is given by Williams et al.~\cite{williams2024}%., while its application to free-surface turbulence sensing was previously demonstrated in Ref.~\cite{moen2026}.

Figure \ref{fig:shred_flowchart} illustrates SHRED applied to the reconstruction problem detailed in §\ref{sec:reconstruction_problem}. We let $\mathbf{s}_t = \left[i_{\kappa}(t) i_s(t)\right]^T$ denote the combined vector of surface observables. Rather than reconstructing the subsurface state from the instantaneous surface observation alone, SHRED uses a time-lagged embedded sequence of the preceding $n_h$ observations, $\mathbf{S}_t =\left\{\mathbf{s}_{t-n_h+1},\ldots,\mathbf{s}_{t}\right\}$.
We set $n_h=100$ time steps, which is of $\mathcal{O}(T_\infty)$ for all datasets ($\Delta t$ between snapshots is 0.08 and 0.06 for $\mathrm{Re_L}$ and $\mathrm{Re_H}$ datasets, respectively). The temporal sequence is encoded by a recurrent network $\mathcal{R}_{\theta}$ into a latent representation, $\mathbf{Z}_t = \mathcal{R}_{\theta}\left(\mathbf{S}_t\right)$,
which is subsequently mapped by a shallow decoder $\mathcal{D}_{\phi}$ onto the reconstructed output, $    \hat{\mathbf{y}}_t = \mathcal{D}_{\phi}\left( \mathbf{Z}_t \right)$. In the present implementation, $\mathcal{R}_{\theta}$ consists of a two-layer gated recurrent unit (GRU) network with a hidden-state dimension of 64. The latent representation $\mathbf{Z}_t$ is taken from the final GRU output state of the input sequence. Moreover, we denote the surface and subsurface components of the fixed decoder $\mathcal{D}_{\phi}$ by $\mathcal{D}_s$ and
$\mathcal{D}_q$, respectively. % 
The latent state is passed to a shallow fully connected decoder with hidden layer widths of 350 and 400, ReLU activations, and a dropout probability of 0.1.

Separate SHRED models are trained for MS horizontal divergence, MS vertical vorticity, and mean TKE, yet each individual model reconstructs all depth planes simultaneously for its target output. The decoder output is written as $\hat{\mathbf{y}}_t = \left[ \hat{\mathbf{s}}_t^T,\,\hat{q}(z_0,t),\,\ldots,\,\hat{q}(z_{N_z-1},t)\right]^T$, 
where $\hat{\mathbf{s}}_t$ contains reconstructions of the surface observables and $\hat{q}(z,t)$ denotes the reconstructed turbulence statistic at the sampled depths $z_0,\ldots,z_{N_z-1}$, where $N_z$ is the number of planes. The surface-observable outputs serve as auxiliary training targets and enable latent-state correction during DA-SHRED deployment, as detailed below.%described in 

\subsubsection{Data-assimilation SHRED}
\label{sec:da_shred_theoretical}

After training and validation, the SHRED network is  deployed to reconstruct the time series of MS horizontal divergence, MS vertical vorticity and mean TKE, from the time-lagged sequence of MS curvature and area fraction of surface structures. Although the network parameters remain fixed at the deployment stage, surface observations may be used to correct the reconstructed statistics. We apply latent-state correction by the data-assimilation (DA) SHRED approach, introduced by
Bao and Kutz \cite{Bao2025}. 

In the present implementation, the surface observables are included among the decoder outputs as $\hat{\mathbf{s}}_t$, as described above. The mismatch between the true surface observables $\mathbf{s}_t$ and $\hat{\mathbf{s}}_t$ guides the correction of the latent state. The same latent state is used to reconstruct the subsurface statistics, so the correction also modifies the subsurface reconstruction during the deployment step.

This correction is applied at each time step $t$, starting from the latent state $\mathbf{Z}_{0,t}$ produced by the GRU from the observation history $\mathbf{S}_t$. We adjust this state to improve the agreement between the decoded surface quantities and the current observations, while penalizing large deviations from $\mathbf{Z}_{0,t}$. The trained network parameters remain fixed. The corrected latent state is computed from the optimization

\begin{equation}
    \mathbf{Z}^{*}_t
    =
    \underset{\mathbf{Z}}{\arg\min}
    \left[
    \operatorname{MSE}
    \left(
        \mathcal{D}_{s}(\mathbf{Z}),
        \mathbf{s}_t
    \right)
    +
    \lambda_{\mathrm{DA}}
    \operatorname{MSE}
    \left(
        \mathbf{Z},
        \mathbf{Z}_{0,t}
    \right)
    \right],
    \label{eq:da_shred}
\end{equation}
where $\mathcal{D}_{s}(\mathbf{Z})$ denotes the components of the trained decoder corresponding to the reconstructed surface observables, $\mathbf{s}_t$ denotes the observed surface state at time $t$, and $\lambda_{\mathrm{DA}}$ regularizes the correction by penalizing large deviations from the uncorrected latent state $\mathbf{Z}_{0,t}$. In the present implementation, Eq.~(\ref{eq:da_shred}) is approximately minimized at each time step using 100 Adam iterations, with learning rate and $\lambda_{\mathrm{DA}}$ set by hyperparameter tuning on in-case validation data (details in App.~\ref{app:da_shred}).
The optimized state $\mathbf{Z}^{*}_t$ is subsequently passed
through the fixed decoder to reconstruct the target
turbulence-statistic profile, $
%\begin{equation}
    \hat q_{\mathrm{DA}}(z,t)
    =
    \left[\mathcal{D}_{q}(\mathbf{Z}^{*}_t)\right]_z,
%\end{equation}
$ where $\mathcal{D}_{q}$ denotes the decoder outputs that corresponds
to the target statistic, and $[\cdot]_z$ selects the output
at depth $z$.

%%%%%%%%%%%%%%%%%%%%%%%%%%%%%%%%%%%%%%%%%%%%%%%%%%%%%%%%%%%%%%%%%%%%%

%%%%%%%%%%%%%%%%%%%%%%%%%%%%%%%%%%%%%%%%%%%%%%%
%%%%%% TRAINING STRATEGY AND LOSS FUNCTION %%%%
%%%%%%%%%%%%%%%%%%%%%%%%%%%%%%%%%%%%%%%%%%%%%%%
%\FloatBarrier

\subsection{Model training and deployment}
\label{sec:training_strategy}

%\subsubsection{Temporal segmentation and preprocessing}

\begin{figure}[h!]
        \centering
        \includegraphics[width=0.9\linewidth]{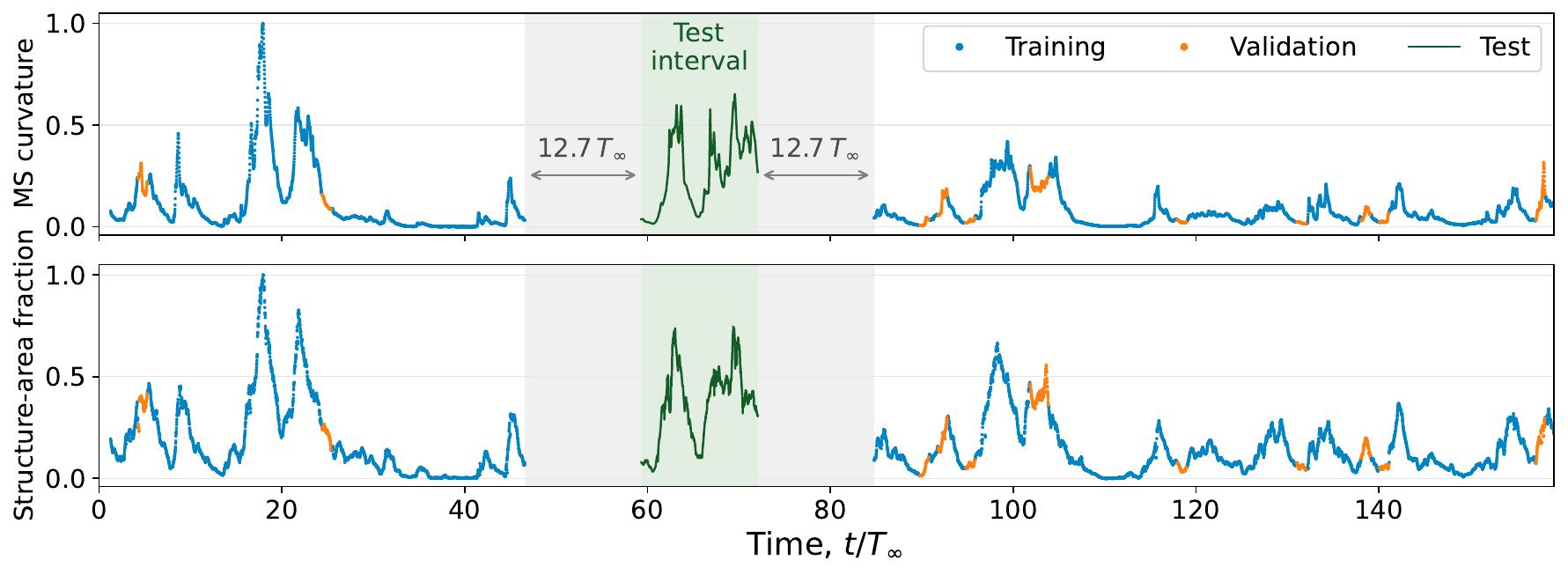}
        \caption{Surface input time series and example data partition for in-case reconstruction of \caseHH. % at $\mathrm{Re}=2500$ and $\mathrm{We}=\infty$. 
        Upper and lower panels show the min–max scaled MS curvature and structure-area fraction, respectively. Interspersed training and validation samples are shown in blue and orange, while the test interval is shown in green. Grey bands indicate exclusion intervals.}% of duration $10T_{\mathrm{\infty}}$ on either side of the test interval, which separate the test data from the training and validation data.}
        \label{fig:input_segmentation}
\end{figure}

%Each flow case corresponds to one combination of Reynolds and Weber numbers. 
We train and validate SHRED models on (parts of) each of the DNS datasets, one dataset at a time. The validation consists of randomly selected blocks with time duration $T_\infty$: Blocks from the training/validation data are interspersed, with each block having a probability of 0.15 of being selected as a validation data block.
After training and validation we apply two different strategies: (1) \textit{in-case} and (2) \textit{across-case} reconstruction. In-case reconstruction is our reference, and consists in reconstructing the output statistics for a held-out test interval of the same dataset that the model is trained and validated on. Importantly, the test interval is temporally separated from the training and validation data---an example of such a data split is shown in Fig. \ref{fig:input_segmentation}. Each test interval spans 1000 snapshots ($\approx$13\,$T_\infty$--18$\,T_\infty$) and is separated from the training and validation data by exclusion intervals of the same duration. Note the input observables and all output statistics are temporally aligned and the data is min--max scaled using parameters fitted to the training data only.  
We use multiple train--validation--test splits and train an ensemble of networks for each split, to counter the arbitrariness inherent in the models from both the data split
and from network training. Within each ensemble, the data remain fixed, while stochastic training procedures result in different training realizations. The in-case reconstruction performance is evaluated across $N_M=11$ data splits and $N_K = 10$ training ensemble realizations per split. Reconstruction metrics are first calculated for each trained realization and held-out test data independently, before averaging across the ensemble. This procedure samples both variability associated with the turbulent events included in the data partition and variability associated with the training procedure.

Across-case reconstruction consists of training and validating models on one dataset (henceforth denoted the source data), and deployment on a different dataset (target data), that is, on a flow case with different parameters, without retraining. Surface inputs and subsurface outputs are min--max scaled using reference ranges fitted to the training portion of the target-flow split, excluding the test interval in focus; the same output ranges are used to convert predictions back to physical units. Calibrating the output ranges requires subsurface information from the target flow, hence, the across-flow reconstructions assume that the range of subsurface statistics in the new flow is known, for instance from a separate measurement campaign. The range calibration is a linear transformation of each reconstructed time series and does not affect the temporal correlation, but sets the level and range and therefore affects the NMSE and log variance ratio. We return to this point in \S\ref{sec:zero_shot_main}.

For each source flow, the same 11 data splits used in the in-case analysis are retained, with each data split yielding one trained model. 
Each model is evaluated on 11 target intervals of the same lengths as the test intervals of the in-case testing, with the same target intervals used for all source models. 
Repeated trainings of the same source data split form an optimization ensemble, yet evaluating all realizations across every target dataset is computationally costly. We therefore retain one representative trained realization for each of the 11 source-data splits for the across-case reconstruction experiments, where the representative realization is the one with the median validation loss.

A second version of the across-case reconstruction uses DA-SHRED, as an inference-time adaptation of the standard across-case reconstruction described above. For each reconstruction, DA-SHRED starts from the same pretrained source model and is applied to the same target-flow test interval as the corresponding across-case SHRED reconstruction, such that the across-case performance of SHRED and DA-SHRED may be directly compared. %is therefore paired for every source-model--target-interval combination.
During deployment, only the surface observations available from the target flow are assimilated according to the procedure described in \S\ref{sec:da_shred_theoretical}. %No subsurface target information is used for the latent-state correction, and all pretrained network parameters remain fixed. 
Assimilation is performed independently for each target interval, and the corrected latent states are not transferred between test intervals. Each target interval starts with the same pretrained network parameters and a zero-initialized GRU hidden state. We refer to this procedure as DA-assisted across-case reconstruction.

During the training step of the models, network parameters are updated by minimizing the composite loss function
\begin{equation}
    \mathcal{L}
    =
    \mathcal{L}_{\mathrm{out}}
    + \lambda_{\mathrm{in}}\mathcal{L}_{\mathrm{in}}
    + \lambda_{\Delta t}\mathcal{L}_{\Delta t}
    = 
    \left(\sum_{j=0}^{N_z-1}\operatorname{MSE}\left(\hat{\mathbf{y}}_j,\mathbf{y}_j \right)\right)
    +     \lambda_{\mathrm{in}}\operatorname{MSE} \left(\hat{\mathbf{s}},\mathbf{s}\right)
    +     \lambda_{\Delta t}\operatorname{MSE} \left(\Delta_t \hat{\mathbf{y}},\Delta_t \mathbf{y}    \right) \, .
    %+ \lambda_{\mathrm{reg}}\mathcal{L}_{\mathrm{reg}},
    \label{eq:loss_total}
\end{equation}
Here, $\mathcal{L}_{\mathrm{out}}$ is the primary reconstruction loss, aggregated after independent evaluation at each plane $j$, $\mathcal{L}_{\mathrm{in}}$ is the surface observable reconstruction loss, and $\mathcal{L}_{\Delta t}$ is the first difference loss, with $\Delta_t \mathbf{y}_t = \mathbf{y}_{t+1} - \mathbf{y}_t $, included to promote accurate reconstruction of the temporal evolution of the outputs.
The loss weights are set to $\lambda_{\mathrm{in}}=0.025$ and $\lambda_{\Delta t}=0.05$.
For the network optimization, we use AdamW with a learning rate of $5\times10^{-4}$, weight decay of $0.01$, and a batch size of 128. Training runs for at most 3000 epochs, with early stopping after 10 consecutive epochs without improvement in the validation loss.

%%%%%%%%%%%%%%%%%%%%%%%%%%%%%%%%%%%%%%%%%%%%%%%
%%%%%% RIDGE %%%%
%%%%%%%%%%%%%%%%%%%%%%%%%%%%%%%%%%%%%%%%%%%%%%%
\subsection{Ridge regression baseline}
\label{sec:ridge_method}
We evaluate the performance of SHRED against target statistics reconstructed by ridge regression on the same datasets. Ridge regression is fitted to the same input and output as SHRED, and makes use of surface observables $\mathbf{S}_t$ with a time history $n_h=100$. Hence, the reconstructed statistics are obtained as a linear combination of the lagged inputs: %Stacking this history of observables yields the vector $\mathbf{x}_t \in \mathbb{R}^{2n_h}$, from which the reconstructed statistics at all planes are obtained as a linear combination:
%inputs, 
\begin{equation}
    \hat{\mathbf{y}}_t = \mathbf{W}\mathbf{x}_t + \mathbf{b} \, ,
\end{equation}
where $x_t$ is the observation history $\mathbf{S}_t$ flattened into a single vector.
The weights are computed (for each target statistic) by least squares with a magnitude penalty:
\begin{equation}
    \min\limits_{\mathbf{W},\mathbf{b}} \sum_t \lVert \mathbf{y}_t - \mathbf{W}\mathbf{x}_t - \mathbf{b} \rVert^2 + \alpha \lVert \mathbf{W} \rVert^2 \, ,
\end{equation}
where the penalty (Tikhonov regularization) stabilizes the fit when the lagged inputs are strongly correlated with each other. The regularization parameter $\alpha$ is selected from the set ${0, 10^{-8},10^{-7.5},\dots,10^{6}}$ by the mean squared error on validation data, and is shared by all depth-planes. Inputs and outputs are min–max scaled the same way as for SHRED, and the same data splits and reference ranges are used. 

%%%%%%%%%%%%%%%%%%%%%%%%%%%%%%%%%%%%%%%%%%%%%%%%%%%%%%%%%%%%%%
%%%%%%%%%%% P E R F O R M A N C E    M E T R I C S %%%%%%%%%%%
%%%%%%%%%%%%%%%%%%%%%%%%%%%%%%%%%%%%%%%%%%%%%%%%%%%%%%%%%%%%%%

\subsection{Performance and statistical analysis}
\label{sec:performance_main}

\subsubsection{Reconstruction performance metrics}
\label{sec:reconstruction_metrics}

Time-averaged reconstruction performance is evaluated at each depth using three complementary metrics. The normalized mean-square error, NMSE, the Pearson correlation coefficient, $\rho$, and the logarithmic variance ratio, $R_\sigma$, measure the overall reconstruction error, temporal agreement, and fluctuation amplitude, respectively. The metrics are defined as
\begin{align}
    \label{eq:NMSE}
    \mathrm{NMSE}(z) = \frac{\overline{\left(q-\hat{q}\right)^2}}{\overline{\left(q - \overline{q}\right)^2}}\, , \qquad
    \rho(z) = \frac{\overline{\left(q-\overline{q}\right)\left(\hat{q}-\overline{\hat{q}}\right)}}{\sigma_q \sigma_{\hat{q}}}\, , \qquad 
    %\label{eq:fluct_amp}
    R_{\sigma}(z) = \ln\left(\frac{\sigma_{\hat{q}}^2}{\sigma_q^2}\right) \, ,
\end{align}
where $\overline{(\cdots)}$ denotes temporal averaging, $q = q(t,z)$ and $\hat{q} = \hat{q}(t,z)$ are true and reconstructed values, respectively, while $\sigma_q = \sigma_q(z)$ and $\sigma_{\hat{q}} = \sigma_{\hat{q}}(z)$ are the standard deviations of $q$ and $\hat{q}$. When only the size of the variance mismatch matters, we use the variance-fidelity error, $E_{\sigma}(z) = \left| R_{\sigma}(z) \right|$. An exact reconstruction is identified by $\left(\mathrm{NMSE}, \rho, R_\sigma, E_\sigma \right) = \left(0,1, 0, 0\right)$. For $R_\sigma \neq 0$ negative and positive values denote under- and overprediction of variance.
Each metric is first calculated independently for every test interval, trained-model realization, and depth plane. Averaging across test intervals and model realizations is subsequently performed as appropriate for each analysis.

We can rewrite the NMSE to emphasize the relation between the three metrics, as 
\begin{equation}
    \label{eq:NMSE_decomp}
    \mathrm{NMSE} = \left(1 - \rho^2\right) + \left(\frac{\sigma_{\hat{q}}}{\sigma_q} - \rho\right)^2
    + \left(\frac{\overline{\hat q}-\overline{q}}{\sigma_q}\right)^2 \, ,
\end{equation}
where $\sigma_{\hat{q}}/\sigma_q = e^{R_\sigma /2}$. The first term depends only on the temporal correlation, the second depends on temporal correlation and the ratio of standard deviations, while the third---which we denote the \textit{offset} term---is the squared difference between the temporal means of the reconstructed and true series, normalized by the standard deviation of the true series. Two points we wish to emphasize from this formulation of the NMSE are: (1) reconstructed statistics can have a close temporal agreement with the true data and a small log variance ratio, yet have large NMSE due to their offset; (2) a log variance ratio of zero does not guarantee better reconstruction in terms of NMSE (the second term is zero for $\sigma_{\hat{q}}/\sigma_q  = \rho$, i.e., for matched variance only when the true and reconstructed signals are perfectly correlated).
%\subsubsection{Depth-resolved and temporal analysis}

In addition to the depth-resolved performance metrics, we examine time--depth diagrams of the true and reconstructed target quantities, %, $q(t,z)$ and $\hat{q}(t,z)$. 
which show the spatiotemporal performance of SHRED with respect to larger turbulent events. %the timing and depth dependence of turbulent events and how well SHRED reproduces them. 
We use the max-normalized absolute reconstruction error, $E(t,z) = \left|q(t,z)-\hat{q}(t,z)\right|/ \text{max|E|}$ and the lagged Pearson correlation, $C_z(\tau)= \mathrm{corr}\left[q(t,z_s),q(t+\tau,z)\right]$ between the surface plane ($z=z_s$) and subsurface data,
to quantify the spatiotemporal reconstruction accuracy and temporal offset within each target output, respectively. For the latter, we use temporal lags of $\pm 100$ snapshots%\tau\in[-1.5T_{\infty},1.5T_{\infty}]$
, where only overlapping portions of the two time series are used at each lag. The convention used here indicates that the surface events lead (lag) the subsurface events for $\tau>0$ ($\tau<0$). 
The peak-correlation lag, $\tau_\mathrm{peak} (z)$, is calculated at each depth, as the argmax of $C_z(\tau)$, 
computed separately for the true and reconstructed signals---once per test interval for the true signals; one for each interval and trained-model realization for the reconstructions. The resulting peak lags are averaged across the respective samples. 

%%%%%%%%%%%%%%%%%%%%%%%%%%%%%%%%%%%%%%%%%%%%%%%%%%%%%%%%%%%%%%%%%%%

\subsubsection{Additional across-case performance metrics}% variability}
\label{sec:zero_shot_decomp_method}

The performance of across-case reconstructions varies: some models transfer better overall, some target intervals are more difficult to reconstruct than others, and particular source--target combinations perform better or worse than their average tendencies suggest. To assess the performance of the across-case deployment, we make use of an array of different metrics aimed at specific parts of the variability in across-case performance. 

First, we decompose the reconstruction error into a source-model--target-interval matrix for each reconstructed quantity, performance metric, and transfer case, as $E_{ij} = \overline{E} + M_i + T_j + I_{ij}$, for
%Each entry $E_{ij}$ contains the performance of source model 
source model $i$ on target interval $j$. %. We write
%\begin{equation}
%E_{ij} = \overline{E} + M_i + T_j + I_{ij},
%\label{eq:performance_decomposition}
%\end{equation}
The decomposition consists of $\overline{E}$, the grand mean over all source models and target
intervals, $M_i = \overline{E}_{i\cdot} - \overline{E}$, the measurement of whether a model performs systematically above or below the grand mean across target intervals;
$T_j = \overline{E}_{\cdot j} - \overline{E}$, the complementary measurement of whether a target interval produces systematically higher or lower metric values across source models,
and $I_{ij}=E_{ij}-\overline{E}_{i\cdot}-\overline{E}_{\cdot j}+\overline{E}$, the  remaining departure from the performance expected by adding these two contributions to the grand mean; here $\overline{E}_{i\cdot}$ is the performance of source model $i$ averaged over all target intervals and $\overline{E}_{\cdot j}$ is the performance on target interval $j$ averaged over all source models.

% Sum of squares decomposition of variability 

We next quantify how much of the variation across source-model--target-interval pairs is associated with each contribution. For a complete matrix containing $N_S$ source models and $N_T$ target intervals, the sums of squared contributions and total sum of squares are
\begin{equation}
SS_M=N_T \sum_i M_i^2 \, , \qquad
SS_T=N_S \sum_j T_j^2\, , \qquad
SS_I=\sum_{i,j} I_{ij}^2 \, , \qquad 
SS_{\mathrm{tot}}=\sum_{i,j}
\left(E_{ij}-\overline{E}\right)^2 \, .
\label{eq:ss_components}
\end{equation}
Hence, the fractional contribution of component $k\in\{M,T,I\}$ is %
$f_k = SS_k / SS_{\mathrm{tot}}$, such that $f_M+f_T+f_I=1.$
These fractions describe how the variation in performance is distributed between source-model differences, target-interval differences, and their interaction. (They do not describe fractions of the reconstruction error itself.)

%%%%%%%%%%%%%%%%%%%%%%%%%%%%%%%%%%%%%%%%%%%%%%%%%%%%%%%%%%%%%%%%%%%%%%%%%

To investigate which statistical properties are associated with successful or challenging across-case reconstruction, we relate the performance to the statistical descriptors %signal descriptors introduced in
from \S\ref{sec:data_description}, that is, to relative means, CV, and $P_{99}/P_{95}$. Whereas Table~\ref{tab:global_dataset_statistics} characterizes the complete datasets, the descriptors here assess individual source and target intervals.
For the analysis of target intervals, we investigate whether intervals that differ from the typical behavior of their own flow case are easier or harder to reconstruct. Each physical target interval is counted once, with performance averaged across source models and, where applicable, source flows. Before pooling intervals from different datasets, the descriptor values are standardized per dataset to facilitate comparison according to the relative position of the intervals within their own dataset, by $D^{*}_{cj} = \left(\left(D_{cj}-\overline{D}_{c}\right) / \sigma_{D,c}\right)$,% each case:
where $D_{cj}$ is the descriptor value for interval $j$ of dataset $c$, and  $\overline{D}_c$ and $\sigma_{D,c}$ are its mean and standard deviation across intervals of that dataset.% Thus, $D^*_{cj}=1$ denotes a value one standard deviation above the case mean. 

%Spearman correlations
We use Spearman's rank correlation coefficient, $\rho_s$, to assess whether larger descriptor values tend to accompany higher or lower reconstruction performance. This measure captures monotonic relationships without requiring a linear relationship or normally distributed variables. The analysis is performed separately for each reconstructed quantity, performance metric, and descriptor. Statistical significance is assessed using permutation tests, in which the descriptor--performance pairing is randomized and Spearman's coefficient is recalculated.

The final set of metrics we use is a set of %To quantify the effect of data assimilation relative to the corresponding zero-shot SHRED reconstruction, we define a 
signed performance gain metrics $G$, which quantify the effect of using DA-SHRED over the standard SHRED method for across-case deployment. These are % reconstruction, we define a for each error metric (NMSE, correlation, variance fidelity) such that positive values consistently indicate improvement with DA-SHRED. For NMSE we orient the gain such that,
%\begin{equation}
%\begin{align}
    $G_{\mathrm{NMSE}}=\mathrm{NMSE}_{\mathrm{SHRED}}-\mathrm{NMSE}_{\mathrm{DA}} ,
    %\label{eq:gain_nmse}
    G_{\rho}=\rho_{\mathrm{DA}}-\rho_{\mathrm{SHRED}} ,
    %\label{eq:gain_corr}
    G_{\mathrm{var}}=\left|R_{\sigma,\mathrm{SHRED}}\right|-\left|R_{\sigma,\mathrm{DA}}\right|  ,
    $
with positive gains $G$ indicating better performance (lower NMSE, higher temporal correlation, or a smaller variance mismatch) while negative gains indicate deterioration.

We used the language models listed in the AI Use Disclosure primarily to help write the data analysis code and examine the reconstruction results, and occasionally to debug the training code. We checked the results by comparing reconstructed and reference time series and depth profiles.

\FloatBarrier

%%%%%%%%%%%%%%%%%%%%%%%%%%%%%%%%%%%%%%%%%%%%%%%%%%%%%%%%%%%%%%%%%%%%
%%%%%%%%%%%%%% R E S U L T S  %%%%%%%%%%%%%%
%%%%%%%%%%%%%%%%%%%%%%%%%%%%%%%%%%%%%%%%%%%%%%%%%%%%%%%%%%%%%%%%%%%%
\section{Results and discussion}
\label{sec:resultsAnddiscussion}
\FloatBarrier
In the following, we present and discuss the results of SHRED applied to sensing turbulence statistics from surface observables only. First, we present the results for surface-plane time series and depth-resolved reconstruction of all three turbulence statistics for datasets \caseHH and \caseLH, followed by time-depth diagrams of true and reconstructed MS divergence for the same datasets. %baseline flow cases of RE1000 and RE2500. 
Second, we investigate across-case reconstruction, i.e. SHRED models trained on one flow and deployed to a different flow.
Third, we compare the results with a linear method, using ridge regression for the exact same time-history and train-validation-test splits as our baselines to evaluate potential performance gains with SHRED. Fourth, we quantify the sources of variability in across-case reconstruction.
Finally, we show how the data-assimilation (DA) SHRED affects reconstruction performance for the transfer to different flows. 
% Finally, we investigate the transferability of the SHRED model from one pre-trained flow case to new flows, essential to real-world applications.

%%%%%%%%%%%%%%%%%%%%%%%%%%%%%%%%%%%%%%%%%%%%%%%
%%%%%%%%%%% BASELINE RESULTS %%%%%%%%%%%%%%%%%%
%%%%%%%%%%%%%%%%%%%%%%%%%%%%%%%%%%%%%%%%%%%%%%%
\subsection{In-case reconstruction performance}% of turbulence statistics}
\label{sec:baseline_results}
%%%%%%%%%%%%%%%%%%%%%%%%%%%%%%%%%%%%%%%%%%%%%%%
%%%%%%%%%%% BASELINE RESULTS %%%%%%%%%%%%%%%%%%
%%%%%%%%%%%%%%%%%%%%%%%%%%%%%%%%%%%%%%%%%%%%%%%

\begin{figure}[htb!]
  \centering
  \begin{overpic}[width=\textwidth]{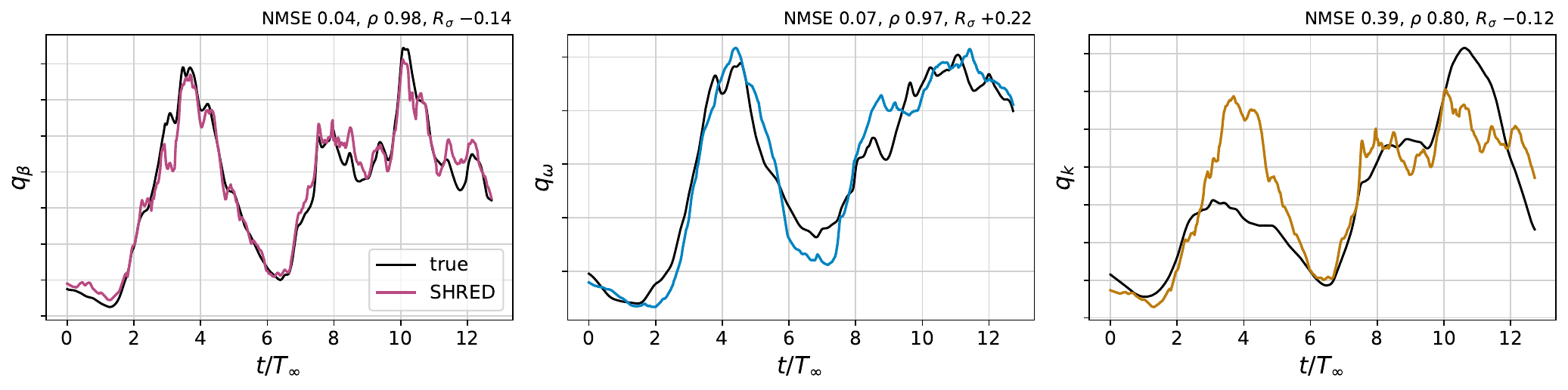}
    \put(4,20.4){(a)}
    \put(37,20.4){(b)}
    \put(71,20.4){(c)}
  \end{overpic}
  \caption{True (black) and SHRED-reconstructed target statistics (colored by quantity) at the surface plane of (a) MS surface divergence, (b) MS vertical vorticity and (c) mean TKE for one test interval of $\mathrm{Re}_H\,\mathrm{We}_\infty$.}
  \label{fig:fig04_surface_timeseries}
\end{figure}

As reference cases to demonstrate the capabilities of SHRED, we present the in-case reconstruction of turbulence statistics for one high- and one low-$\Reynolds$ case with Weber numbers of infinity (results for the remaining cases are included in App.~\ref{app:in-case-reconstruction}). The models are trained and performance is assessed according to the strategies outlined in \S \ref{sec:training_strategy} and \ref{sec:reconstruction_metrics}, which, in particular, means that all reconstructions are temporally separated from the training and validation data by several turbulent turnover times. 
A first indication of model performance is depicted in Fig.~\ref{fig:fig04_surface_timeseries}, which displays the true and reconstructed time series of the surface statistics defined in  Eq.~\eqref{eq:output_statistics}. The results are plotted for one representative test interval for the dataset \caseHH{}, selected as representative using the interval with the median NMSE of the 11 splits and median validation loss among the models (this method for selecting representative intervals is used throughout the study).
The reconstruction of MS horizontal divergence ($q_\beta$) and MS vertical vorticity ($q_\omega$) closely follow the true temporal evolution, including capturing major intermittent events. However, reconstruction of the mean TKE ($q_k$) is more challenging. The reconstructed mean TKE correlates well with truth, although weaker than the other two, but additionally overshoots significantly in amplitude and oscillations for this interval, affecting in particular its NMSE.

\begin{figure}[htb!]
  \centering
  \begin{overpic}[width=\textwidth]{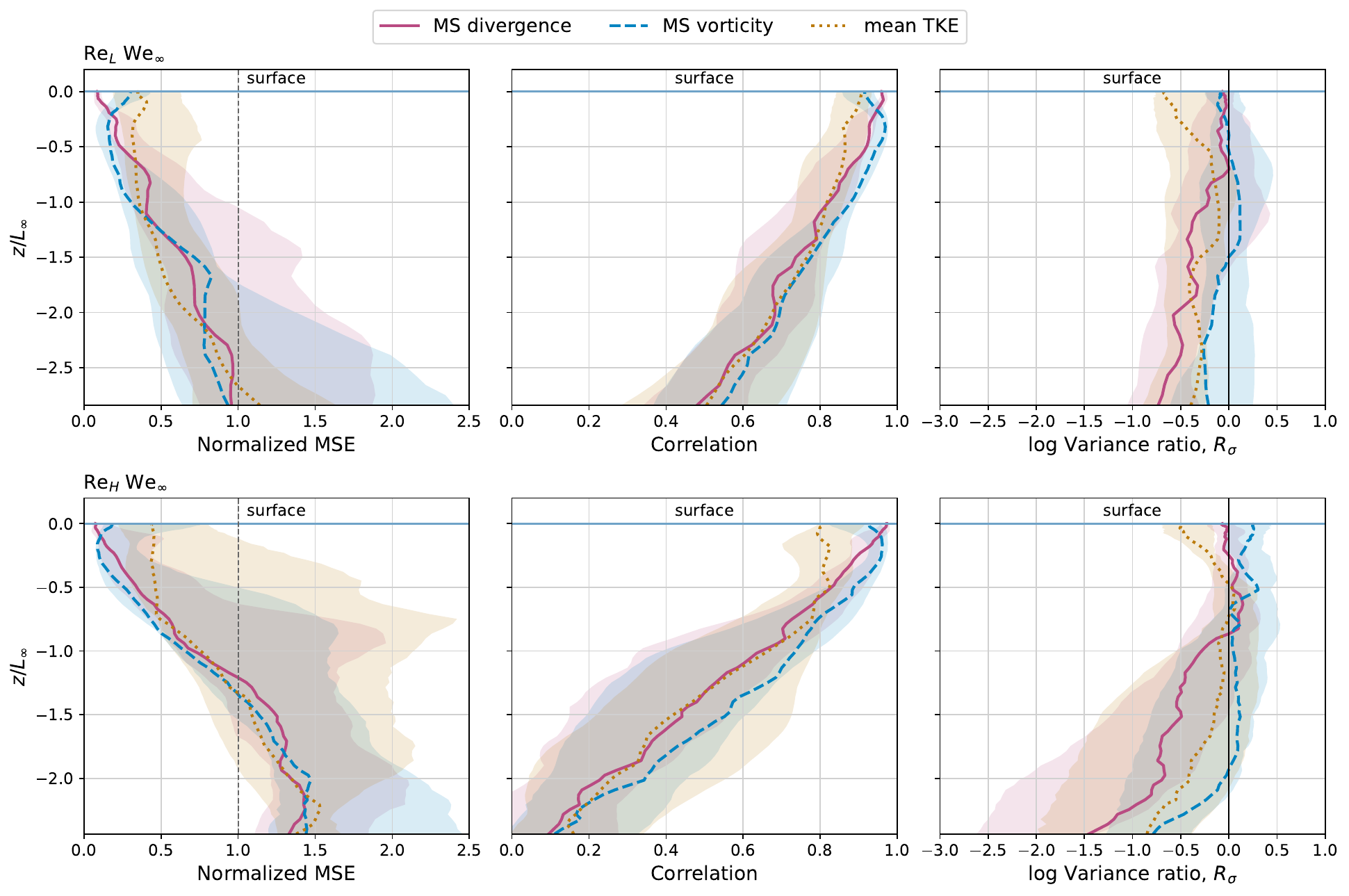}
    \put(31,57.5){(a)}
    \put(39,57.5){(b)}
    \put(71,57.5){(c)}
    \put(31,25){(d)}
    \put(39,25){(e)}
    \put(71.1,25){(f)}
  \end{overpic}
  \caption{Depth-resolved in-case reconstruction performance for (a)--(c) \caseLH{} and (d)--(f) \caseHH{}. Columns show NMSE, the temporal correlation and the log variance ratio. The profiles are medians over the 11 data splits of the per-split mean over 10 training realizations; shading spans the interquartile range over splits. For NMSE, performance below 1.0 (dashed vertical line) represents better NMSE than that of a constant time series at the average. For log variance ratio, matched variance is found at zero (vertical full line). }
  \label{fig:fig05_incase_depth_profiles}
\end{figure}

To gain further insight into SHRED performance, we compute and summarize the results in terms of the depth-dependent performance metrics NMSE, temporal correlation, and logarithmic variance ratio for datasets \caseLH{} and \caseHH{} in Fig.~\ref{fig:fig05_incase_depth_profiles}. The plotted curves represent the median result for MS horizontal divergence (solid magenta), MS vertical vorticity (dashed blue), and mean TKE (dotted ocher) with shading representing the interquartile range over data splits. Medians are preferred over means, as the intervals with low variance have large adverse impacts on the mean NMSE due to our definition using variance in the normalization, somewhat obscuring the actual performance.
%summarizes NMSE, temporal
%correlation, and logarithmic variance ratio across depth. 

The curves in Fig.~\ref{fig:fig05_incase_depth_profiles} corroborate that the models perform similarly for the output statistics MS horizontal divergence and MS vertical vorticity for dataset \caseHH{}, extending this result also to the low--Reynolds number dataset \caseLH{} and to subsurface performance. For both datasets the NMSE (correlation) is very low (high) near the surface and increases (decreases) with depth, while the logarithmic variance ratio is small for $z \gtrsim -L_\infty$, increasing in absolute value at depth. 
The observations are consistent with the well-known weakening local surface--subsurface correlation with depth \cite{aarnes2025,kjellevold2026,babiker2026,ruth2026}, which puts fundamental limits on the possibility of predicting one from the other. The depth deterioration holds for both datasets, yet is noticeably stronger for the high Reynolds number case. %, although reconstruction errors alone do not establish an observability limit. 
The models have a harder time reconstructing the near-surface mean TKE. For both datasets it has worse performance than the other output statistics on all three performance metrics in the near-surface region, and a significantly larger spread among intervals, particularly for the NMSE. However, the mean TKE reconstruction below $z=-L_\infty$ is similar to horizontal divergence and vertical vorticity, with a near-monotonic deterioration with depth in the NMSE and correlation.

\subsubsection{Time--depth reconstruction analysis}

\newcommand{\panel}[1]{{\setlength{\fboxsep}{0.6pt}\colorbox{white}{\small(#1)}}}
\begin{figure}[htb]
  \centering
  \begin{overpic}[width=\textwidth]{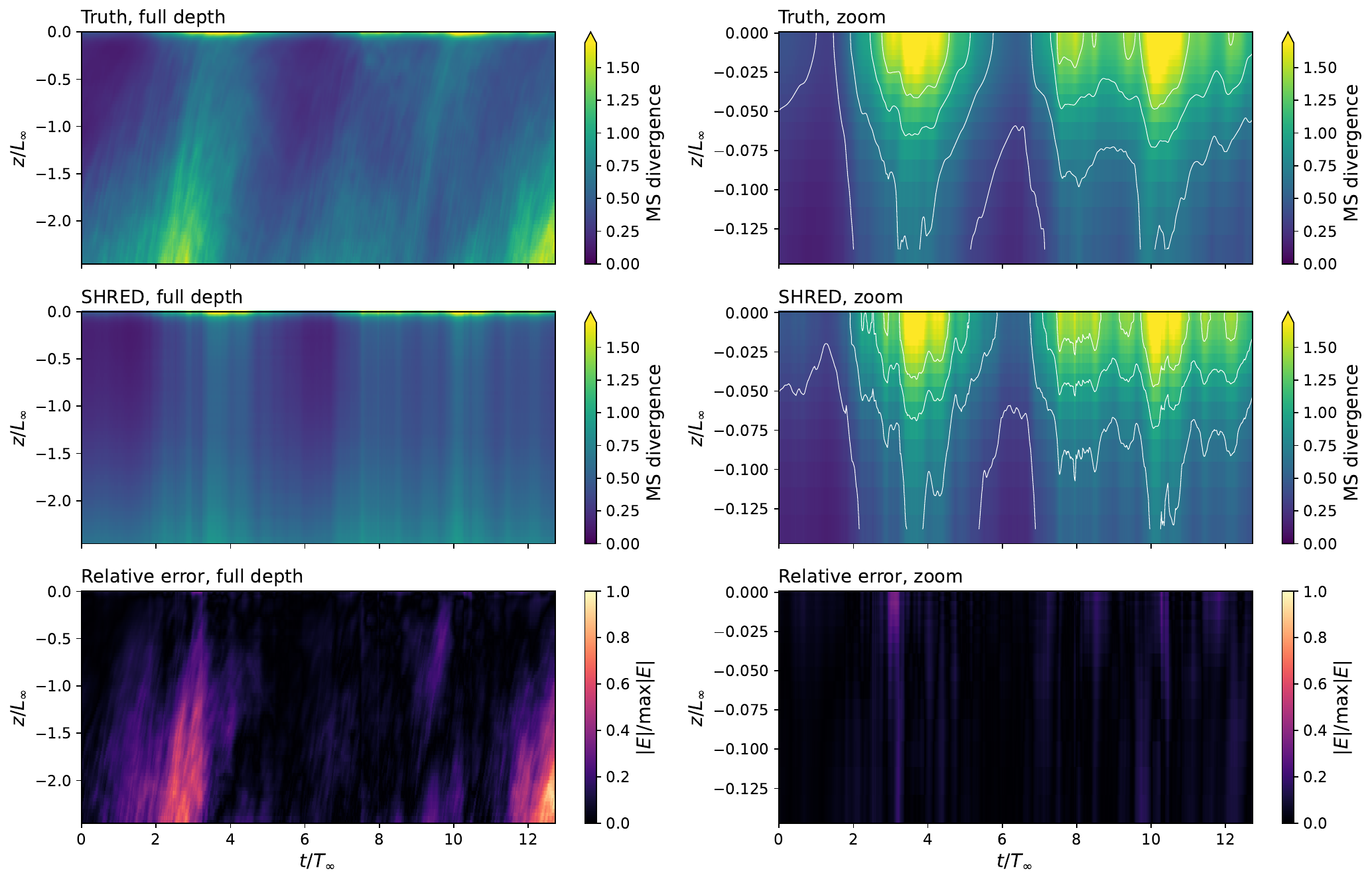}
    \put(6.5,61.7){\makebox(0,0)[lt]{\panel{a}}}
    \put(57.6,61.7){\makebox(0,0)[lt]{\panel{b}}}
    \put(6.5,41.2){\makebox(0,0)[lt]{\panel{c}}}
    \put(57.6,41.2){\makebox(0,0)[lt]{\panel{d}}}
    \put(6.5,20.7){\makebox(0,0)[lt]{\panel{e}}}
    \put(57.6,20.7){\makebox(0,0)[lt]{\panel{f}}}
  \end{overpic}
  \caption{Time--depth diagrams of (a, b) true and (c, d) reconstructed MS horizontal divergence of \caseHH{} for a representative test interval. Columns show full depth (left) and near-surface region (right). Coloring saturates at 80\,\% of the maximum of the truth horizontal divergence to highlight structures. Bottom row shows normalized absolute reconstruction error, normalized using the maximum absolute error.}
  \label{fig:hovmoller_RE2500_div}
\end{figure}
To analyze the temporal coherence of reconstructions, we plot the time-depth (Hovmöller) diagrams of test intervals and the corresponding normalized absolute error. The time-depth diagram of MS horizontal divergence for a representative target interval is shown in Fig.~\ref{fig:hovmoller_RE2500_div}, for the full depth (left) and zoomed in at the near surface region (right), with color scaled to saturate at 80\,\% of the maximum horizontal divergence to highlight the depth-dependence of flow events. We observe that the model captures the vertically coherent events associated with surface-observed dynamics, while events in the deep are not fully  captured.
%yet does not fully capture events before they reach the surface. %, even though SHRED works with a time-history of $n_h = 100$ frames ($= 1.257$\,$T_\infty$ for \caseHH). 
As mentioned, since turbulence is generated deep beneath the surface and diffuses upward, the turbulent features can only be detected when they get sufficiently close to the surface, hence the visible lag between bulk and surface visible in panel (a) of Fig.~\ref{fig:hovmoller_RE2500_div}, while (c) shows how SHRED is unable to infer past events in a `causal' reconstruction where past (lagged) information is used to reconstruct at time $t$, while future information is not used for this reconstruction scheme. 
Allowing later time steps in the surface observation window could improve retrospective reconstruction, but is not examined here.

Consider, e.g., the highly energetic event appearing at the deepest plane at $t/T_\infty = 2$--3 in the true field. The event is captured at the surface by SHRED, yet correctly timed only at and in the immediate vicinity of the surface. In the reconstructed field (panel c), the same event appears nearly simultaneously across depths, which results in the reconstructions lagging the actual event increasingly more with depth. This lagged reconstruction shows up as a significant source of reconstruction error in the normalized absolute error plot (bottom row), which for this interval is further emphasized by the amplitude of the reconstructed MS horizontal divergence being dampened at depth. 
However, in the zoomed-in views (b and d), the contours show that the near-surface evolution is reproduced more closely in both timing and magnitude, with smaller normalized absolute errors in panel (f), qualitatively reasonable since the shallower the depth, the greater the range of turbulent eddies that are large enough to leave detectable surface imprints.
For Hovm\"{o}ller diagrams for MS vertical vorticity and mean TKE, see App.~\ref{app:Hovmöller}.

\begin{figure}[htb]
  \centering
  \begin{overpic}[width=0.7\textwidth, trim={0bp 0bp 0bp 8.5bp},
        clip]{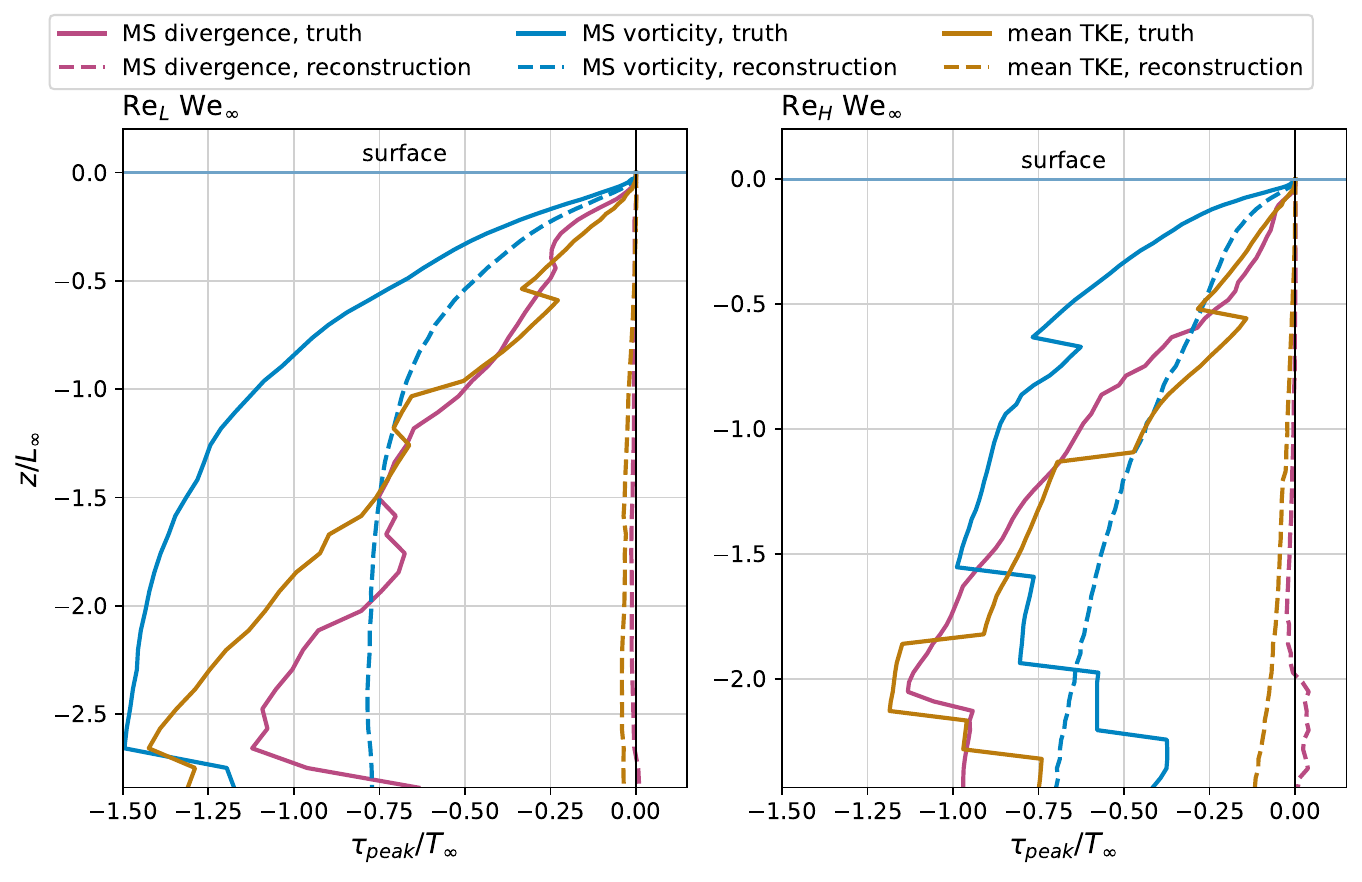}
    \put(9.5,50){\makebox(0,0)[lt]{\panel{a}}}
    \put(58,50){\makebox(0,0)[lt]{\panel{b}}}
  \end{overpic}
  \caption{Surface-to-subsurface peak-correlation lag $\tau_{peak}$ for the true (solid) and reconstructed (dashed) statistics for (a) dataset \caseLH{} and (b) dataset \caseHH, averaged over intervals. Negative lag indicates subsurface dynamics leads surface dynamics, while zero-lag indicates simultaneous peak-correlation.}
  \label{fig:correlation_lag}
\end{figure}

To quantify the average temporal shift between the surface signatures and the flow events at depth, we calculate the lag of maximum correlation, $\tau_\mathrm{peak}(z)$ between each turbulence statistic at depth $z$ and the same statistic at the surface, which we plot in Fig.~\ref{fig:correlation_lag}. 
We observe that for all measured statistics, $\tau_\mathrm{peak}\leq0$ below the surface, i.e., the surface lags the subsurface flow. Although the exact value of the lag varies between the measured statistics, we see a clear pattern: the magnitude of the lag generally increases with depth, reaching $\mathcal{O}(T_\infty)$ at $z=-2L_\infty$. For the reconstructed data we observe---consistent with the results in Fig.~\ref{fig:hovmoller_RE2500_div}---that the time shift between the surface and the subsurface is not captured by SHRED. The exception is the vertical vorticity, where the reconstructions partly reflect the true time-depth signal dependency. 
It is physically interesting that while hotspots of horizontal divergence---corresponding to large upwelling events---approach the surface at a constant rate since $\tau_\mathrm{peak}(z)\propto z$, peaks in vertical vorticity---corresponding to surface-attached ``bathtub'' vortices near $z=0$---exhibit a different vertical motion, gradually slowing down as they rise.

\FloatBarrier

%%%%%%%%%%%%%%%%%%%%%%%%%%%%%%%%%%%%%%%%%%%%%%%%%%%%%%%%%%
%%%%%%%%%%%  Zero-shot across Re and We %%%%%%%%%%%%%%%%%%
%%%%%%%%%%%%%%%%%%%%%%%%%%%%%%%%%%%%%%%%%%%%%%%%%%%%%%%%%%

\subsection{Across-case reconstruction performance}%Deployment across Reynolds and Weber numbers}
\label{sec:zero_shot_main}

The results for in-case testing establish that SHRED achieves strong near-surface reconstruction, with overall reconstruction performance deteriorating with greater depths. We next investigate whether the learned surface-to-subsurface mapping generalizes to new flow conditions. To do this, we train and validate models on one dataset and deploy them to different Reynolds- and Weber-number flows without retraining, using reference ranges of the target flow (a limitation we return to at the end of this section). We first consider models trained on dataset \caseHH{}, deployed on low-Re datasets, followed by an analysis of the reverse direction with models trained on dataset \caseLH{} deployed to high-Re datasets.% We examine both directions of Reynolds-number transfer.

\subsubsection{Model transferability from higher to lower Reynolds numbers}
\begin{figure}[htb!]
  \centering
  \begin{overpic}[width=\textwidth]{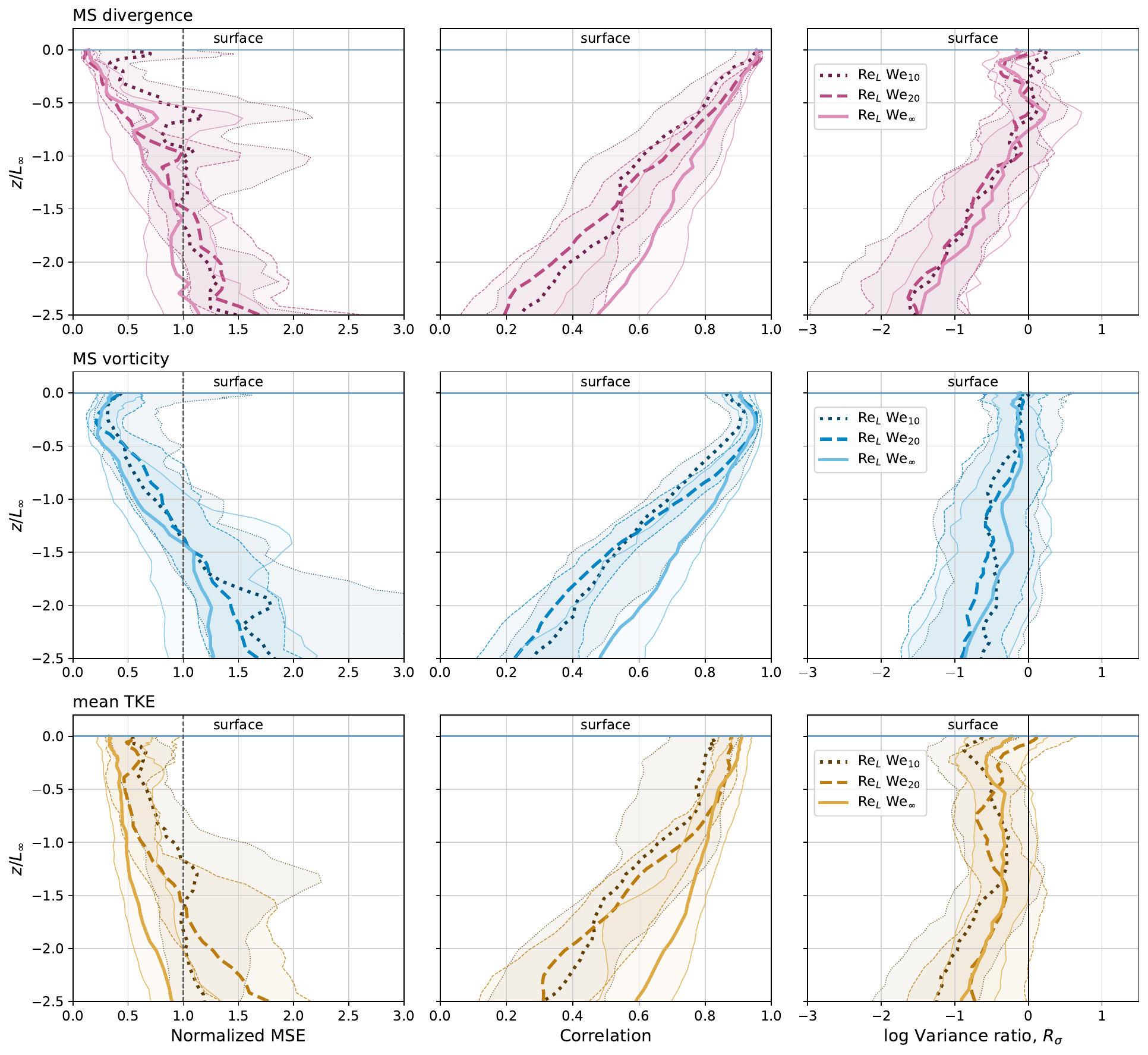}
    \put(31,87){\makebox(0,0)[lt]{\panel{a}}}
    \put(38.9,87){\makebox(0,0)[lt]{\panel{b}}}
    \put(98.6,87){\makebox(0,0)[rt]{\panel{c}}}
    \put(31,57){\makebox(0,0)[lt]{\panel{d}}}
    \put(38.9,57){\makebox(0,0)[lt]{\panel{e}}}
    \put(98.6,57){\makebox(0,0)[rt]{\panel{f}}}
    \put(31,27.3){\makebox(0,0)[lt]{\panel{g}}}
    \put(38.9,27.3){\makebox(0,0)[lt]{\panel{h}}}
    \put(98.6,27.3){\makebox(0,0)[rt]{\panel{i}}}
  \end{overpic}
  \caption{Depth-wise reconstruction performance for across-case deployment of models trained on dataset \caseHH{} applied to the three lower-$\Reynolds$ datasets, for (a--c) MS horizontal divergence, (d--f) MS vertical vorticity and (g--i) TKE; column-wise split into (left) NMSE, (middle) temporal correlation and (right) logarithmic variance ratio.  Target flows are distinguished by line style and shade, with color identifying output statistics. Thick lines are medians, thin lines of the same style bound their interquartile range. Inputs and outputs are min--max scaled with the target split's training ranges.}
  \label{fig:transfer_high2low_main}
\end{figure}

Figure~\ref{fig:transfer_high2low_main} shows model performance using the dataset \caseHH{} as source data, deployed on the three low--Reynolds number datasets from Table \ref{tab:flowProp} (results for transfer to other high--Reynolds number datasets are included in App.~\ref{app:across-case-reconstruction}). The color coding is according to reconstructed quantity (matching the color scheme of in-case reconstruction quantities), with different hues for the three target datasets and shading for interquartile ranges (outlined by thin curves to better separate them). From the figure, we note that the across-case performance compares favorably with the in-case results presented above. Importantly, the results show that within the conditions considered, the learned mapping remains useful beyond the training flow, especially near the surface. %(Fig.~\ref{fig:baseline_depth_errors}(b)), particularly near the surface. 

We observe that the across-case reconstructions show similar trends to that of in-case, where SHRED performance was most accurate for MS horizontal divergence and MS vertical vorticity near the surface. For the across-case reconstruction, the median reconstruction performance of TKE near the surface is better than in-case for all but the transfer to \caseLL{}. For the latter, MS horizontal divergence is slightly less accurate, and shows large accuracy-spread over intervals and models also near the surface. For the datasets considered here, there is, however, not enough of a pattern to establish any Weber-number dependency.

%%%%%%%%%%%%%%%%%%%%%%%%%%%%%%%%%%%%%%%%%%%%%%%%%%%%%%%%%%%%%%%%%%%%%%%%%%

\subsubsection{Model transferability from lower to higher Reynolds numbers}
% TRANSFERABILITY FROM LOW TO HIGH REYNOLDS NUMBER:
\begin{figure}[htb!]
  \centering
  \begin{overpic}[width=\textwidth]{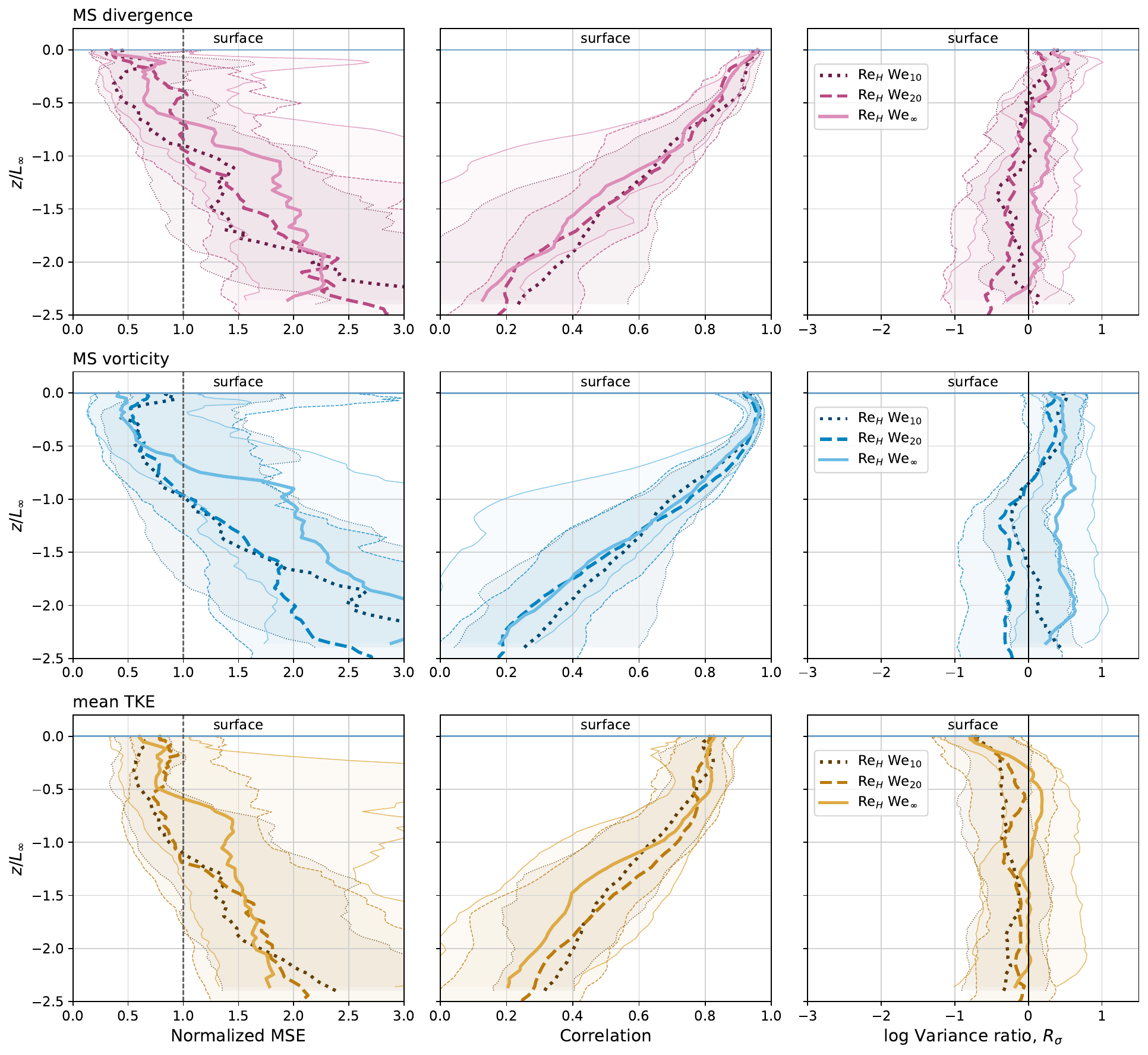}
    \put(6.9,89.1){\makebox(0,0)[lt]{\panel{a}}}
    \put(38.9,89.1){\makebox(0,0)[lt]{\panel{b}}}
    \put(71.0,89.1){\makebox(0,0)[lt]{\panel{c}}}
    \put(6.9,59.2){\makebox(0,0)[lt]{\panel{d}}}
    \put(38.9,59.2){\makebox(0,0)[lt]{\panel{e}}}
    \put(71.0,59.2){\makebox(0,0)[lt]{\panel{f}}}
    \put(6.9,29.2){\makebox(0,0)[lt]{\panel{g}}}
    \put(38.9,29.2){\makebox(0,0)[lt]{\panel{h}}}
    \put(71.0,29.2){\makebox(0,0)[lt]{\panel{i}}}
  \end{overpic}
  \caption{Same as Fig.~\ref{fig:transfer_high2low_main}, for models trained on dataset \caseLH{} and deployed to the higher--Reynolds number datasets.}
  \label{fig:transfer_low2highRE}
\end{figure}

\begin{figure}[tbp!]
  \centering
  \begin{overpic}[width=\textwidth]{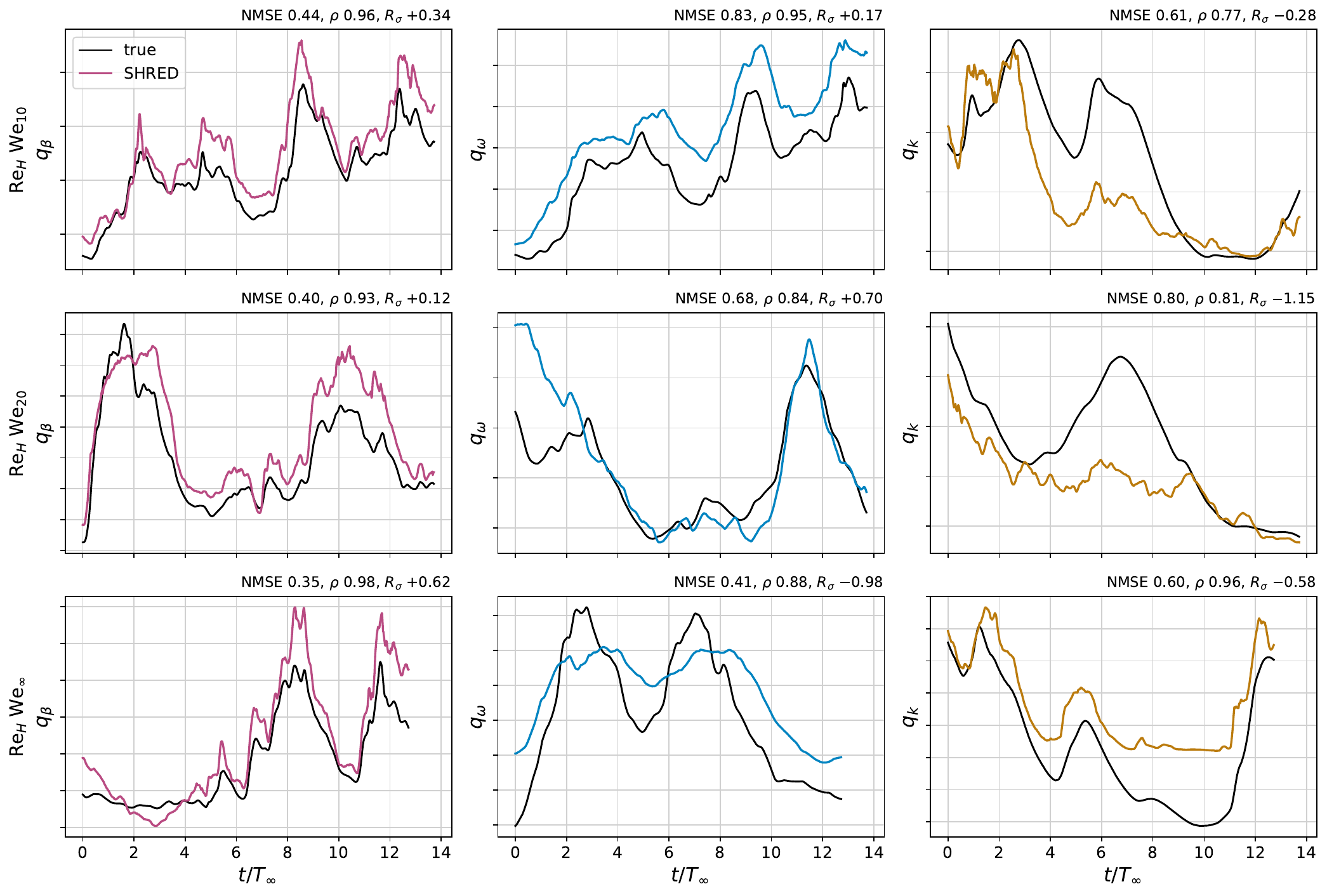}
    \put(5.6,64.5){\makebox(0,0)[lt]{\panel{a}}}
    \put(38.2,64.5){\makebox(0,0)[lt]{\panel{b}}}
    \put(70.9,64.5){\makebox(0,0)[lt]{\panel{c}}}
    \put(5.6,43){\makebox(0,0)[lt]{\panel{d}}}
    \put(38.2,43){\makebox(0,0)[lt]{\panel{e}}}
    \put(70.9,43){\makebox(0,0)[lt]{\panel{f}}}
    \put(5.6,21.3){\makebox(0,0)[lt]{\panel{g}}}
    \put(38.2,21.3){\makebox(0,0)[lt]{\panel{h}}}
    \put(70.9,21.3){\makebox(0,0)[lt]{\panel{i}}}
  \end{overpic}
  \caption{True (black) and SHRED-reconstructed target statistics (colored by quantity)  at the surface plane of MS horizontal divergence (magenta), MS vertical vorticity (blue) and mean TKE (ocher), for models trained on \caseLH{} and deployed to one test interval of (a--c) \caseHL{}, (d--f) \caseHM{} and (g--i) \caseHH{}.}
  \label{fig:fig09b_across_timeseries_low_to_high_Re}
\end{figure}

Having seen that SHRED generalizes well from higher to lower Reynolds numbers, we next investigate the opposite direction, and consider whether models trained on low Reynolds number dataset successfully generalize to a higher Reynolds number flow. Since the target flow has a larger scale separation than the source flow, we expect this deployment to be more challenging.
%than the high Re source to low Re target deployment presented above. 
In particular, the high Re target flow will contain small scale structures that are less than half the size of the smallest scales in the source flow (compare $L_K$ across Reynolds numbers in Table \ref{tab:flowProp}).

The deployment to higher Reynolds number than the model is trained on is of great practical interest: Realistically, fully resolved training can only be performed on DNS data with low or moderate Reynolds numbers; even LES or lab data will likely be well below the regimes for real world application, in addition to introducing limits on the training resolution and (for lab data) the planes that may be captured simultaneously. This is an example of the `sim2real' challenge in machine learning (e.g., \cite{Miao2023}). Admittedly, the Reynolds number spread is limited in our present data, but it may nevertheless indicate tendencies which hold across larger training and deployment discrepancy. %Such transfer is also of practical interest, as high-Reynolds-number datasets are often considerably more expensive to generate or acquire than lower-Reynolds-number data. 
In our setup, models are trained on the \caseLH{} dataset and deployed to the three high Reynolds number datasets. We present the median depth-profiles of this transfer direction in Fig.~\ref{fig:transfer_low2highRE}. We also include time series for representative target intervals at the surface for reconstruction of \caseHH{}, \caseHM{} and \caseHL{} in Fig.~\ref{fig:fig09b_across_timeseries_low_to_high_Re}.% (additional results in the appendix, Fig. \ref{fig:all_transfer_errors_low2high}).

In Fig.~\ref{fig:transfer_low2highRE}, we first notice the larger NMSE in the reconstruction of high-$\Reynolds$ statistics from low-$\Reynolds$ datasets, than for the in-case reconstruction and across-case reconstruction of low Re from high Re dataset. The median curves cross one at a shallower depth ($z$ between $- 0.4L_\infty$ and $-L_\infty$) than the results in Fig.~\ref{fig:transfer_high2low_main} ($z$ between $- 0.6L_\infty$ and $-2.2L_\infty$). 
%is larger than one at the surface, meaning that the error is larger than the variance of the true signal. 
However, the correlations between the true and reconstructed time series remain high near the surface, particularly for MS horizontal divergence and MS vertical vorticity. What this translates to in time-resolved statistics, can be seen in Fig.~\ref{fig:fig09b_across_timeseries_low_to_high_Re}: The larger events in the time-dependent statistics are well captured for MS horizontal divergence (panels a and d) and MS vertical vorticity (panels b and e) although reconstructed series are offset from the true ones (e.g., Fig.~\ref{fig:fig09b_across_timeseries_low_to_high_Re}b). %the amplitude-tracking is less accurate than we have seen for in-case reconstructions (Fig.~\ref{fig:fig04_surface_timeseries}). 
The same is true to some extent for the mean TKE (panels c and f), yet with only partial correlation between the large events in the true and reconstructed signal. 

The increase in NMSE for the transfer from low to high Reynolds number is further investigated in terms of the decomposition of NMSE in Eq.~\eqref{eq:NMSE_decomp}. The decomposition shows that $76$\% of the increase in NMSE stems from the offset term and nearly all the rest from the second term in Eq.~\eqref{eq:NMSE_decomp}. The median temporal correlation changes by less than 0.029 (MS horizontal divergence) and 0.056 (MS vertical vorticity) at any plane, and less than 0.02 when the three target datasets are pooled. Relative to in-case reconstruction, the temporal correlation changes little; most of the additional NMSE comes from a bias in the reconstructed interval mean.%In short, the reconstructions follow the temporal evolution of the target flow, but at a misplaced mean. 

\begin{table}[]
    \centering
    \caption{Change in NMSE (across-case minus in-case) for the same target intervals, for MS horizontal divergence / MS vertical vorticity; medians over source models and target intervals of depth-averaged NMSE, with each plane weighted by the depth interval it represents.} %. Models trained on dataset in each row and deployed to dataset in each column.}
\begin{tabular}{lcccccc}
\hline
 & \multicolumn{6}{c}{Target data}\\
\cline{2-7}
Source data
 & $\mathrm{Re_L We_{10}}$ & $\mathrm{Re_L We_{20}}$ & $\mathrm{Re_L We_{\infty}}$ & $\mathrm{Re_H We_{10}}$ & $\mathrm{Re_H We_{20}}$ & $\mathrm{Re_H We_{\infty}}$ \\
\hline
$\mathrm{Re_L We_{10}}$         & --             & 0.18 / 0.40    & 0.08 / 0.15    & 0.03 / 0.34    & 0.12 / 0.19    & 0.37 / 0.88    \\
$\mathrm{Re_L We_{20}}$         & 0.21 / 0.32    & --             & 0.26 / 0.41    & 0.09 / 0.25    & 0.30 / 0.30    & 0.22 / 0.43    \\
$\mathrm{Re_L We_{\infty}}$     & 0.25 / 0.23    & 0.35 / 0.52    & --             & 0.43 / 0.45    & 0.56 / 0.43    & 0.60 / 0.67    \\
$\mathrm{Re_H We_{10}}$         & 0.20 / 0.12    & 0.15 / 0.18    & 0.11 / 0.17    & --             & 0.10 / 0.08    & 0.31 / 0.43    \\
$\mathrm{Re_H We_{20}}$         & 0.16 / 0.15    & 0.15 / 0.28    & 0.05 / 0.20    & 0.02 / 0.12    & --             & 0.27 / 0.75    \\
$\mathrm{Re_H We_{\infty}}$     & 0.30 / 0.34    & 0.17 / 0.32    & 0.15 / 0.30    & 0.09 / 0.20    & 0.21 / 0.23    & --             \\
\hline
\end{tabular}
    \label{tab:across_case_all}
\end{table}

So far, we have considered only one source dataset for each transfer direction. To examine whether the larger NMSE observed for low-to-high transfer is also found for other choices of source dataset, we deploy models trained on each of the six datasets to the other five. 
%To separate the two, we deploy models trained on each of the six datasets to every other dataset. 
The resulting increase in NMSE (as compared to in-case reconstructions) for MS horizontal divergence and MS vertical vorticity are listed in Table \ref{tab:across_case_all}. 
The NMSE is averaged over depth, with each plane weighted by the depth interval it represents, to avoid overemphasizing the closely spaced near-surface planes. Tables \ref{tab:transfer_msdiv}--\ref{tab:transfer_tke} in App.~\ref{app:across-case-reconstruction}, give the same results separated into near surface and for greater depths, including also the NMSE increase for mean TKE (for which we observe no systematic effect from transfer direction). 

We observe that the increase in NMSE can vary as much between source datasets at the same Reynolds number as between the two transfer directions. In particular, models trained on \caseLH{} show a relatively large increase in NMSE, which helps explain the larger errors in Fig.~\ref{fig:transfer_low2highRE}. %The large NMSE in Fig.~\ref{fig:transfer_low2highRE} stems mainly from the source dataset \caseLH{} yielding a particularly large NMSE increase. 
The target data also matters: for most source datasets, the largest increase occurs when the models are deployed to \caseHH{}. Despite this variation, all nine dataset pairs show the same pattern for near-surface MS vertical vorticity: NMSE increases more for low-to-high transfer than for the reverse direction (Table~\ref{tab:transfer_msvort}). Finally, for most source-target pairs, the offset between the reconstructed and true interval means remains the dominant contribution to the increase in NMSE. Meanwhile, the temporal correlation changes by less than 0.05 at 93\% of the depth range over all pairs. We return to the influence of target-interval and source-models in \S\ref{sec:zero_shot_characteristics}.

\subsubsection{Mean offsets and calibration across flows}

We next ask whether comparable transfer performance can be obtained when the depth-dependent output ranges of the target flow are truly unavailable. As stated in \S\ref{sec:training_strategy}, output ranges from the target datasets are used to calibrate the across-case reconstructions. As such information may not always be available in practice, we consider the impact of the calibration: Using only the output ranges from the source data (i.e., no calibration), the deployment between datasets of the same Reynolds numbers yields reconstruction accuracy with comparable NMSE to the one with calibration. For deployment on different Reynolds number than the models are trained on, however, the NMSE increases significantly, since the reconstructed statistics are offset by a factor approximately equal to the ratio of the temporal means of the target statistics in the training and deployment datasets. For MS horizontal divergence and MS vertical vorticity, this factor is close to the ratio of the source Reynolds number to the target Reynolds number. When models trained on \caseHH{} are applied to \caseLH{}, the reconstructed means are 2.8--3.0 times the true means. In the reverse direction, the reconstructed means are about one-third of the true means.%this factor is approximately equal to the ratio of Reynolds numbers, yielding overpredicted statistics for models trained on \caseHH{} and deployed to \caseLH{} by 2.8-3.0 and underprediction by the inverse factor when going the other way. 

An alternative calibration is to rescale the source output ranges using only the turbulence properties in Table \ref{tab:flowProp}. 
In homogeneous turbulence, $\varepsilon/\nu$ provides a characteristic scale for mean-square velocity
gradients, where $\varepsilon$ is the mean kinetic energy dissipation rate per unit mass and  $\nu$ is the kinematic viscosity \cite{tennekes1972first}. This motivates using the target-to-source ratio $(\varepsilon/\nu)_{\mathrm{target}}/(\varepsilon/\nu)_{\mathrm{source}}$ to rescale the output ranges for MS horizontal divergence and MS vertical vorticity.
Using the bulk estimate $\varepsilon \sim u_{\mathrm{rms}}^3/\ell$, we can approximate this ratio by $(\mathrm{Re}_\infty/T_\infty^2)_{\mathrm{target}}/
 (\mathrm{Re}_\infty/T_\infty^2)_{\mathrm{source}}$ (for mean TKE, the scaling is $\left(L_\infty /T_\infty\right)^2_{\rm target} / \left(L_\infty /T_\infty\right)^2_{\rm source}$, since it scales with the square of the velocity, not its gradients).
With this single factor per output statistics---which will typically be available from flow characterization in a laboratory setting, but not from the surface alone---the median NMSE for transfer across Reynolds numbers is close to the target range calibration (1.19 against 1.11 for MS horizontal divergence, 1.00 against 1.19 for MS vertical vorticity, and 1.14 against 1.01 for mean TKE), and within 0.3 of it for both MS divergence and MS vorticity in two-thirds of
the pairs. These results suggest that the error without calibration can indeed be corrected using bulk turbulence scales.

In sum, the across-case reconstructions retain the temporal correlation and much of the depth-dependent structure observed in the in-case reconstruction, yet with larger NMSE, which stems mainly from the offset. Moreover, these errors vary substantially between quantities, target intervals, and source-model realizations, as the interquartile ranges in Fig.~\ref{fig:transfer_low2highRE} demonstrate. 
We return to this in \S\ref{sec:zero_shot_characteristics}, after first assessing the accuracy of SHRED by comparing it to a simpler reconstruction method.

%\label{sec:zero_shot_main}
\FloatBarrier

%%%%%%%%%%%%%%%%%%%%%%%%%%%%%%%%%%%%%%%%%%%%%%%%%%%%%%%%%%
%%%%%%%%%%%  Baselines %%%%%%%%%%%%%%%%%%
%%%%%%%%%%%%%%%%%%%%%%%%%%%%%%%%%%%%%%%%%%%%%%%%%%%%%%%%%%
\subsection{Comparison with baseline results}
\label{sec:baselines}
%%%%%%%%%%%%%%%%%%%%%%%%%%%%%%%%%%%%%%%%%%%%%%%
%%%%%%%%%%% BASELINES 
%%%%%%%%%%%%%%%%%%%%%%%%%%%%%%%%%%%%%%%%%%%%%%%
\begin{figure}[htb!]
  \centering
  \begin{overpic}[width=\textwidth]{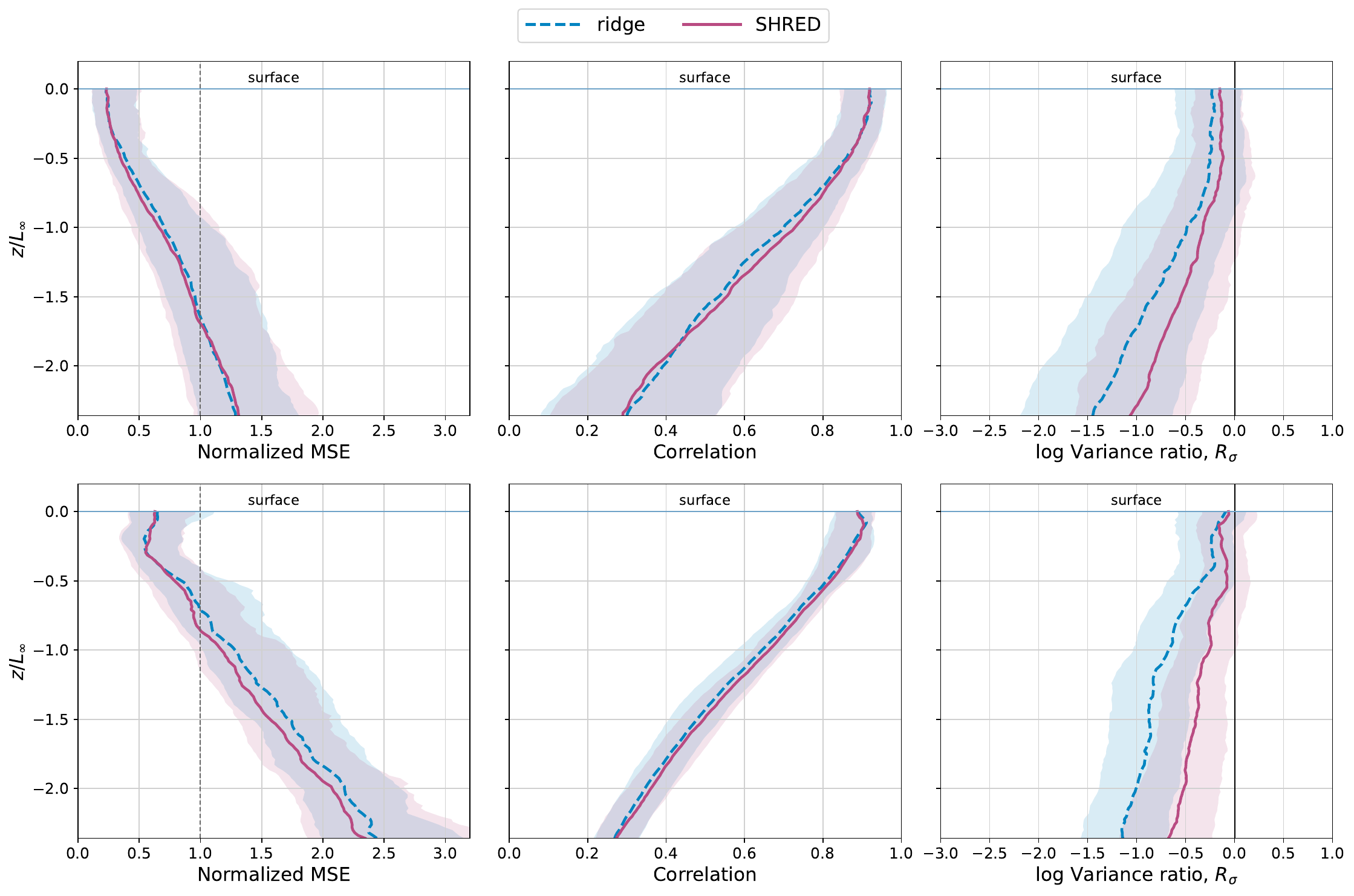}
    \put(6.3,61.3){\makebox(0,0)[lt]{\panel{a}}}
    \put(38.2,61.3){\makebox(0,0)[lt]{\panel{b}}}
    \put(70.1,61.3){\makebox(0,0)[lt]{\panel{c}}}
    \put(6.3,30.1){\makebox(0,0)[lt]{\panel{d}}}
    \put(38.2,30.1){\makebox(0,0)[lt]{\panel{e}}}
    \put(70.1,30.1){\makebox(0,0)[lt]{\panel{f}}}
  \end{overpic}
  \caption{Depth profiles of (a, d) NMSE, (b, e) correlation and (c, f) log variance ratio for SHRED (solid) and a ridge regression on the same $n_h=100$ frame input history (dashed). (Top row) In-case reconstruction for all datasets and output statistics; (bottom row) across-case reconstruction for models trained on \caseHH{} and \caseLH{}, each deployed to the five other datasets. Lines are medians and shading the interquartile range over all pooled profiles.}
  \label{fig:baseline_comparison}
\end{figure}

To evaluate the performance of SHRED as a reconstruction method, we compare it with the ridge regression baseline of \S \ref{sec:ridge_method}. %the method to a ridge-regression baseline. 
The ridge regression method uses the same $n_h$ step history and is trained, validated and tested on the identical data splits as SHRED. We compute both in-case and across-case reconstructions with ridge regression, using the same reference ranges of the target flow in the latter case as SHRED does. Figure~\ref{fig:baseline_comparison} shows the median across datasets and target statistics of NMSE, temporal correlation and log variance ratio for both models, for in-case (top row) and across-case reconstruction (bottom row).

Near the surface, the two models perform almost identically. For $z \gtrsim -0.5L_\infty$
both reach correlations above 0.9, and the NMSE of the ridge baseline and SHRED results are hardly distinguishable. Deeper beneath the surface, the temporal correlations are nearly equal, in-case and across-case. For the across-case reconstructions, SHRED slightly outperformed ridge regression with respect to NMSE for the datasets we consider. According to the decomposition Eq.~\eqref{eq:NMSE_decomp}, the difference lies mainly in the offset term. Ridge regression underpredicts the variance more strongly at depth (Fig.~\ref{fig:baseline_comparison}c and f). 
In sum, SHRED and the ridge regression reconstruct the same temporal information, with SHRED performing slightly better, mainly due to how it sets the mean level of the reconstruction. 

\FloatBarrier
%%%%%%%%%%%%%%%%%%%%%%%%%%%%%%%%%%%%%%%%%%%%%%%%%%%%%%%%%%%%%%%
%%%%%%%%%%%  Reconstruction characterization %%%%%%%%%%%%%%%%%%
%%%%%%%%%%%%%%%%%%%%%%%%%%%%%%%%%%%%%%%%%%%%%%%%%%%%%%%%%%%%%%%

%\subsection{Characteristics of robust and challenging zero-shot reconstructions}
\subsection{Sources of variability in across-case reconstruction performance}
\label{sec:zero_shot_characteristics}

The preceding results demonstrate that SHRED can reconstruct turbulence statistics across flow conditions. 
The depth-dependent performance profiles are the result of considering medians and interquartile ranges of ensembles over source-models and target-intervals (note that source-model and target-interval refer to a model trained on one data split of the source data and one interval from the target dataset, respectively).
Although the performance is promising, spread across source models and target intervals indicates substantial performance variation across ensembles.
To examine where this variation arises, we next decompose depth-averaged reconstruction performance into contributions associated with the source-model, the target-interval, and their interaction. As described in \S\ref{sec:zero_shot_decomp_method}, we decompose the error into $M_i$, $T_j$ and $I_{ij}$ that indicate whether a source-model, target-interval or interaction between the two dominates the performance variation.

\begin{figure}[htb!]
  \centering
  \begin{overpic}[width=\textwidth]{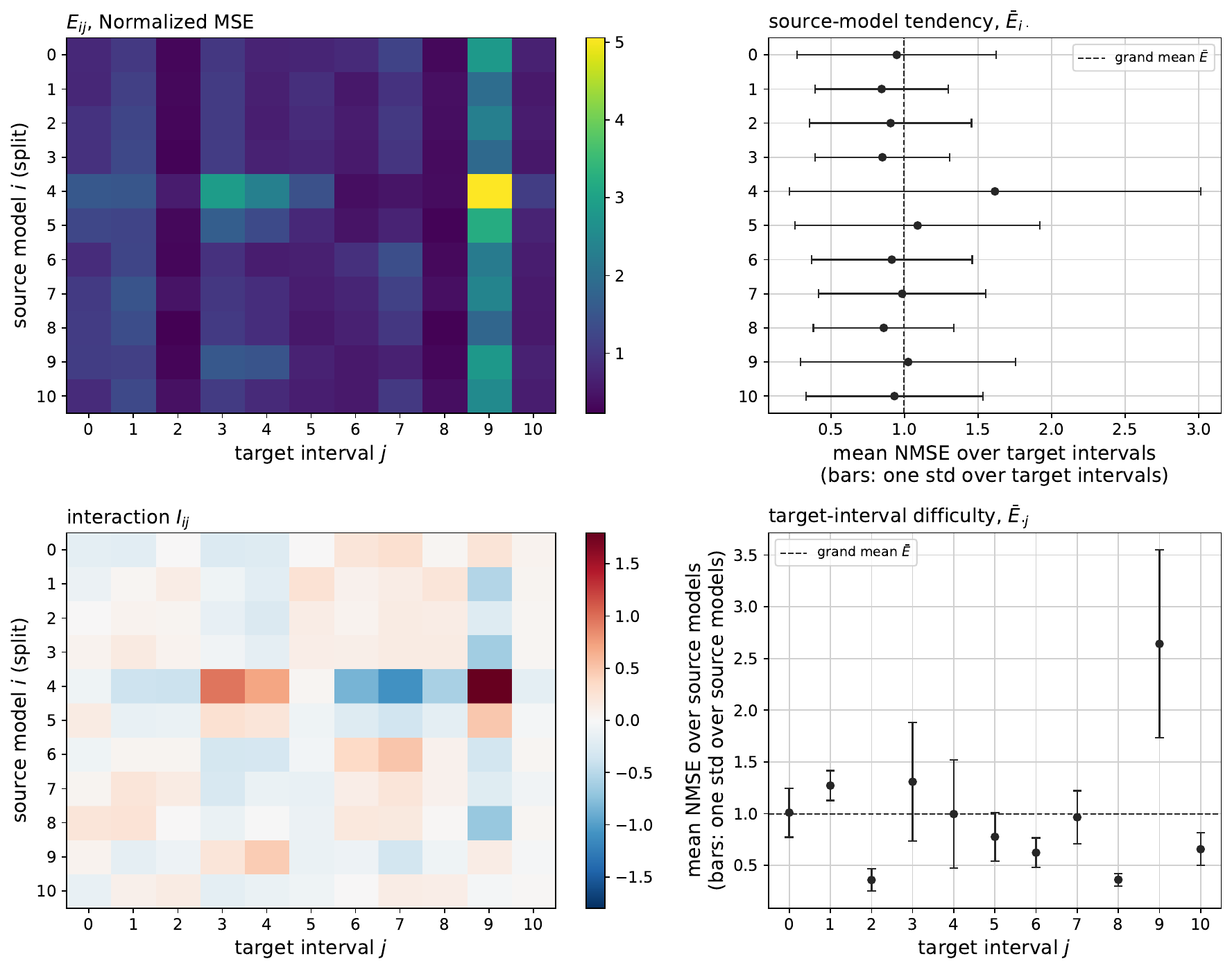}
    \put(6.3,75.2){\makebox(0,0)[lt]{\panel{a}}}
    \put(63.1,75.2){\makebox(0,0)[lt]{\panel{b}}}
    \put(6.3,35.1){\makebox(0,0)[lt]{\panel{c}}}
    \put(98.6,35.1){\makebox(0,0)[rt]{\panel{d}}}
  \end{overpic}
  \caption{Decomposition of the depth-averaged NMSE of MS horizontal divergence for models trained on \caseHH{} and deployed to \caseLH{}: (a) NMSE $E_{ij}$ of source model $i$ on target interval $j$; (b) source-model tendency, the mean over target intervals $\bar E_{i\cdot}$ (points) with one standard deviation over target intervals (bars) around the grand mean $\bar E$ (dashed); (c) interaction $I_{ij}$, the error left after removing the source-model and target-interval means; (d) target-interval difficulty, the mean over source models $\bar E_{\cdot j}$ with one standard deviation over source models, around $\bar E$.}
  \label{fig:fig11_decomposition_example}
\end{figure}

\begin{figure}
    \centering
    \begin{overpic}[width=\textwidth]{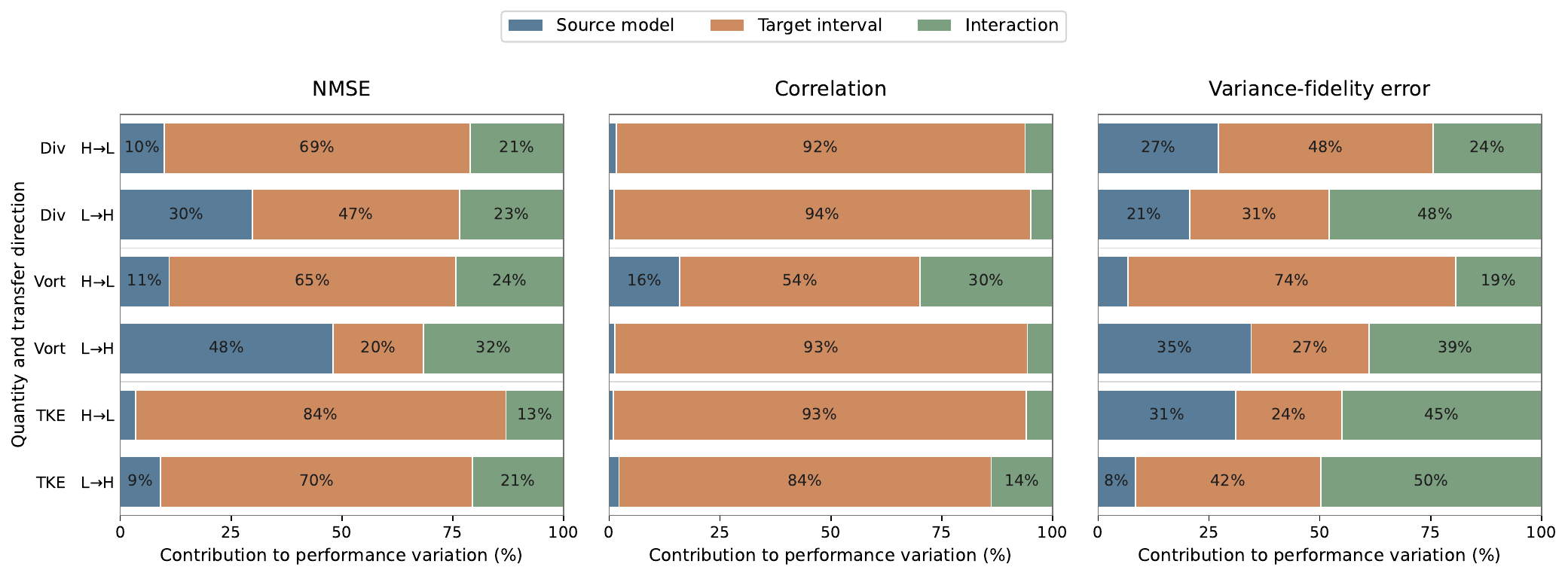}
    \put(7.2,32.2){\makebox(0,0)[lt]{\panel{a}}}
    \put(38.3,32.2){\makebox(0,0)[lt]{\panel{b}}}
    \put(69.8,32.2){\makebox(0,0)[lt]{\panel{c}}}
    \end{overpic}
    \caption{Fractions of across-case performance variation associated with the source model, target interval, and their interaction for (a) NMSE, (b) temporal correlation, and (c) variance error. Fractions expressed per transfer direction (H$\to$L: trained on \caseHH{}, deployed to $\mathrm{Re}_L$ flows; L$\to$H: trained on \caseLH{}, deployed to $\mathrm{Re}_H$ flows) and output statistic (Div: MS horizontal divergence; Vort: MS vertical vorticity, TKE: mean TKE).}
    \label{fig:contrib_to_variations}
\end{figure}

Figure~\ref{fig:fig11_decomposition_example} illustrates this
decomposition of the NMSE-based error for MS horizontal divergence, using models trained on \caseHH{}
and deployed to \caseLH{}. From this decomposition we observe the difficulty of reconstructing specific intervals, resulting in large NMSE for specific targets. The fractional influence of each term is found by the sum of squares of $M_i, T_j$ and $I_{ij}$ (Eq.~\eqref{eq:ss_components}) and plotted in Fig.~\ref{fig:contrib_to_variations}. 
For this specific case, models trained on \caseHH{} and deployed on \caseLH{}, the source-model accounts for 9\% of the total variation, the target-interval for 72\% and their interaction for 19\%, with one interval (interval 9) being the hardest reconstruction task for every source-model (mean NMSE 2.6, against 0.36--1.31 for the other intervals).

Figure \ref{fig:contrib_to_variations}  shows that the balance between source-model, target-interval, and interaction contributions depends on the reconstruction metric, output quantity and transfer direction. Target intervals account for the largest share of performance variation in 13 of the 18 combinations of metric, quantity and transfer direction, including all temporal-correlation comparisons where their contributions range from 54\% to 94\%. 
For NMSE and variance-fidelity, the contributions are more mixed. Source-model differences are particularly important for low-to-high Reynolds-number transfer of MS vertical vorticity, accounting for 48\% of NMSE variation and 35\% of variance-fidelity variation. For low-to-high transfer of MS horizontal divergence, source models also contribute substantially to NMSE variation (30\%, against 47\% from target intervals). Interactions account for the largest share of variance-fidelity variation for low-to-high transfer of MS horizontal divergence and for mean TKE in both transfer directions, and approximately equal to the source-model share for vorticity---this interaction term is the part of the variation not explained by isolating the source-model and target-interval effects. 

\begin{figure}
    \centering
    \begin{overpic}[
        width=0.85\linewidth,
        trim={0bp 20bp 0bp 0bp},
        clip, htb!
    ]{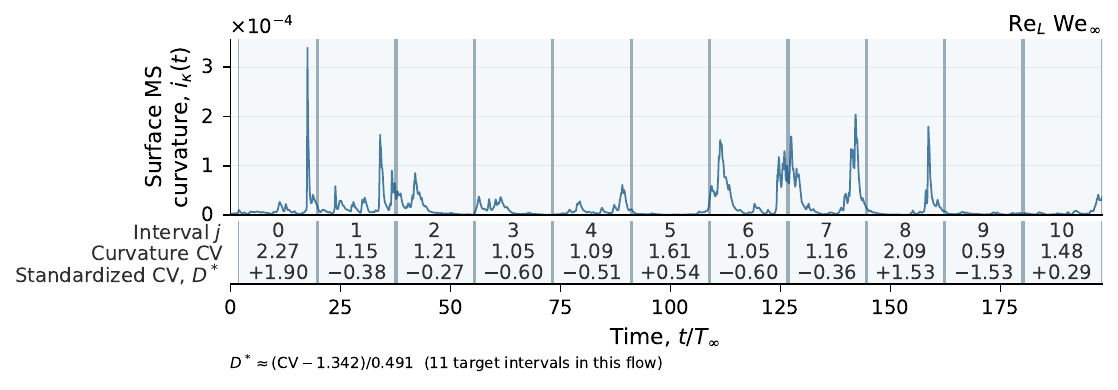}
        \put(10,31){\makebox(0,0)[lt]{\panel{a}}}
    \end{overpic}

    \vspace{0.5em}
    \begin{overpic}[width=\linewidth]
        {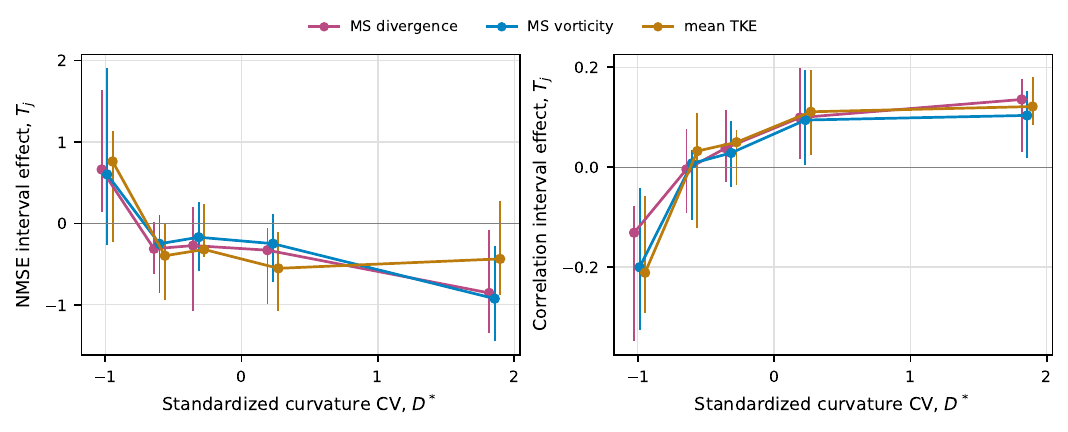}
        \put(11,34.1){\makebox(0,0)[lt]{\panel{b}}}
        \put(60,34.1){\makebox(0,0)[lt]{\panel{c}}}
    \end{overpic}
    \caption{%
        (a) Surface mean-square curvature $i_\kappa(t)$ for the
        $\mathrm{Re}_L\mathrm{We}_\infty$ target flow.
        The rows beneath the trace give each interval's index $j$,
        curvature CV, and standardized curvature CV $D^*$.
        Standardization uses the mean and standard deviation of the 11
        intervals of this flow.
        (b,c) Target-interval effect $T_j$ of SHRED on (b) depth-averaged NMSE and (c) depth-averaged correlation $\rho$, against $D^*$ for MS horizontal divergence, MS vertical vorticity and mean TKE. $D^*$ is split into quintiles, with markers and bars denoting quintile medians and interquartile ranges, respectively. Positive $T_j$ indicates above-average NMSE in (b), hence a worse reconstruction, but above-average correlation in (c), hence better temporal agreement.%
    }
    \label{fig:target_CV}
\end{figure}

Because target intervals account for the largest share of performance variation in all six temporal-correlation comparisons (54–94\%), we next ask whether surface observations can help identify which intervals are easier to reconstruct. We therefore investigate target influence by relating the target-interval effect $T_j$ to the statistical descriptors in \S\ref{sec:data_description} (with per interval descriptors described in \S\ref{sec:zero_shot_decomp_method}), where we find the clearest association with the coefficient of variation ($\mathrm{CV}=\sigma / \mu$) of the intervals. 
Here, we consider the coefficient of variation, CV, of the MS surface curvature, which is available from surface observations during deployment. When computed per interval, a high CV indicates that MS surface curvature fluctuates strongly relative to its interval mean, while a low CV is associated with relatively weak variations from the mean. 
%a high CV indicates intervals where large, intermittent events dominate, while a low CV is associated with relatively calm intervals. 
Since CV differs between datasets (see Table~\ref{tab:global_dataset_statistics}), we standardize it as $D^*=(\mathrm{CV}-\langle\mathrm{CV}\rangle)/\sigma_{\mathrm{CV}}$,
where the mean and standard deviation are taken over the eleven intervals of each dataset. Thus, $D^*=0$ corresponds to the dataset's mean CV, while $D^*=1$ corresponds to a CV one standard deviation above that mean. Figure~\ref{fig:target_CV}(a) illustrates this standardization for the $\mathrm{Re}_L\mathrm{We}_\infty$ target flow considered in Fig.~\ref{fig:fig11_decomposition_example}. As an example, interval~9 has the lowest curvature CV ($\mathrm{CV}=0.59$, $D^*=-1.53$). This is also the interval with the largest source-model-averaged NMSE for MS horizontal divergence in Fig.~12(d). Figure~\ref{fig:target_CV}(b,c) relates $D^*$ to the target-interval effect $T_j$ on depth-averaged NMSE and temporal correlation, pooling the 66 target intervals across all six flows.%Figure \ref{fig:target_CV} shows the target-interval effect $T_j$ for depth-averaged NMSE and temporal correlation $\rho$ for all three output statistics, pooling across-case results from models trained on \caseLH{} and \caseHH{}.

Intervals with higher MS surface-curvature CV generally show more favorable temporal correlation in reconstructions for all three quantities, with the association leveling off at higher CV. It is an intuitive result, as intervals with more turbulent surface activity contain more information that SHRED can use for the learned mapping to the subsurface, than intervals with comparatively few turbulent events. 
The NMSE interval effect also generally decreases with increasing curvature CV, although the trend is less regular. These results suggest that surface-curvature CV may help indicate which intervals are more favorable for reconstruction, particularly in terms of temporal agreement. Nevertheless, intervals with similar curvature CV can still differ considerably in reconstruction performance, even while trends persist.

We also examined whether source-model effects were associated with the statistical properties of the source data, and whether interaction effects were associated with differences between source and target statistics. These associations varied across quantities and transfer configurations, and no consistent explanatory relationship emerged. Moreover, models that performed better on average also tended to perform more consistently across different target flows. This tendency was clearest for horizontal divergence and vertical vorticity when transferring from lower to higher Reynolds numbers, but was not observed in every case. The statistical properties of the source data that we examined did not explain why some models performed better and more consistently than others.

%\label{sec:zero_shot_characteristics}
\FloatBarrier

%%%%%%%%%%%%%%%%%%%%%%%%%%%%%%%%%%%%%%%%%%%%%%%
%%%%%%%%%%% DA-SHRED Results %%%%%%%%%%%%%%%%%%
%%%%%%%%%%%%%%%%%%%%%%%%%%%%%%%%%%%%%%%%%%%%%%%
\subsection{Deployment across Reynolds and Weber numbers with data-assimilation SHRED}% for cross-fltestow corrections}
\label{sec:da_results}

The across-case results, particularly those from models trained on the high Reynolds number dataset \caseHH{}, show substantial reconstruction skill across flow regimes with temporal correlations between prediction and truth exceeding $0.8$ near-surface, yet they are sensitive to the target interval, source model and depth. We now examine whether data assimilation at inference time can 
improve accuracy of the deployed models. 
We deploy the same pretrained models used in the preceding section, to target intervals from the other datasets, but now with DA-SHRED rather than SHRED (see \S\ref{sec:da_shred_theoretical} for the difference between the two). %We limit this part of the study to models pretrained on dataset \caseHH{}, deployed to the other five datasets (ten intervals for each dataset). 
Before deploying the models with DA-SHRED, we perform short hyperparameter tuning runs to select the appropriate learning rate, regularization weight $\lambda_{DA}$ and state carry, where tuning is performed strictly on the in-case validation data (details on the hyperparameter tuning are found in App.~\ref{app:da_shred}). In short, we use learning rate $10^{-3}$, $\lambda_{DA} = 100$ and do not carry the corrected state to the next observation window. The two latter choices have the effect of only allowing small correction from the data assimilation to the pre-trained SHRED model. 

\begin{table}%[htbp]
  \centering\footnotesize\setlength{\tabcolsep}{4pt}
  \caption{
  Gain in NMSE from data assimilation, $G_{\rm NMSE}=\mathrm{NMSE}_{\rm SHRED}-\mathrm{NMSE}_{\rm DA}$, per deployment (one source model applied to one target interval), averaged over depth with each plane weighted by the depth interval it represents. Each entry gives the median and (mean) over deployments and the percentage of deployments with $G_{\rm NMSE}>0$.} %The number of deployments is
%220 for $\mathrm{Re_H}\to\mathrm{Re_H}$ and $\mathrm{Re_L}\to\mathrm{Re_L}$, and %330 for transfer
%between Reynolds numbers} 
  \label{tab:da_summary}
\begin{tabular}{lcccccc}
\hline
 & \multicolumn{2}{c}{MS horizontal divergence} & \multicolumn{2}{c}{MS vertical vorticity} & \multicolumn{2}{c}{Mean TKE}\\
\cline{2-3}\cline{4-5}\cline{6-7}
Transfer & median (mean) & $G>0$ & median (mean) & $G>0$ & median (mean) & $G>0$ \\
\hline
$\mathrm{Re}_H\to\mathrm{Re}_H$ & 0.001 (0.005) & 55\% & 0.003 (0.006) & 61\% & 0.002 (0.007) & 66\% \\
$\mathrm{Re}_H\to\mathrm{Re}_L$ & 0.002 (0.005) & 57\% & 0.005 (0.008) & 63\% & 0.004 (0.009) & 67\% \\
$\mathrm{Re}_L\to\mathrm{Re}_H$ & 0.036 (0.162) & 77\% & 0.042 (0.130) & 82\% & 0.009 (0.038) & 71\% \\
$\mathrm{Re}_L\to\mathrm{Re}_L$ & 0.005 (0.044) & 63\% & 0.011 (0.045) & 71\% & 0.007 (0.021) & 75\% \\
\hline
\end{tabular}
\end{table}

\begin{figure}[htbp]
  \centering
  \includegraphics[width=\textwidth]{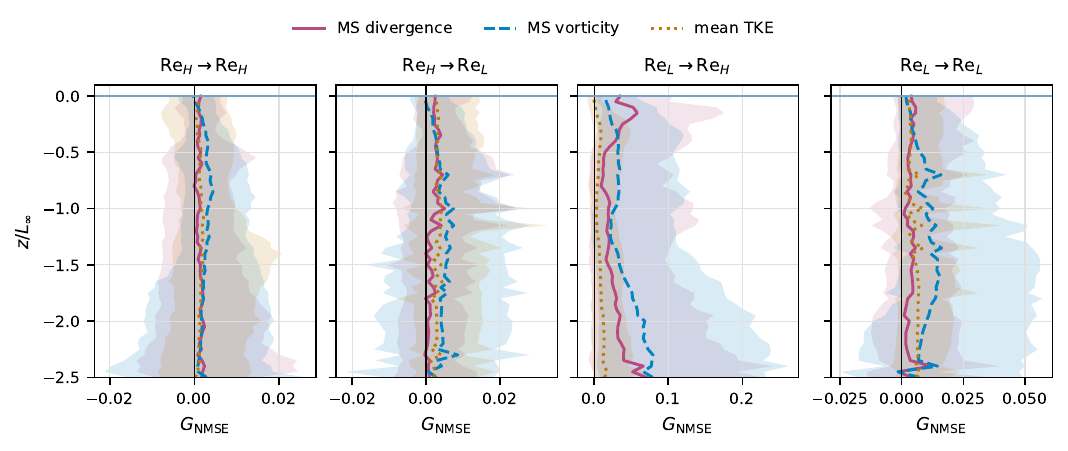}
  \caption{Depth profiles of $G_{\mathrm{NMSE}}$ for models trained on \caseHH{} or \caseLH{} and deployed to high ($\mathrm{Re}_H$) and low ($\mathrm{Re}_L$) Reynolds number flows. Lines are medians over deployments; shading spans the interquartile range.}
  \label{fig:da_profiles}
\end{figure}

Table \ref{tab:da_summary} and Fig.~\ref{fig:da_profiles} summarize the effect of DA-SHRED on the NMSE of the reconstructed statistics, with tabulated values averaged across depth while the plotted data shows the depth-dependence. All gains are computed such that positive means a favorable effect from the data assimilation. For all three quantities, DA-SHRED lowers the depth-averaged NMSE in a majority of deployments in each transfer configuration, and displays no systematic depth-dependent effect. For models trained on dataset \caseHH{} improvements are minor, with median gains of at most 0.005, improving 55\%--67\% of deployments. % below 1\% of the NMSE for SHRED reconstructions and generally in the region of 0.1--0.5\%. 

The clearest gains occur for models trained on \caseLH{} and deployed to the high Reynolds number datasets. %, particularly for the deployment to the high Reynolds number datasets, where SHRED performance was least accurate. 
For those cases,  DA-SHRED lowers the NMSE in 77\% (MS horizontal divergence) and 82\% (MS vertical vorticity) of the deployments, with median gains of 0.036 and 0.042. The corresponding mean gains, 0.162 and 0.130, respectively, are larger, indicating that some deployments benefit considerably more than the medians suggest. Approximately 80--85\% of the gain stems from the offset term, while improvements to the temporal correlations are minor
---the effect of the data assimilation is thus not an improvement of the timing, but a shift of the mean towards the true mean.

\FloatBarrier

%%%%%%%%%%%%%%%%%%%%%%%%%%%%%%%%%%%%%%%%%%%%%%%%%%%%%%%%%%%%%%%%%%%%
%%%%%%%%%%%%%% C O N C L U S I O N  %%%%%%%%%%%%%%%%%%%%%%%%%%%%%%%%
%%%%%%%%%%%%%%%%%%%%%%%%%%%%%%%%%%%%%%%%%%%%%%%%%%%%%%%%%%%%%%%%%%%%

\section{Conclusion}
\label{sec:conclusions}
%%%%%%%%%%%%%%%%%%%%%%%%%%%%%%%%%%%%%%%%%%%%%%%%%%%%%%%%%%%%%%%%%%%%
%%%%%%%%%%%%%% C O N C L U S I O N  %%%%%%%%%%%%%%%%%%%%%%%%%%%%%%%%
%%%%%%%%%%%%%%%%%%%%%%%%%%%%%%%%%%%%%%%%%%%%%%%%%%%%%%%%%%%%%%%%%%%%

%What has been done

This study investigated whether the lightweight surface-based SHRED models can reconstruct plane-averaged turbulence statistics at and below a free surface, and how the learned surface-to-subsurface mapping generalizes across different Reynolds numbers ($\Reynolds$) and Weber numbers ($\Weber$). We evaluated reconstruction from the surface down past two integral length scales below the surface, using six direct numerical simulations that span two Reynolds numbers and three Weber numbers.

% overall turbulence statistics reconstruction performance

SHRED accurately reconstructs mean-square (MS) horizontal divergence and MS vertical vorticity from MS surface curvature and the area fraction of detected surface structures, achieving particularly high accuracy near the surface. Mean TKE shows somewhat weaker performance: in both reference flows, its temporal correlation is lower and its NMSE varies more across test intervals. The distinct surface signatures of upwellings and surface-attached vortices may help explain why divergence and vorticity are more readily reconstructed from models using only surface observations as input. The performance holds for in-case testing with test data temporally separated from the training and validation data, and for across-case deployment to different flow conditions. Compared to ridge regression, as a performance baseline, SHRED reconstruction accuracy was comparable or better. Both methods retain the same temporal correlation, with SHRED having a lower offset from the true statistics and hence reaching slightly lower NMSE at depth.
In general, reconstruction performance deteriorates with depth, as is expected, since turbulent structures further beneath the surface will have weaker surface signatures. Time-depth analysis shows that SHRED does not fully recover the temporal lead of deeper turbulent events relative to their later surface manifestations. This behavior is consistent with weaker surface--subsurface coupling at depth, although the present analysis does not isolate physical observability from model limitations.

% zero-shot as a step towards deployment reality
The models generalize well across the different datasets, for the limited number of datasets and Reynolds- and Weber-number ranges tested here. The temporal correlation and the depth dependence remained comparable to that of the in-case reconstructions, while the NMSE generally increased relative to in-case reconstruction on the same target intervals, mainly due to the offset between the reconstructed and true mean. The size of the offset depended on the pair of flows and on the output calibration: in the main analysis, we scaled each predicted statistic using its observed range in the new flow outside the test interval. With calibration, transfer from lower to higher Reynolds number increased NMSE more than the reverse in the reference cases, but this pattern did not hold for every pair of flows tested.
Using training scale ranges from the source flow instead left temporal correlation unchanged and had little effect on NMSE at fixed Reynolds number, but increased the offsets substantially when transferring between Reynolds numbers. A separate test suggested an alternative: rescaling the training ranges using bulk turbulence properties gave median NMSE close to that obtained with the target-flow calibration.  
The effect of source-model (the specific interval of the source dataset that a model was trained on) and target interval (the specific interval of the dataset the pretrained model was deployed to) were investigated in detail, showing that the target-interval had the largest effect on performance variation of the reconstructed statistics. In particular, target-intervals with little variation in the surface curvature (low CV) proved a challenge for the models, likely due to the models depend solely on information from surface curvature and flow structures visible at the surface at inference time.

% data assimilation at inference to correct near-surface dynamics
To further improve the performance of the SHRED models, we apply data assimilation (DA-SHRED) to correct the latent state at inference using surface observations while keeping the trained network parameters fixed. With the correction settings chosen using validation data from the training flow, DA-SHRED reduced depth-averaged NMSE in a majority of deployments, although the typical gain was small. The clearest gains were for MS horizontal divergence and MS vertical vorticity when models trained on lower Reynolds number flow and deployed to higher Reynolds number flow. In those cases, the correction mainly reduced the offset between the reconstructed and true interval means, while temporal correlation changed little.

Taken together, the results show that surface observations can be used to track how plane-averaged turbulence statistics evolve at different depths. We demonstrated this on held-out intervals of the training flow and, with calibration, in other flows across the Reynolds and Weber numbers studied. 
Nevertheless, we identify three main limitations of the present work, which should be addressed in future research. First, the calibration used in the main across-case results requires prior information about the range of the subsurface statistics in the new flow. Bulk turbulence scales offer an alternative, but may be difficult to estimate from surface observations alone. 
Second, the across-case reconstructions cover a limited parameter space. In particular, the Reynolds numbers in the DNS are limited, and extending the investigation closer to real-world flow conditions is a key next step. Third, we have assumed accurate access to derived surface quantities. Particularly measuring the MS curvature may pose a challenge in practice. Once the two surface signals are available, a trained model can update the full depth profile of its target statistic in a single evaluation. This could allow subsurface turbulence statistics to be monitored in real time, without waiting for the full measurement record.
\FloatBarrier

%%%%%%%%%%%%%%%%%%%%%%%%%%%%%%%%%%%%%%%%%%%%%%%%%%%%%%%%%%%%%%%%%%%%
%%%%%%%%%%%%%% Data availability & acknowledgements  %%%%%%%%%%%%%%%
%%%%%%%%%%%%%%%%%%%%%%%%%%%%%%%%%%%%%%%%%%%%%%%%%%%%%%%%%%%%%%%%%%%%

\section*{Acknowledgements}

DNS data was generously shared with us by Prof.\ Lian Shen and Dr.\ Anqing Xuan at the University of Minnesota. 
The work of JRA, KSM and S\AA E was co-funded by the Research Council of Norway (\emph{iMOD}, grant 325114) and the European Union (ERC CoG, \emph{WaTurSheD}, grant 101045299). 
Views and opinions expressed are however those of the authors only and do not necessarily reflect those of the European Union or the European Research Council. Neither the European Union nor the granting authority can be held responsible for them.
The work of JNK was supported in part by the US National Science Foundation (NSF) AI Institute for Dynamical Systems (dynamicsai.org), grant 2112085. JNK further acknowledges support from the Air Force Office of Scientific Research  (FA9550-24-1-0141).

KSM designed the study. KSM and JRA performed the data analysis and wrote the first draft of the manuscript. The work was jointly supervised by JRA and S{\AA}E. All authors contributed to concept development, discussions and writing of the final manuscript.

\section*{Data availability}
No new direct numerical simulations were performed for this study. The DNS data were generated for Ref.~\cite{aarnes2025} and are available, with supporting codes, from \cite{aarnes2025b}. The code to reproduce all results will be made available on Github by the time of publication.

\section*{AI use disclosure}
We used Claude (Anthropic, Opus 5 and Opus 5.5) and OpenAI's GPT-5.6 and GPT-6 Astra to assist with coding and data analysis. We conceived the study and wrote the manuscript, using language models to polish the text. We thoroughly inspected any AI-assisted analysis to verify the legitimacy of the results. We take full responsibility for the content of the manuscript.

%%%%%%%%%%%%%%%%%%%%%%%%%%%%%%%%%%%%%%%%%%%
%%%%%%%%%%%%%% A P P E N D I X %%%%%%%%%%%%
%%%%%%%%%%%%%%%%%%%%%%%%%%%%%%%%%%%%%%%%%%%

\appendix

\counterwithin{figure}{section}
\counterwithin{table}{section}

\renewcommand{\thefigure}{\thesection\arabic{figure}}
\renewcommand{\thetable}{\thesection\arabic{table}}

\section{Additional results for in-case reconstruction}

\subsection{In-case reconstruction performance}
\label{app:in-case-reconstruction}
\begin{figure}[tbp]
  \centering
  \resizebox{0.95\textwidth}{!}{%
  \begin{overpic}[width=\textwidth]{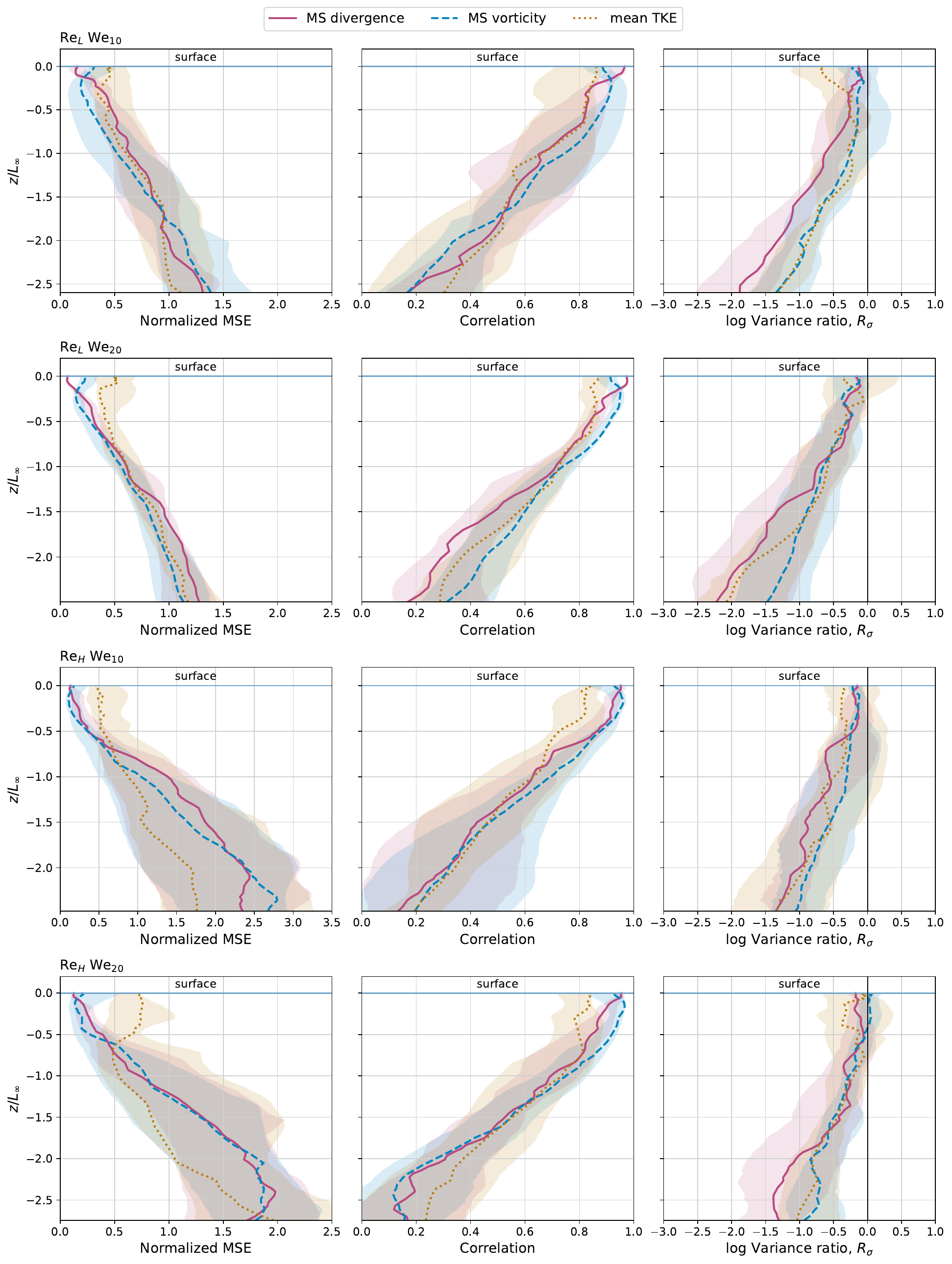}
    \put(5.3,95.5){\makebox(0,0)[lt]{\panel{a}}}
    \put(29.1,95.5){\makebox(0,0)[lt]{\panel{b}}}
    \put(53.0,95.5){\makebox(0,0)[lt]{\panel{c}}}
    \put(5.3,71.1){\makebox(0,0)[lt]{\panel{d}}}
    \put(29.1,71.1){\makebox(0,0)[lt]{\panel{e}}}
    \put(53.0,71.1){\makebox(0,0)[lt]{\panel{f}}}
    \put(5.3,46.7){\makebox(0,0)[lt]{\panel{g}}}
    \put(29.1,46.7){\makebox(0,0)[lt]{\panel{h}}}
    \put(53.0,46.7){\makebox(0,0)[lt]{\panel{i}}}
    \put(5.3,22.3){\makebox(0,0)[lt]{\panel{j}}}
    \put(29.1,22.3){\makebox(0,0)[lt]{\panel{k}}}
    \put(53.0,22.3){\makebox(0,0)[lt]{\panel{l}}}
  \end{overpic}
  }
  \caption{As Fig.~\ref{fig:fig05_incase_depth_profiles}, for the flows with finite Weber number: depth-resolved in-case reconstruction performance for (a--c) $\mathrm{Re}_L\,\mathrm{We}_{10}$, (d--f) $\mathrm{Re}_L\,\mathrm{We}_{20}$, (g--i) $\mathrm{Re}_H\,\mathrm{We}_{10}$ and (j--l) $\mathrm{Re}_H\,\mathrm{We}_{20}$.}%, for MS divergence (solid), MS vertical vorticity (dashed) and mean TKE (dotted). Columns show NMSE, the temporal correlation $\rho$ and the log variance ratio $R_\sigma$. Lines are medians over the 11 data splits of the per-split mean over 10 training realisations; shading spans the interquartile range over splits. }
  \label{fig:figA1_incase_depth_profiles_finite_we}
\end{figure}

We presented the in-case reconstruction performance in \S\ref{sec:baseline_results}, using datasets \caseHH{} and \caseLH{} as reference cases. Figure \ref{fig:figA1_incase_depth_profiles_finite_we} displays the in-case reconstruction performance for the four datasets with finite Weber number, detailed in Table \ref{tab:flowProp}. We observe that the trends match those for infinite Weber number.

\subsection{Hovm\"{o}ller diagrams for vertical vorticity and TKE}

\label{app:Hovmöller}
\begin{figure}[htb!]
  \centering
  \begin{overpic}[width=\textwidth]{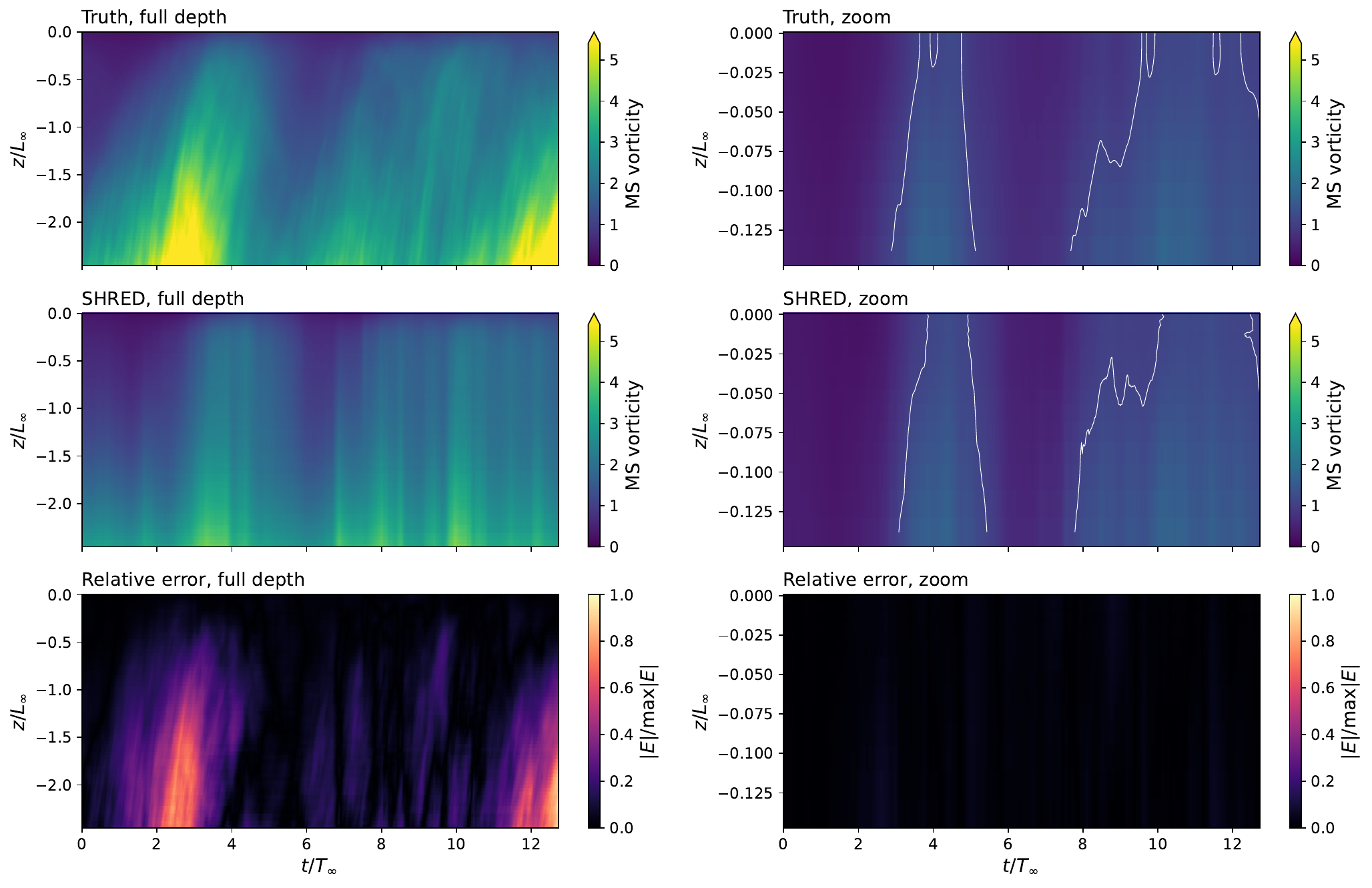}
    \put(6.5,61.7){\makebox(0,0)[lt]{\panel{a}}}
    \put(57.6,61.7){\makebox(0,0)[lt]{\panel{b}}}
    \put(6.5,41.2){\makebox(0,0)[lt]{\panel{c}}}
    \put(57.6,41.2){\makebox(0,0)[lt]{\panel{d}}}
    \put(6.5,20.7){\makebox(0,0)[lt]{\panel{e}}}
    \put(57.6,20.7){\makebox(0,0)[lt]{\panel{f}}}
  \end{overpic}
  \caption{As Fig.~\ref{fig:hovmoller_RE2500_div}, for MS vertical vorticity: time--depth diagrams of (a, b) true and (c, d) reconstructed MS vertical vorticity of \caseHH{} for the representative test interval, full depth (left) and $z/L_\infty\geq-0.15$ (right), with (e, f) the absolute error relative to its maximum.}
  \label{fig:figA2_timedepth_vorticity}
\end{figure}

\begin{figure}[htb!]
  \centering
  \begin{overpic}[width=\textwidth]{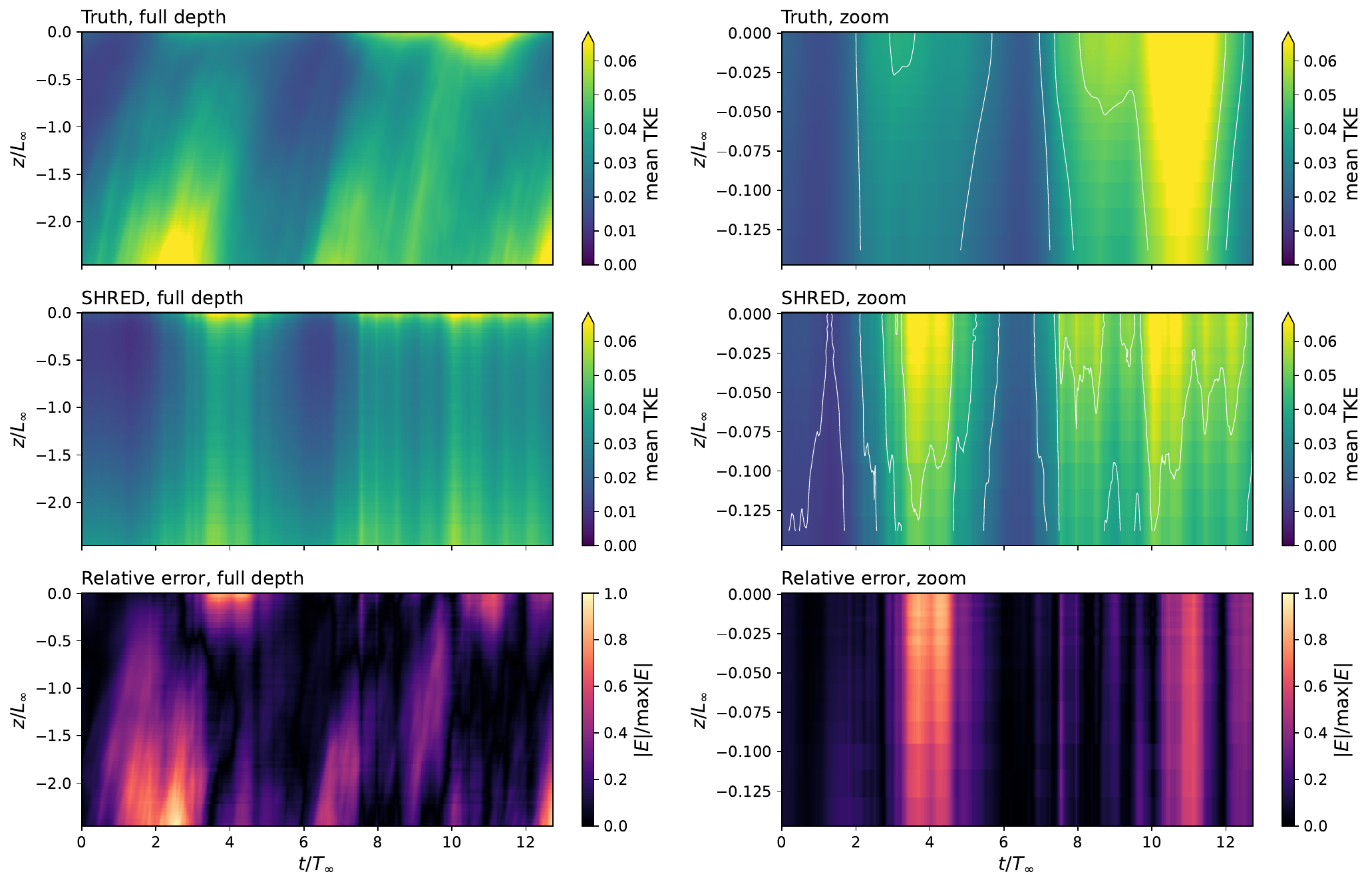}
    \put(6.5,61.6){\makebox(0,0)[lt]{\panel{a}}}
    \put(57.6,61.6){\makebox(0,0)[lt]{\panel{b}}}
    \put(6.5,41.1){\makebox(0,0)[lt]{\panel{c}}}
    \put(57.6,41.1){\makebox(0,0)[lt]{\panel{d}}}
    \put(6.5,20.7){\makebox(0,0)[lt]{\panel{e}}}
    \put(57.6,20.7){\makebox(0,0)[lt]{\panel{f}}}
  \end{overpic}
  \caption{As Fig.~\ref{fig:hovmoller_RE2500_div} and \ref{fig:figA2_timedepth_vorticity}, for mean TKE: time--depth diagrams of (a, b) true and (c, d) reconstructed mean TKE of $\mathrm{Re}_H\,\mathrm{We}_\infty$ for the representative test interval, full depth (left) and $z/L_\infty\geq-0.15$ (right), with (e, f) the absolute error relative to its maximum, $|E|/\max|E|$.}
  \label{fig:figA3_timedepth_mean_tke}
\end{figure}

We analyzed the time--depth (Hovm\"{o}ller) diagram of true and reconstructed MS horizontal divergence in \S\ref{sec:baseline_results}, Fig.~\ref{fig:hovmoller_RE2500_div}. Here we extend this analysis to vertical vorticity and TKE for the same interval we investigated previously. 

Figures \ref{fig:figA2_timedepth_vorticity} and \ref{fig:figA3_timedepth_mean_tke} display the results for MS vertical vorticity and mean TKE, respectively. The color scale in the top two rows of both figures saturates at 80\,\% of the maximum true value to emphasize structures throughout the depth. The Hovm\"{o}ller diagram for MS vertical vorticity shows a structural resemblance to the equivalent diagram for MS horizontal divergence. In the true signal, fluctuations at depth precede those at the surface, whereas in the reconstruction they appear more nearly simultaneous. For this interval, strong near-surface events are less evident in the MS vertical vorticity diagram than in the MS horizontal divergence diagram. One possible explanation is that dimples associated with vertical-vorticity hotspots occupy only a small fraction of the surface area~\cite{aarnes2025}, making their contribution less apparent in the plane average. %Again, we see that strong surface events are reconstructed as though they were close to instantaneous in time with their concomitant subsurface turbulent events, while in the true signal the surface lags the depth---with the main difference between the MS horizontal divergence and the MS vertical vorticity diagrams being that we do not find obvious evidence of strong vertical vorticity events at the surface in the Hovm\"{o}ller diagram, as we do for the MS horizontal divergence, a likely explanation for which being that the surface signatures of vertical-vorticity hotspots---dimples \cite{aarnes2025}---cover a very small portion of the surface area. 
The Hovm\"{o}ller diagram for mean TKE shows the same overall effect of near instantaneous reconstruction, yet with a more pronounced difference between the true and reconstructed statistics. The latter is particularly visible in the near-surface region, where the reconstructions overshoot on patterns for $t/T_\infty\approx$3--5. %As for MS horizontal divergence and MS vertical vorticity, the subsurface events are reconstructed as occurring simultaneously as the surface events, yet with much larger errors between the true and reconstructed fields. We even observe that in contrast to the other output statistics where the near-surface region had very low errors, the reconstructed mean TKE show some large error patterns in the direct vicinity of the surface.

\FloatBarrier
\section{Additional results for across-case reconstruction}
\label{app:across-case-reconstruction}
\begin{figure}[tbh]
  \centering
  \begin{overpic}[width=\textwidth]{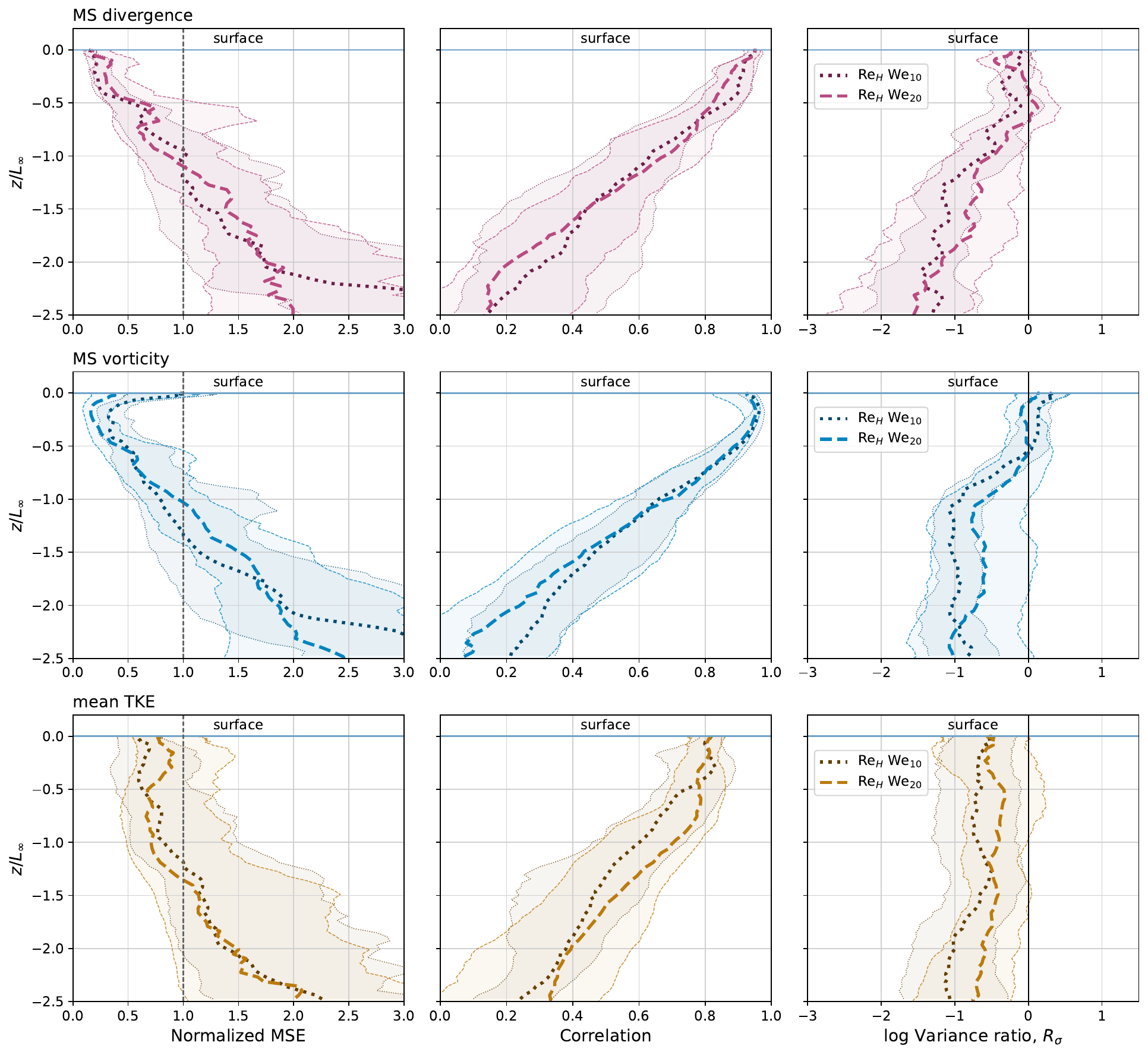}
    \put(6.9,89.1){\makebox(0,0)[lt]{\panel{a}}}
    \put(38.9,89.1){\makebox(0,0)[lt]{\panel{b}}}
    \put(71.0,89.1){\makebox(0,0)[lt]{\panel{c}}}
    \put(6.9,59.2){\makebox(0,0)[lt]{\panel{d}}}
    \put(38.9,59.2){\makebox(0,0)[lt]{\panel{e}}}
    \put(71.0,59.2){\makebox(0,0)[lt]{\panel{f}}}
    \put(6.9,29.3){\makebox(0,0)[lt]{\panel{g}}}
    \put(38.9,29.3){\makebox(0,0)[lt]{\panel{h}}}
    \put(71.0,29.3){\makebox(0,0)[lt]{\panel{i}}}
  \end{overpic}
  \caption{As Fig.~\ref{fig:transfer_high2low_main}, for models trained on \caseHH{} and deployed to datasets \caseHM{} and \caseHL{}, for (a--c) MS horizontal divergence, (d--f) MS vertical vorticity and (g--i) mean TKE; columnwise split into (left) NMSE, (middle) temporal correlation and (right) log variance ratio.}
  \label{fig:figB1_across_same_Re_from_high_Re}
\end{figure}

\begin{figure}[t]
  \centering
  \begin{overpic}[width=\textwidth]{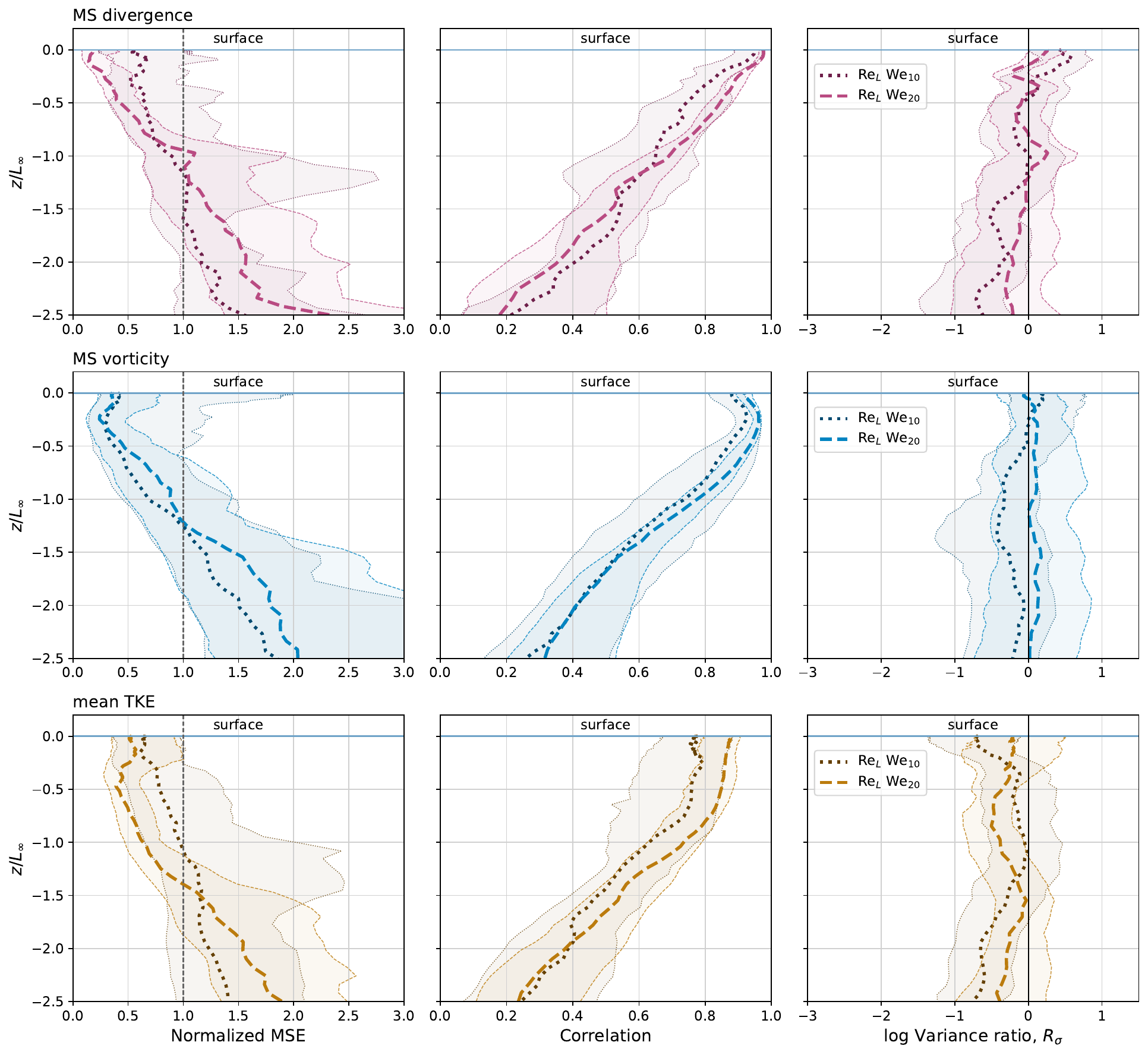}
    \put(6.9,89.1){\makebox(0,0)[lt]{\panel{a}}}
    \put(38.9,89.1){\makebox(0,0)[lt]{\panel{b}}}
    \put(71.0,89.1){\makebox(0,0)[lt]{\panel{c}}}
    \put(6.9,59.2){\makebox(0,0)[lt]{\panel{d}}}
    \put(38.9,59.2){\makebox(0,0)[lt]{\panel{e}}}
    \put(71.0,59.2){\makebox(0,0)[lt]{\panel{f}}}
    \put(6.9,29.3){\makebox(0,0)[lt]{\panel{g}}}
    \put(38.9,29.3){\makebox(0,0)[lt]{\panel{h}}}
    \put(71.0,29.3){\makebox(0,0)[lt]{\panel{i}}}
  \end{overpic}
  \caption{As Fig.~8 and \ref{fig:figB1_across_same_Re_from_high_Re}, for models trained on \caseLH{} and deployed to datasets \caseLM{} and \caseLL{}.}
  \label{fig:figB2_across_same_Re_from_low_Re}
\end{figure}

For the across-case reconstruction results in \S\ref{sec:zero_shot_main}, we presented figures only for models trained on high/low Reynolds number and deployed on datasets low/high Reynolds numbers. Here we include the results for across-case reconstructions for models deployed to flows with the same nominal Reynolds numbers, but with different Weber numbers (note that, when converted to dimensional units for comparison with experiments, the various descriptive Reynolds numbers are not entirely identical, as summarized in Table \ref{tab:flowProp}). Results for the high-Re flows are displayed in Fig.~\ref{fig:figB1_across_same_Re_from_high_Re}; the low-Re flows in Fig.~\ref{fig:figB2_across_same_Re_from_low_Re}.

The extended results for increase in all target statistics for across-case reconstructions compared to the in-case reconstructions can be found in Tables \ref{tab:transfer_msdiv}--\ref{tab:transfer_tke}. In each table, the increase in NMSE for target statistics computed for across-case reconstructions as compared to those for in-case reconstructions are presented, averaged for near the surface ($z\geq-0.5L_\infty$) and at depth ($z<-0.5L_\infty$). When depth-averages are computed, planes are weighted by the depth range they span, to ensure that regions where the grid is more dense are not counted more towards the average.

\begin{table*}[h]
\caption{Increase in NMSE of the MS divergence for across-case reconstruction relative to in-case reconstruction of the same target intervals, for models trained on the dataset in each row (source data) and deployed to the dataset in each column (target data). Each entry gives the value near the surface ($z\geq-0.5L_\infty$) / at depth ($z<-0.5L_\infty$). Values are medians over source models and target intervals of the NMSE averaged over the planes in each depth range, with planes weighted by the depth range they span. The temporal correlation changes by less than 0.05 at 100\% of the depth range.}
\label{tab:transfer_msdiv}
\begin{tabular}{lcccccc}
\hline
 & \multicolumn{6}{c}{Target data}\\
\cline{2-7}
Source data
 & $\mathrm{Re_L We_{10}}$ & $\mathrm{Re_L We_{20}}$ & $\mathrm{Re_L We_{\infty}}$ & $\mathrm{Re_H We_{10}}$ & $\mathrm{Re_H We_{20}}$ & $\mathrm{Re_H We_{\infty}}$ \\
\hline
$\mathrm{Re_L We_{10}}$         & --             & 0.05 / 0.21    & 0.07 / 0.08    & 0.03 / 0.02    & 0.19 / 0.09    & 0.22 / 0.41    \\
$\mathrm{Re_L We_{20}}$         & 0.29 / 0.18    & --             & 0.10 / 0.29    & 0.13 / 0.08    & 0.35 / 0.26    & 0.20 / 0.21    \\
$\mathrm{Re_L We_{\infty}}$     & 0.28 / 0.22    & 0.07 / 0.43    & --             & 0.26 / 0.51    & 0.33 / 0.48    & 0.38 / 0.68    \\
$\mathrm{Re_H We_{10}}$         & 0.14 / 0.17    & 0.06 / 0.18    & 0.10 / 0.12    & --             & 0.16 / 0.07    & 0.07 / 0.38    \\
$\mathrm{Re_H We_{20}}$         & 0.16 / 0.13    & 0.02 / 0.16    & 0.02 / 0.06    & 0.04 / 0.03    & --             & 0.07 / 0.33    \\
$\mathrm{Re_H We_{\infty}}$     & 0.24 / 0.31    & 0.05 / 0.20    & 0.07 / 0.17    & 0.03 / 0.10    & 0.10 / 0.23    & --             \\
\hline
\end{tabular}
\end{table*}

\begin{table*}[htb!]
\caption{Same as Table \ref{tab:transfer_msdiv}, but for MS vertical vorticity. The temporal correlation changes by less than 0.05 at 87\% of the depth range.}
\label{tab:transfer_msvort}
\begin{tabular}{lcccccc}
\hline
 & \multicolumn{6}{c}{Target data}\\
\cline{2-7}
Source data
 & $\mathrm{Re_L We_{10}}$ & $\mathrm{Re_L We_{20}}$ & $\mathrm{Re_L We_{\infty}}$ & $\mathrm{Re_H We_{10}}$ & $\mathrm{Re_H We_{20}}$ & $\mathrm{Re_H We_{\infty}}$ \\
\hline
$\mathrm{Re_L We_{10}}$         & --             & 0.13 / 0.32    & 0.13 / 0.16    & 0.72 / 0.12    & 0.32 / 0.19    & 0.48 / 0.89    \\
$\mathrm{Re_L We_{20}}$         & 0.19 / 0.21    & --             & 0.23 / 0.43    & 0.46 / 0.18    & 0.51 / 0.30    & 0.22 / 0.45    \\
$\mathrm{Re_L We_{\infty}}$     & 0.07 / 0.27    & 0.12 / 0.56    & --             & 0.37 / 0.44    & 0.23 / 0.44    & 0.21 / 0.82    \\
$\mathrm{Re_H We_{10}}$         & 0.17 / 0.08    & 0.17 / 0.15    & 0.28 / 0.15    & --             & 0.16 / 0.01    & 0.14 / 0.43    \\
$\mathrm{Re_H We_{20}}$         & 0.10 / 0.15    & 0.04 / 0.32    & 0.11 / 0.22    & 0.19 / 0.13    & --             & 0.06 / 0.92    \\
$\mathrm{Re_H We_{\infty}}$     & 0.16 / 0.36    & 0.06 / 0.36    & 0.08 / 0.33    & 0.22 / 0.16    & 0.07 / 0.27    & --             \\
\hline
\end{tabular}
\end{table*}

\begin{table*}[htb!]
\caption{Same as Table \ref{tab:transfer_msdiv}, but for mean TKE. The temporal
correlation changes by less than 0.05 at 98\% of the depth range.}
\label{tab:transfer_tke}
\begin{tabular}{lcccccc}
\hline
 & \multicolumn{6}{c}{Target data}\\
\cline{2-7}
Source data
 & $\mathrm{Re_L We_{10}}$ & $\mathrm{Re_L We_{20}}$ & $\mathrm{Re_L We_{\infty}}$ & $\mathrm{Re_H We_{10}}$ & $\mathrm{Re_H We_{20}}$ & $\mathrm{Re_H We_{\infty}}$ \\
\hline
$\mathrm{Re_L We_{10}}$         & --             & 0.00 / 0.22    & 0.01 / 0.02    & 0.06 / $-$0.04 & 0.10 / 0.12    & 0.22 / 0.30    \\
$\mathrm{Re_L We_{20}}$         & 0.23 / 0.22    & --             & 0.06 / 0.16    & 0.12 / $-$0.11 & 0.30 / 0.08    & 0.14 / 0.08    \\
$\mathrm{Re_L We_{\infty}}$     & 0.18 / 0.30    & 0.03 / 0.31    & --             & 0.09 / 0.05    & 0.21 / 0.40    & 0.16 / 0.29    \\
$\mathrm{Re_H We_{10}}$         & 0.43 / 0.39    & 0.03 / 0.16    & 0.09 / 0.08    & --             & 0.17 / 0.46    & 0.17 / 0.32    \\
$\mathrm{Re_H We_{20}}$         & 0.10 / 0.11    & 0.06 / 0.13    & 0.10 / 0.11    & 0.01 / $-$0.12 & --             & 0.19 / 0.24    \\
$\mathrm{Re_H We_{\infty}}$     & 0.14 / 0.18    & $-$0.01 / 0.17 & 0.04 / 0.07    & 0.11 / $-$0.07 & 0.22 / 0.11    & --             \\
\hline
\end{tabular}
\end{table*}
\FloatBarrier

\FloatBarrier
\section{DA-SHRED tuning and extended results}
\label{app:da_shred}
Tables \ref{tab:DA_G_rho} and \ref{tab:DA_G_var} list the reconstruction accuracy gains in temporal correlation and variance fidelity with DA-SHRED compared to the standard SHRED version we have used elsewhere. As for the corresponding values for NMSE gains in Table \ref{tab:da_summary}, the tables list the depth averaged gain with DA-SHRED for models trained on \caseHH{} (top two rows) and \caseLH{} (bottom two rows), per target interval, expressed as the median (mean) over deployments, and the percentage of deployments with positive gain. As we saw for NMSE in \S\ref{sec:da_results}, using DA-SHRED has a small positive impact, largest for the models trained on \caseLH{}.

\begin{table}[]
    \centering
    \caption{Gain in temporal correlation from data assimilation, $G_\rho=\rho_{\rm DA}-\rho_{\rm SHRED}$, in
    units of $10^{-3}$, aggregated as in Table~\ref{tab:da_summary}: median (mean) over deployments and the
    percentage of deployments with $G_\rho>0$.}
    \begin{tabular}{lcccccc}
    \hline
 & \multicolumn{2}{c}{MS divergence} & \multicolumn{2}{c}{MS vorticity} & \multicolumn{2}{c}{Mean TKE}\\
\cline{2-3}\cline{4-5}\cline{6-7}
Transfer & median (mean) & $G>0$ & median (mean) & $G>0$ & median (mean) & $G>0$ \\
\hline
$\mathrm{Re}_H\to\mathrm{Re}_H$ & 0.5 (0.9) & 62\% & 2.2 (3.4) & 74\% & 1.2 (1.5) & 69\% \\
$\mathrm{Re}_H\to\mathrm{Re}_L$ & 0.5 (0.8) & 64\% & 2.8 (5.7) & 80\% & 1.3 (3.5) & 77\% \\
$\mathrm{Re}_L\to\mathrm{Re}_H$ & 0.4 (4.1) & 56\% & 0.9 (1.0) & 64\% & 0.9 (3.7) & 64\% \\
$\mathrm{Re}_L\to\mathrm{Re}_L$ & 0.2 (6.3) & 53\% & 1.4 (2.0) & 70\% & 1.2 (8.2) & 72\% \\
    \hline
    \end{tabular}
    \label{tab:DA_G_rho}
\end{table}

\begin{table}[]
    \centering
    \caption{Same as Table \ref{tab:DA_G_rho}, but for DA-SHRED gains in variance fidelity,  $G_{\rm var}=|R_\sigma|_{\rm SHRED}-|R_\sigma|_{\rm DA}$}
    \begin{tabular}{lcccccc}
    \hline
 & \multicolumn{2}{c}{MS divergence} & \multicolumn{2}{c}{MS vorticity} & \multicolumn{2}{c}{Mean TKE}\\
\cline{2-3}\cline{4-5}\cline{6-7}
Transfer & median (mean) & $G>0$ & median (mean) & $G>0$ & median (mean) & $G>0$ \\
\hline
$\mathrm{Re}_H\to\mathrm{Re}_H$ & $-$1.2 ($-$3.5) & 48\% & 0.5 (3.3) & 54\% & 2.3 (4.9) & 58\% \\
$\mathrm{Re}_H\to\mathrm{Re}_L$ & 0.4 ($-$0.8) & 51\% & 5.2 (17.3) & 65\% & 5.6 (15.5) & 65\% \\
$\mathrm{Re}_L\to\mathrm{Re}_H$ & 1.3 (6.0) & 54\% & 1.9 (7.6) & 55\% & 0.8 (1.7) & 53\% \\
$\mathrm{Re}_L\to\mathrm{Re}_L$ & 1.7 (14.5) & 60\% & 4.8 (10.8) & 62\% & 1.8 (13.6) & 62\% \\
\hline
    \end{tabular}
    \label{tab:DA_G_var}
\end{table}

Before the results reported in Tables \ref{tab:da_summary}, \ref{tab:DA_G_rho} and \ref{tab:DA_G_var} were computed, we performed a sweep over three hyperparameters (learning rate, regularization constant, state carry) for DA-SHRED to select the best settings: we ran every combination of the learning rate $= (10^{-4}, 3 \times 10^{-4}, 10^{-3}, 3 \times 10^{-3}$), the regularization constant $\lambda_{DA} = 
(10^{-4}, 10^{-2}, 1, 10, 100)$ and state carry (on or off; where state carry on means that the DA-SHRED corrected state is carried to the next time step). This hyperparameter selection was done separately for models trained on datasets \caseHH{} and \caseLH{}, with all parameter tuning performed on the in-case validation data. The same hyperparameter sweeps are performed on the test data intervals from the other datasets (11 intervals from each of the five datasets) to assess the influence of the different parameters, yet without letting the test data performance influence our final choice of hyperparameter value. 

\begin{figure}[htb!]
  \centering
  \includegraphics[width=\textwidth]{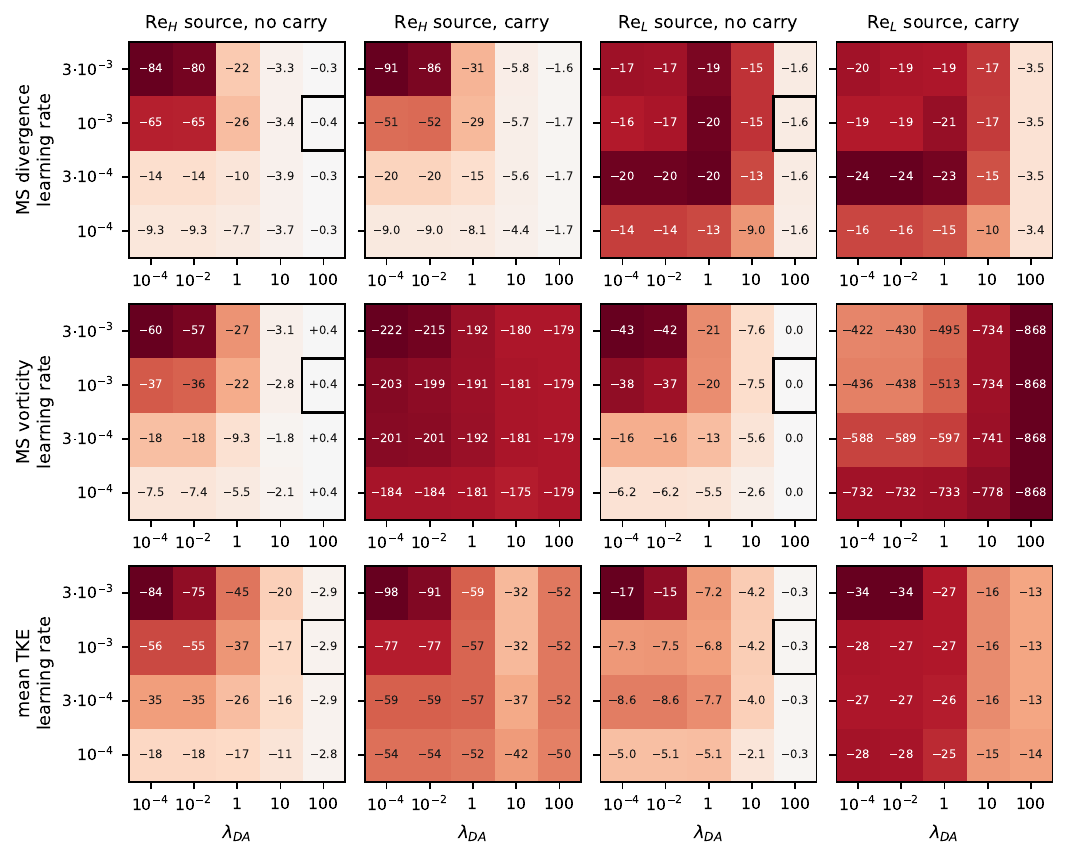}
  \caption{Median depth-averaged $G_{\mathrm{NMSE}}$ against learning rate and $\lambda_{DA}$, on the in-case validation data, with and without state carry, for models trained on \caseHH{} and \caseLH{} (denoted $Re_H$ source and $Re_L$ source, respectively). Numerical values reported as $G_{\mathrm{NMSE}} \times10^3$. Optimal hyperparameter selection marked with black square.}
  \label{fig:app_da_hyperparam}
\end{figure}

\begin{figure}[htb!]
  \centering
  \includegraphics[width=\textwidth]{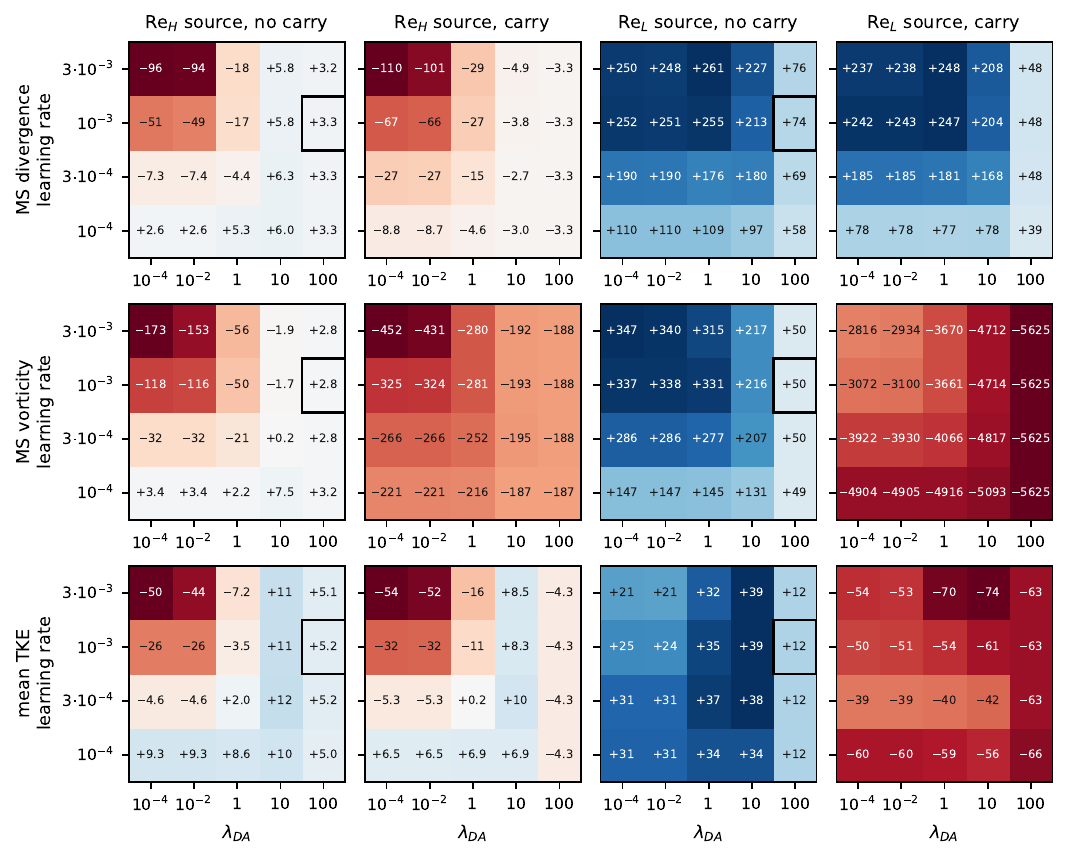}
  \caption{Median depth-averaged $G_{\mathrm{NMSE}}$ against learning rate and $\lambda_{DA}$, on the target data, with and without state carry, for models trained on \caseHH{} and \caseLH{} (denoted $Re_H$ source and $Re_L$ source, respectively). Numerical values reported as $G_{\mathrm{NMSE}} \times10^3$. Hyperparameter selection from validation data marked with black square.}
  \label{fig:app_da_hyperparam_test}
\end{figure}

Figure \ref{fig:app_da_hyperparam} shows the results of the hyperparameter sweep in the validation data; the results of the same hyperparameter sweep in the target intervals are found in Fig.~\ref{fig:app_da_hyperparam_test}. We observe that DA-SHRED generally does not improve the results in the validation data, regardless of parameter combination, and tuning therefore seeks to minimize the impact of DA-SHRED resulting in setting $\lambda_{DA}$ to the upper limit. For the models trained on \caseHH{}, the hyperparameter tuning is, nevertheless, not far off from setting optimal values for the deployment to target intervals (although a smaller learning rate and regularization constant there would improve the latter). For the models trained on \caseLH{}, however, DA-SHRED yields significantly better results on the target intervals with smaller regularization constant value (i.e., where the impact of the data assimilation part on the reconstruction is larger). As this is not known at the time of training and validation, we use the conservative parameter selection (marked with a black square in the figures). 

\begin{figure}[htbp]
  \centering
  \includegraphics[width=\textwidth]{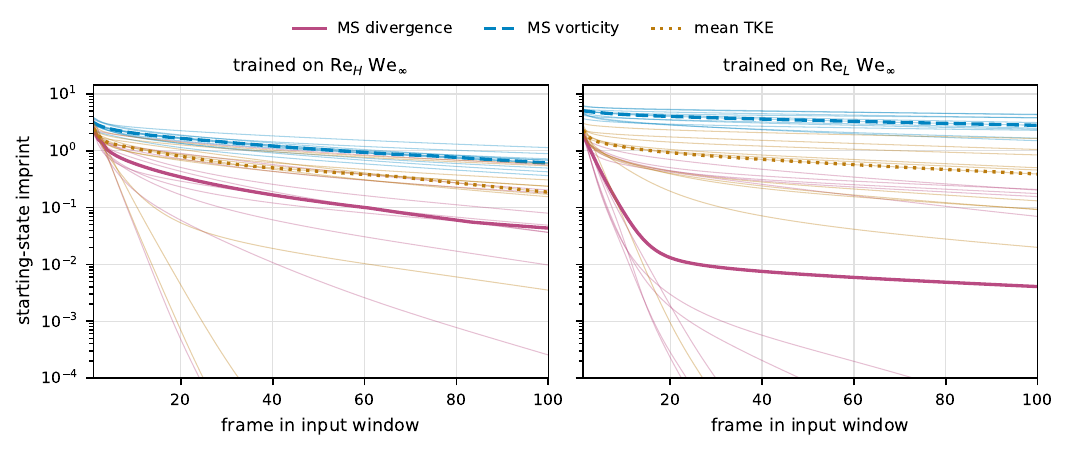}
  \caption{Imprint of the starting state of the GRU along the $n_h=100$ frame input window for models trained on \caseHH{} (left) and \caseLH{} (right): difference between the latent state obtained from a carried state and from a zero state, relative to the variability of the latent state over time. Thin lines: all 11 source models; thick lines: the median source model.}
  \label{fig:da_memory}
\end{figure}

An important result from the hyperparameter sweep is that carrying the state degrades reconstructions regardless of how learning rate and $\lambda_{DA}$ are set. The impact is particularly large for MS vertical vorticity for the models trained on \caseHH{}, and for MS vertical vorticity and mean TKE for models trained on \caseLH{}. During the training stage, the GRU always starts each $n_h=100$ frame input window from a zero state. With state carry on, DA-SHRED starts from the state of the previous time step. Figure~\ref{fig:da_memory} shows the influence of the starting state on each of the target statistics: MS horizontal divergence largely ``forgets'' the starting state after a few frames. For MS vertical vorticity, on the other hand, the starting state has a large influence on the GRU hidden state over the entire input window, and for mean TKE it is in between the two. Hence, any detrimental effect on the reconstructions from carrying the state will have the largest impact on MS vertical vorticity and a smaller effect on MS horizontal divergence, which is what we observed from the hyperparameter-tuning results.

%As reported in \S\ref{sec:zero_shot_main}, models trained on dataset \caseLH{} had high correlation with the true data when deployed to the $Re_H$ flows, yet were unable to accurately capture the magnitude of the target statistics. Deploying these models to higher Reynolds number flows, using DA-SHRED rather than SHRED, improves the reconstructed MS horizontal divergence for almost all target intervals but degrades the MS vertical vorticity significantly. The full results are displayed in Table \ref{tab:app_DA}. We also include a diagnostic variant in which the corrected GRU state is not carried between windows (each window starts from a zero state, as in training). We observe that the MS vertical vorticity error is smaller for this case, but so is the improvement on MS horizontal divergence and TKE.

\FloatBarrier

\bibliographystyle{unsrt}
\bibliography{references}

\end{document}